# On the existence of the relative equilibria of a rigid body in the $J_2$ problem

Yue Wang*, Shijie Xu

*Room B1024, New Main Building, Department of Guidance, Navigation and Control, School of Astronautics, Beijing University of Aeronautics and Astronautics, 100191, Beijing, China*

Liang Tang

*Science and Technology on Space Intelligent Control Laboratory, 100190, Beijing China;*
*Beijing Institute of Control Engineering, 100190, Beijing China*

---

## Abstract

The motion of a point mass in the $J_2$ problem has been generalized to that of a rigid body in a $J_2$ gravity field for new high-precision applications in the celestial mechanics and astrodynamics. Unlike the original $J_2$ problem, the gravitational orbit-rotation coupling of the rigid body is considered in the generalized problem. The existence and properties of both the classical and non-classical relative equilibria of the rigid body are investigated in more details in the present paper based on our previous results. We nondimensionalize the system by the characteristic time and length to make the study more general. Through the study, it is found that the classical relative equilibria can always exist in the real physical situation. Numerical results suggest that the non-classical relative equilibria only can exist in the case of a negative $J_2$, i.e., the central body is elongated; they cannot exist in the case of a positive $J_2$ when the central body is oblate. In the case of a negative $J_2$, the effect of the orbit-rotation coupling of the rigid body on the existence of the

*Corresponding author. Tel.: +86 10 8233 9751.

*E-mail addresses:* ywang@sa.buaa.edu.cn (Y. Wang), starsjxu@yahoo.com.cn (S. Xu), tl614@hotmail.com (L. Tang)

non-classical relative equilibria can be positive or negative, which depends on the values of $J_2$ and the angular velocity $\Omega_e$. The bifurcation from the classical relative equilibria, at which the non-classical relative equilibria appear, has been shown with different parameters of the system. Our results here have given more details of the relative equilibria than our previous paper, in which the existence conditions of the relative equilibria are derived and primarily studied. Our results have also extended the previous results on the relative equilibria of a rigid body in a central gravity field by taking into account the oblateness of the central body.

**Keywords:** $J_2$ problem; Rigid body; Gravitational orbit-rotation coupling; Classical relative equilibria; Non-classical relative equilibria

---

## 1 Introduction

The $J_2$ problem, also known as the main problem of artificial satellite theory, is one of the most important problems in both celestial mechanics and astrodynamics, as the most significant non-spherical mass distribution of the central celestial body, i.e., the zonal harmonic $J_2$, is taken into account (Broucke 1994, Wang and Xu 2013). In the $J_2$ problem, the motion of a point mass in a gravity field truncated on the zonal harmonic $J_2$ is studied. The $J_2$ problem has broad applications in the orbital dynamics and orbital design of spacecraft, such as the design of the sun synchronization orbits and the $J_2$ invariant relative orbits in the spacecraft formations (Xu et al. 2012). This classical problem has been studied in many works, such as Broucke (1994) and the literatures cited therein.

However, neither natural nor artificial celestial bodies are point masses or have spherical mass distributions. A practical generalization of the point mass model is the assumption that the body considered is perfectly rigid that is precise enough for most applications in celestial mechanics and astrodynamics. The orbital and rotational motions of the rigid body are coupled through the gravity field due to its non-spherical mass distribution. The orbit-rotation coupling may cause qualitative effects in the motion, which are more significant when the ratio of the dimension of the rigid body to the orbit radius is larger, as shown by Wang and Xu (in press a).

The point mass model of the $J_2$ problem has a high precision for an artificial Earth satellite, since the dimension of an artificial Earth satellite is small in comparison with the orbital radius and the orbit-rotation coupling is insignificant. However, when a spacecraft orbiting around an asteroid or an irregular natural satellite around a planet, such as Phobos, is considered, the mass distribution of the considered body is far from a sphere and the dimension of the body is not small anymore in comparison with the orbital radius. In these cases, the orbit-rotation coupling causes significant effects and should be taken into account in the precise motion theories, as shown by Scheeres (2006a), Wang and Xu (in press a). For these new high-precision applications in celestial mechanics and astrodynamics, we have generalized the $J_2$ problem to the motion of a rigid body in a $J_2$ gravity field in Wang and Xu (2013, in press b) with the orbit-rotation coupling considered. This generalized problem is a good model for coupled orbital and rotational motions of a spacecraft orbiting a spheroid asteroid, or an irregular natural satellite around a dwarf planet or planet.

The relative equilibria and their properties in the celestial mechanics and astrodynamics are of great interest, since most natural celestial bodies evolved tidally to the state of relative equilibria, and the relative equilibria can be used as the nominal motion in the mission design in the astrodynamics.

We have studied the relative equilibria of the rigid body in the $J_2$ gravity field in Wang and Xu (2013) under the second-order gravitational potential. We found two types of relative equilibria: one is the classical type, also called Lagrangian relative equilibria, in which the circular orbit of the rigid body is in the equatorial plane of central body; the other is the non-classical type, also called non-Lagrangian relative equilibria, in which the circular orbit of the rigid body is displaced, that is, parallel to, but not in the equatorial plane of central body.

The relative equilibria of the coupled orbital and rotational motions of a general rigid body without symmetry in the central gravity field have been studied by Wang et al. (1991, 1992) and Teixidó Román (2010). Wang et al. (1991) pointed out that under the second-order gravitational potential only the classical relative equilibria can exist in the real physical system; whereas the non-classical relative equilibria can only exist in the case of a very large ratio of the dimension of the rigid body to the orbit radius, which means a very significant orbit-rotation coupling and does not exist in the real physical system. Teixidó Román (2010) has found the classical relative equilibria, called orthogonal equilibria there, and two non-classical types of relative equilibria, called oblique equilibria and parallel equilibria, under the second-order gravitational potential. The parallel equilibria, in which the mass center of the rigid body is fixed above the pole of central body, is the extreme of the oblique equilibria, in which the circular orbit of the rigid body is displaced. Similarly to Wang et al. (1991), neither the two non-classical relative equilibria can exist in the real physical system, since both of them are dependent on a very large ratio of the dimension of the rigid body to the orbit radius.

However, several results have shown that either the higher-order gravitational potential or the symmetry of the rigid body can lead to the existence of the non-classical relative equilibria in the real physical system. Wang et al. (1992) and Teixidó Román (2010) have shown that the non-classical relative equilibria, or called oblique equilibria, can always exist under the exact gravitational potential for an arbitrary ratio of the dimension of the rigid body to the orbit radius. Wang et al. (1992) has also shown that there can be significant changes in orientation of the rigid body away from the classical relative equilibria, although the orbital offset of the non-classical relative equilibria from the attractive central is small. O'Reilly and Tan (2004) and Teixidó Román (2010) pointed out that for an axisymmetric rigid body, the non-classical relative equilibria can exist in the case of a small ratio of the dimension of the rigid body to the orbit radius even under the second-order gravitational potential. This phenomenon is due to the fact that the rotational motion of the rigid body can be balanced at more orientations by the angular momentum about the symmetric axis, and these orientations can provide more options to balance the orbital motion through the orbit-rotation coupling.

We have obtained the non-classical relative equilibria under the second-order gravitational potential for a general rigid body without symmetry in a $J_2$ gravity field in Wang and Xu (2013), which can exist in the real physical system. Notice that our physically realistic non-classical relative equilibria are different from those in Wang et al. (1992), O'Reilly and Tan (2004) and Teixidó Román (2010) mentioned above, since our results are not dependent on either the higher-order gravitational potential or the symmetry of the rigid body. Our physically realistic non-classical relative equilibria are due to the combined effects of the second zonal harmonic of the central body and the orbit-rotation coupling of the rigid body (Wang and Xu 2013).

The relative equilibria of the coupled orbital and rotational motions of rigid bodies have also been studied in many works on the Full Two Body Problem (F2BP), which is the problem of the rotational and orbital motions of two rigid bodies interacting through their mutual gravitational potential. As stated by Maciejewski (1995), the non-classical relative equilibria generically exist in the motions of two generic rigid bodies, whereas the classical relative equilibria can exist only in exceptional cases when the bodies possess some kind of symmetry and the mass distributions satisfy contain conditions. Several works on the sphere-restricted F2BP, in which one body is assumed to be a homogeneous sphere, have obtained the classical relative equilibria, such as the stationary motion in Kinoshita (1972); the long-axis equilibria and short-axis equilibria in Scheeres (2004), Bellerose and Scheeres (2008a, 2008b); and the locally central point in Scheeres (2006b). Aboelnaga and Barkin (1979) and Scheeres (2006b) studied the non-classical relative equilibria of the sphere-restricted F2BP, also called the non-locally central point, by using the exact gravitational potential.

Kinoshita (1970) investigated the classical relative equilibria of an axisymmetric rigid body in the sphere-restricted F2BP by using the exact gravitational potential. Three types of classical relative equilibria, “Arrow” type, “Float” type and “Spoke” type, are obtained under different assumptions of the symmetry of the rigid body. Vereshchagin (2010) has obtained both the classical and non-classical relative equilibria of an axisymmetric rigid body in the sphere-restricted F2BP by using the exact gravitational potential. Two types of classical relative equilibria were given: the cylindrical precession, corresponding to the “Float” type in Kinoshita (1970) and the inclined co-planar precession, including the “Arrow” type and the “Spoke” type in Kinoshita (1970).

We will investigate the existence and properties of both the classical and non-classical relative equilibria obtained in Wang and Xu (2013) in more details in the present paper. Special attention will be paid on the individual effect of the second zonal harmonic $J_2$ and the orbit-rotation coupling of the rigid body on the existence of the non-classical relative equilibria.

## 2 Statement of the Problem

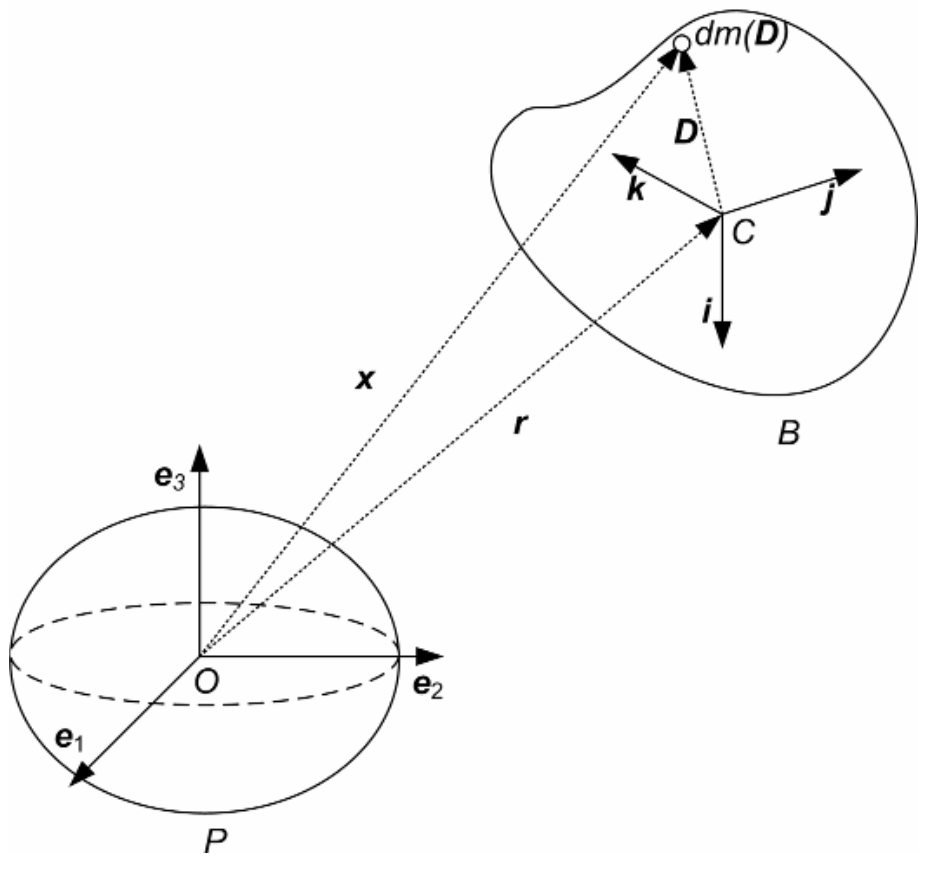

**Fig. 1.** A small rigid body $B$ in the $J_2$ gravity field of a massive axis-symmetrical body $P$

As described in Fig. 1, we consider a small rigid body $B$ in the gravity field of a massive axisymmetrical body $P$. Assume that the body $P$ is rotating uniformly around its axis of symmetry, and the mass center of the body $P$ is stationary in the inertial space, i.e., the body $P$ is in free motion without being affected by the body $B$. The gravity field of the body $P$ is approximated through truncation on the second zonal harmonic $J_2$. The inertial reference frame is defined as $S=\{\boldsymbol{e}_1, \boldsymbol{e}_2, \boldsymbol{e}_3\}$ whose origin $O$ is attached to the mass center of the body $P$. $\boldsymbol{e}_3$ is along the axis of symmetry of the body $P$. The body-fixed reference frame of the body $B$ is defined as $S_b=\{\boldsymbol{i}, \boldsymbol{j}, \boldsymbol{k}\}$ whose origin $C$ is attached to the mass center. The frame $S_b$ coincides with the principal axes reference frame of the body $B$. The attitude matrix of the rigid body $B$ with respect to the inertial frame $S$ is denoted by $\boldsymbol{A}$,

$$\boldsymbol{A}=[\boldsymbol{\alpha},\boldsymbol{\beta},\boldsymbol{\gamma}]^T \in SO(3), \tag{1}$$

where the vectors $\boldsymbol{\alpha}$, $\boldsymbol{\beta}$ and $\boldsymbol{\gamma}$ are coordinates of the unit vectors $\boldsymbol{e}_1$, $\boldsymbol{e}_2$ and $\boldsymbol{e}_3$ expressed in the frame $S_b$, and $SO(3)$ is the 3-dimensional special orthogonal group.

We define $\boldsymbol{r}$ as the position vector of the point $C$ with respect to the point $O$ expressed in the frame $S$. The configuration space of the problem is the Lie group

$$Q = SE(3), \tag{2}$$

known as the special Euclidean group of three space that is the semidirect product of $SO(3)$ and $\mathbb{R}^3$ with elements $(\boldsymbol{A},\boldsymbol{r})$. The elements $\boldsymbol{\Xi}$ of the phase space, the cotangent bundle $T^*Q$, can be written in the following coordinates

$$\boldsymbol{\Xi} = (\boldsymbol{A}, \boldsymbol{r}; \boldsymbol{A}\hat{\boldsymbol{\Pi}}, \boldsymbol{p}), \tag{3}$$

where $\boldsymbol{\Pi}$ is the angular momentum of the rigid body $B$ expressed in the body-fixed frame $S_b$ and $\boldsymbol{p}$ is the linear momentum of the rigid body $B$ expressed in the inertial frame $S$ (Wang and Xu 2012). The hat map $\wedge : \mathbb{R}^3 \to so(3)$ is the usual Lie algebra isomorphism, where $so(3)$ is the Lie Algebras of Lie group $SO(3)$.

The $J_2$ gravity field is axis-symmetrical with axis of symmetry $\boldsymbol{e}_3$. Using this symmetry, we have carried out a reduction, induced a Hamiltonian on the quotient $T^*Q/S^1$, where $S^1$ is the one-sphere, and expressed the dynamics in terms of appropriate reduced variables in Wang and Xu (2013). The reduced variables in $T^*Q/S^1$ can be chosen as

$$\boldsymbol{z} = \left[\boldsymbol{\Pi}^T, \boldsymbol{\gamma}^T, \boldsymbol{R}^T, \boldsymbol{P}^T\right]^T \in \mathbb{R}^{12}, \tag{4}$$

where $\boldsymbol{R} = \boldsymbol{A}^T\boldsymbol{r}$ and $\boldsymbol{P} = \boldsymbol{A}^T\boldsymbol{p}$ are the position vector and the linear momentum of the rigid body $B$ expressed in the frame $S_b$ respectively. The variable $\boldsymbol{\gamma}$, the coordinates of the unit vector $\boldsymbol{e}_3$ expressed in the frame $S_b$, describes the attitude of the rigid body $B$ with respect to the inertial frame $S$.

This system has a non-canonical Hamiltonian structure with the Poisson bracket $\{\cdot,\cdot\}_{\mathbb{R}^{12}}(\boldsymbol{z})$, which can be written in terms of the Poisson tensor as follows:

$$\{f, g\}_{\mathbb{R}^{12}}(\boldsymbol{z}) = \left(\nabla_z f\right)^T \boldsymbol{B}(\boldsymbol{z})\left(\nabla_z g\right). \tag{5}$$

The Poisson tensor $\boldsymbol{B}(\boldsymbol{z})$ is given by:

$$\boldsymbol{B}(\boldsymbol{z}) = \begin{bmatrix} \hat{\boldsymbol{\Pi}} & \hat{\boldsymbol{\gamma}} & \hat{\boldsymbol{R}} & \hat{\boldsymbol{P}} \\ \hat{\boldsymbol{\gamma}} & \mathbf{0} & \mathbf{0} & \mathbf{0} \\ \hat{\boldsymbol{R}} & \mathbf{0} & \mathbf{0} & \mathbf{E} \\ \hat{\boldsymbol{P}} & \mathbf{0} & -\mathbf{E} & \mathbf{0} \end{bmatrix}, \tag{6}$$

where $\mathbf{E}$ is the identity matrix. This Poisson tensor has two independent Casimir functions. One is a geometric integral $C_1(\boldsymbol{z}) = \boldsymbol{\gamma}^T\boldsymbol{\gamma} \equiv 1$, and the other one is $C_2(\boldsymbol{z}) = \boldsymbol{\gamma}^T\left(\boldsymbol{\Pi} + \hat{\boldsymbol{R}}\boldsymbol{P}\right)$, the third component of the angular momentum with respect to the origin $O$ expressed in the inertial frame $S$. $C_2(\boldsymbol{z})$ is the conservative quantity produced by the symmetry of the system, as stated by Noether's theorem.

The Hamiltonian of the problem in the variables $\boldsymbol{z}$ is given as follows:

$$H = \frac{|\boldsymbol{P}|^2}{2m} + \frac{1}{2}\boldsymbol{\Pi}^T \boldsymbol{I}^{-1} \boldsymbol{\Pi} + V \circ \tau_{T*Q}, \tag{7}$$

where $m$ and the diagonal matrix $\boldsymbol{I} = diag\{I_{xx}, I_{yy}, I_{zz}\}$ are the mass and the inertia tensor of the rigid body respectively; $\tau_{T*Q} : T^*Q \to Q$ is the canonical projection. The gravitational potential $V : Q \to \mathbb{R}$ up to the second order is given in terms of moments of inertia of the rigid body as follows (Wang and Xu 2013):

$$V = V^{(0)} + V^{(2)} = -\frac{GM_1 m}{R} - \frac{GM_1}{2R^3}\left[tr(\boldsymbol{I}) - 3\bar{\boldsymbol{R}}^T \boldsymbol{I}\bar{\boldsymbol{R}} + \varepsilon m - 3\varepsilon m(\boldsymbol{\gamma} \cdot \bar{\boldsymbol{R}})^2\right], \tag{8}$$

where $G$ is the Gravitational Constant; $M_1$ is the mass of the body $P$; $\varepsilon$ is defined as $\varepsilon = J_2 a_E^2$ and $a_E$ is the mean equatorial radius of the body $P$. Note that $R = |\boldsymbol{R}|$ and $\bar{\boldsymbol{R}} = \boldsymbol{R}/R$.

The equations of motion of the system can be written in the Hamiltonian form:

$$\dot{\boldsymbol{z}} = \{\boldsymbol{z}, H(\boldsymbol{z})\}_{\mathbb{R}^{12}}(\boldsymbol{z}) = \boldsymbol{B}(\boldsymbol{z})\nabla_z H(\boldsymbol{z}). \tag{9}$$

The explicit equations of motion can be given with the Hamiltonian in Eq. (7) as follows:

$$\begin{aligned}
\dot{\boldsymbol{\Pi}} &= \boldsymbol{\Pi} \times \boldsymbol{I}^{-1}\boldsymbol{\Pi} + \boldsymbol{R} \times \frac{\partial V(\boldsymbol{\gamma}, \boldsymbol{R})}{\partial \boldsymbol{R}} + \boldsymbol{\gamma} \times \frac{\partial V(\boldsymbol{\gamma}, \boldsymbol{R})}{\partial \boldsymbol{\gamma}}, \\
\dot{\boldsymbol{\gamma}} &= \boldsymbol{\gamma} \times \boldsymbol{I}^{-1}\boldsymbol{\Pi}, \\
\dot{\boldsymbol{R}} &= \boldsymbol{R} \times \boldsymbol{I}^{-1}\boldsymbol{\Pi} + \frac{\boldsymbol{P}}{m}, \\
\dot{\boldsymbol{P}} &= \boldsymbol{P} \times \boldsymbol{I}^{-1}\boldsymbol{\Pi} - \frac{\partial V(\boldsymbol{\gamma}, \boldsymbol{R})}{\partial \boldsymbol{R}}.
\end{aligned} \tag{10}$$

In the present paper, we will nondimensionalize the system by the characteristic time $\sqrt{a_E^3/GM_1}$ and the characteristic length $a_E$ to make studies in general cases instead of in specific cases. After nondimensionalization, the equatorial radius $a_E$ and the gravitational constant $GM_1$ of the body *P* are both equal to 1, and the unit of the angular velocity is $\sqrt{GM_1/a_E^3}$ .

## 3 Classical Relative Equilibria

We have found a classical type of relative equilibria based on the equations of motion Eq. (10) under the second-order gravitational potential in Wang and Xu (2013). At this type of relative equilibria, the orbit of the mass center of the rigid body is a circle in the equatorial plane of body *P* with its center coinciding with origin *O*. The rigid body rotates uniformly around one of its principal axes that is parallel to $\boldsymbol{e}_3$ in the inertial frame *S* in angular velocity that is equal to the orbital angular velocity $\boldsymbol{\Omega}_e$. The position vector $\boldsymbol{R}_e$ and the linear momentum $\boldsymbol{P}_e$ are parallel to another two principal axes of the rigid body. When the radius vector $\boldsymbol{R}_e$ is parallel to the principal axes of the rigid body ***i***, ***j***, ***k***, the norm of the orbital angular velocity $\boldsymbol{\Omega}_e$ is given by the following three equations respectively:

$$\Omega_e = \left( \frac{1}{R_e^3} + \frac{3}{2R_e^5} \left[ -2\frac{I_{xx}}{m} + \frac{I_{yy}}{m} + \frac{I_{zz}}{m} + J_2 \right] \right)^{1/2}, \tag{11}$$

$$\Omega_e = \left( \frac{1}{R_e^3} + \frac{3}{2R_e^5} \left[ \frac{I_{xx}}{m} - 2\frac{I_{yy}}{m} + \frac{I_{zz}}{m} + J_2 \right] \right)^{1/2}, \tag{12}$$

$$\Omega_e = \left( \frac{1}{R_e^3} + \frac{3}{2R_e^5} \left[ \frac{I_{xx}}{m} + \frac{I_{yy}}{m} - 2\frac{I_{zz}}{m} + J_2 \right] \right)^{1/2}. \tag{13}$$

The condition of existence of this classical type of relative equilibria is given by the following inequations:

$$\frac{1}{R_e^3} + \frac{3}{2R_e^5} \left[ -2\frac{I_{xx}}{m} + \frac{I_{yy}}{m} + \frac{I_{zz}}{m} + J_2 \right] > 0, \tag{14}$$

$$\frac{1}{R_e^3}+\frac{3}{2R_e^5}\left[\frac{I_{xx}}{m}-2\frac{I_{yy}}{m}+\frac{I_{zz}}{m}+J_2\right]>0, \tag{15}$$

$$\frac{1}{R_e^3}+\frac{3}{2R_e^5}\left[\frac{I_{xx}}{m}+\frac{I_{yy}}{m}-2\frac{I_{zz}}{m}+J_2\right]>0. \tag{16}$$

The norm of the linear momentum $\boldsymbol{P}_e$ is given by:

$$P_e = mR_e\Omega_e. \tag{17}$$

With a given value of $R_e$, there are 24 relative equilibria belonging to this classical type of relative equilibria in total. Without of loss of generality, we will choose one of the relative equilibria as shown by Fig. 2 for detailed studies:

$$\begin{gathered}\boldsymbol{\Pi}_e=\left[0,0,\Omega_e I_{zz}\right]^T, \boldsymbol{\gamma}_e=\left[0,0,1\right]^T, \boldsymbol{R}_e=\left[R_e \quad 0 \quad 0\right]^T, \\ \boldsymbol{P}_e=\left[0 \quad mR_e\Omega_e \quad 0\right]^T, \boldsymbol{\Omega}_e=\left[0 \quad 0 \quad \Omega_e\right]^T.\end{gathered} \tag{18}$$

Other relative equilibria can be converted into this equilibrium by changing the arrangement of the axes of the reference frame $S_b$. At this relative equilibrium, the norm of the orbital angular velocity $\boldsymbol{\Omega}_e$ is given by Eq. (11), and the condition of existence is given by Eq. (14).

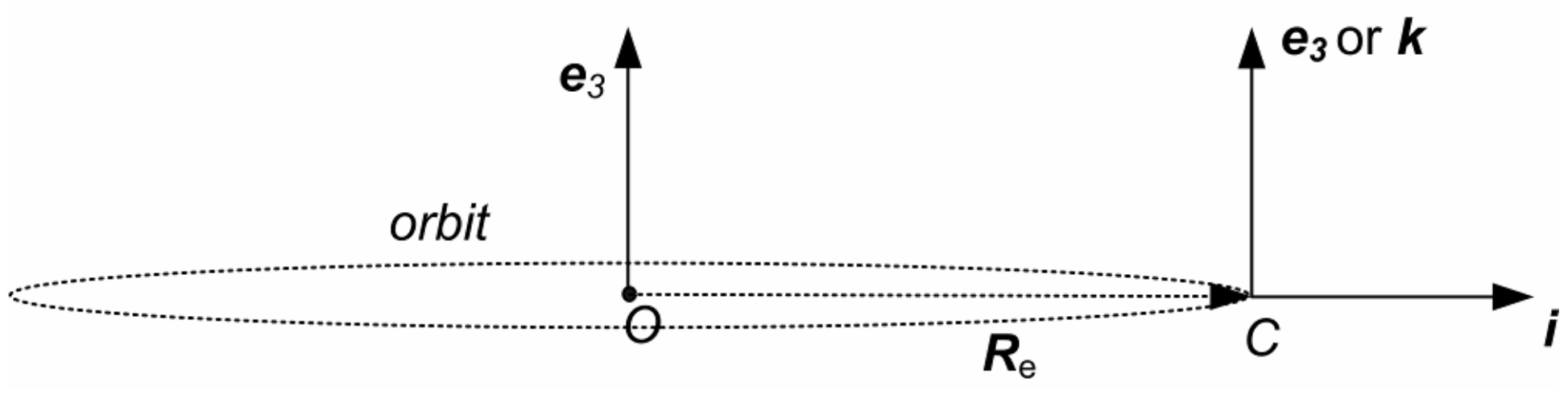

**Fig. 2.** One of the classical type of relative equilibria

The norm of the orbital angular velocity Eq. (11) can written as

$$\begin{aligned}\Omega_e &= \left(\frac{1}{R_e^3}+\frac{3}{2R_e^3}\left[\left(\frac{1}{R_e}\right)^2\left(-2\frac{I_{xx}}{m}+\frac{I_{yy}}{m}+\frac{I_{zz}}{m}\right)+J_2\left(\frac{1}{R_e}\right)^2\right]\right)^{1/2} \\ &= \left(\frac{1}{R_e^3}+\frac{3}{2R_e^3}\left[\left(\frac{1}{R_e}\right)^2\left(\Delta I+J_2\right)\right]\right)^{1/2},\end{aligned} \tag{19}$$

where $\Delta I = -2I_{xx}/m + I_{yy}/m + I_{zz}/m$. The parameter $\Delta I$ is a comprehensive scale of the effect of the orbit-rotation coupling of the rigid body, since it describes both the non-spherical mass distribution and the characteristic dimension of the rigid body, which are two basic elements of the gravitational orbit-rotation coupling. The effect of $\Delta I$ can be considered equivalently as a change of the oblateness of the central body in the sense of the point mass model.

The characteristic dimension of the rigid body $d_C$ can be defined by the following equation:

$$I_{xx} = \frac{1}{2} m d_C^2, \text{ or } \frac{1}{2} d_C^2 = \frac{I_{xx}}{m}. \tag{20}$$

Notice that the characteristic dimension $d_C$ is only an estimation of the dimension of the rigid body, but not the real dimension of the rigid body. Since the characteristic dimension $d_C$ has a very simple relation with $I_{xx}/m$, we will also refer $I_{xx}/m$ as the characteristic dimension in the following.

The mass distribution parameters of the rigid body $\sigma_x$ and $\sigma_y$ can be defined as follows:

$$\sigma_x = \left( \frac{I_{zz} - I_{yy}}{I_{xx}} \right), \quad \sigma_y = \left( \frac{I_{zz} - I_{xx}}{I_{yy}} \right). \tag{21}$$

The parameters $\sigma_y$ and $\sigma_x$ have the following range:

$$-1 \le \sigma_y \le 1, -1 \le \sigma_x \le 1. \tag{22}$$

The bounds of the parameters $\sigma_y$ and $\sigma_x$ are determined by the physical properties of the moments of inertia: $I_{xx} + I_{yy} > I_{zz}$, $I_{xx} + I_{zz} > I_{yy}$ and $I_{yy} + I_{zz} > I_{xx}$.

As shown above, the ratio $I_{xx}/m$ describes the characteristic dimension of the rigid body; the ratios $\sigma_x$ and $\sigma_y$ describe the shape of the rigid body. Solving the definition in Eq. (21), we can have:

$$\frac{I_{yy}}{I_{xx}} = \frac{1-\sigma_x}{1-\sigma_y}, \tag{23}$$

$$\frac{I_{zz}}{I_{xx}} = \frac{1-\sigma_x\sigma_y}{1-\sigma_y}. \tag{24}$$

Then by using Eqs. (23) and (24), the parameter $\Delta I$ can be written in terms of the three ratios $I_{xx}/m$, $\sigma_x$, and $\sigma_y$ as follows:

$$\Delta I = \frac{I_{xx}}{m}\left(-2 + \frac{I_{yy}}{I_{xx}} + \frac{I_{zz}}{I_{xx}}\right) = \frac{I_{xx}}{m} f_\sigma, \tag{25}$$

where $f_\sigma$ is defined as:

$$f_\sigma = \frac{2\sigma_y - \sigma_x - \sigma_x\sigma_y}{1-\sigma_y}. \tag{26}$$

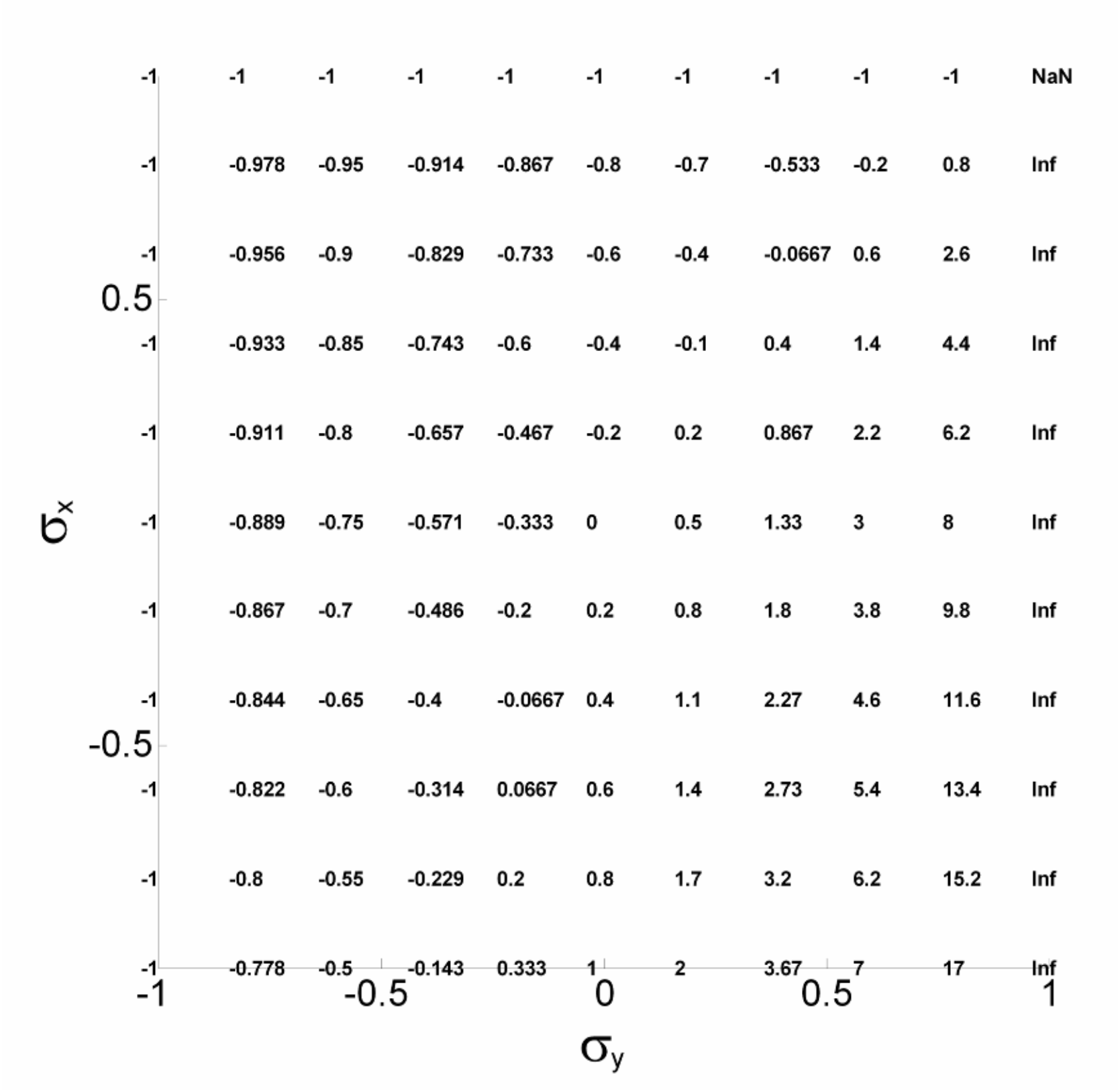

**Fig. 3.** The value of $f_\sigma$ on the $\sigma_y$ - $\sigma_x$ plane

We can estimate the range of the parameter $\Delta I$ through the upper limit of $I_{xx}/m$

and the calculation of $f_\sigma$ on the $\sigma_y$-$\sigma_x$ plane. We assume that the upper limit of $I_{xx}/m$ is equal to 0.125, which means the characteristic dimension of the rigid body $d_C$ is $0.5a_E$ that is the upper limit of the characteristic dimension in our study.

The value of the parameter $f_\sigma$ is shown on the $\sigma_y$-$\sigma_x$ plane in Fig. 3. We can find that the lower limit of $f_\sigma$ is -1, which can be reached in the case of $\sigma_y = -1$. Theoretically, the upper limit of $f_\sigma$ is the positive infinity, which can be reached when $\sigma_y$ approaches 1 that means the mass distribution of the rigid body is a rod along the ***i***-axis. However, in our study we will not consider this extreme case that would not exist in the real physical system. We choose the upper limit of $f_\sigma$ as 16.

Noticing that the upper limit of $I_{xx}/m$ is equal to 0.125, we can obtain the range of the parameter $\Delta I$ as follows:

$$-0.125 < \Delta I < 2 . \tag{27}$$

The range of the second zonal harmonic is chosen as $-0.5 < J_2 < 0.5$ in our study, same as in Broucke (1994), which can cover all the spheroid celestial bodies in our Solar system. Although the body with a negative $J_2$ has not been discovered yet in our Solar System, our studies on the case of a negative $J_2$ are of interest and value for the theoretical studies on the related problems in the celestial mechanics and astrophysics. For a very large rigid body with $I_{xx}/m = 0.125$, when $\sigma_y$ approaches 1, the parameter $\Delta I$ can be larger than $J_2$ that means that the orbit-rotation coupling of the rigid body is dominative and the effect of $J_2$ is insignificant.

According to Eq. (14), the existence condition of the classical relative equilibria can be written as:

$$\Delta I + J_2 > -\frac{2}{3} R_e^2 . \quad (28)$$

The boundary surface of the existence condition are plotted in the three dimensional space $\Delta I$ - $J_2$ - $R_e$ in Fig. 4. The forbidden region, in which the classical relative equilibria do not exist, is below the boundary surface. It is easy to see that the forbidden region is always located in the interval $R_e < 1$.

Notice that $R_e < 1$ means the mass center of the rigid body is within the surface of the central body *P*, which is excluded from the real physical system. Therefore, we can conclude that the classical relative equilibria can always exist in the real physical situation.

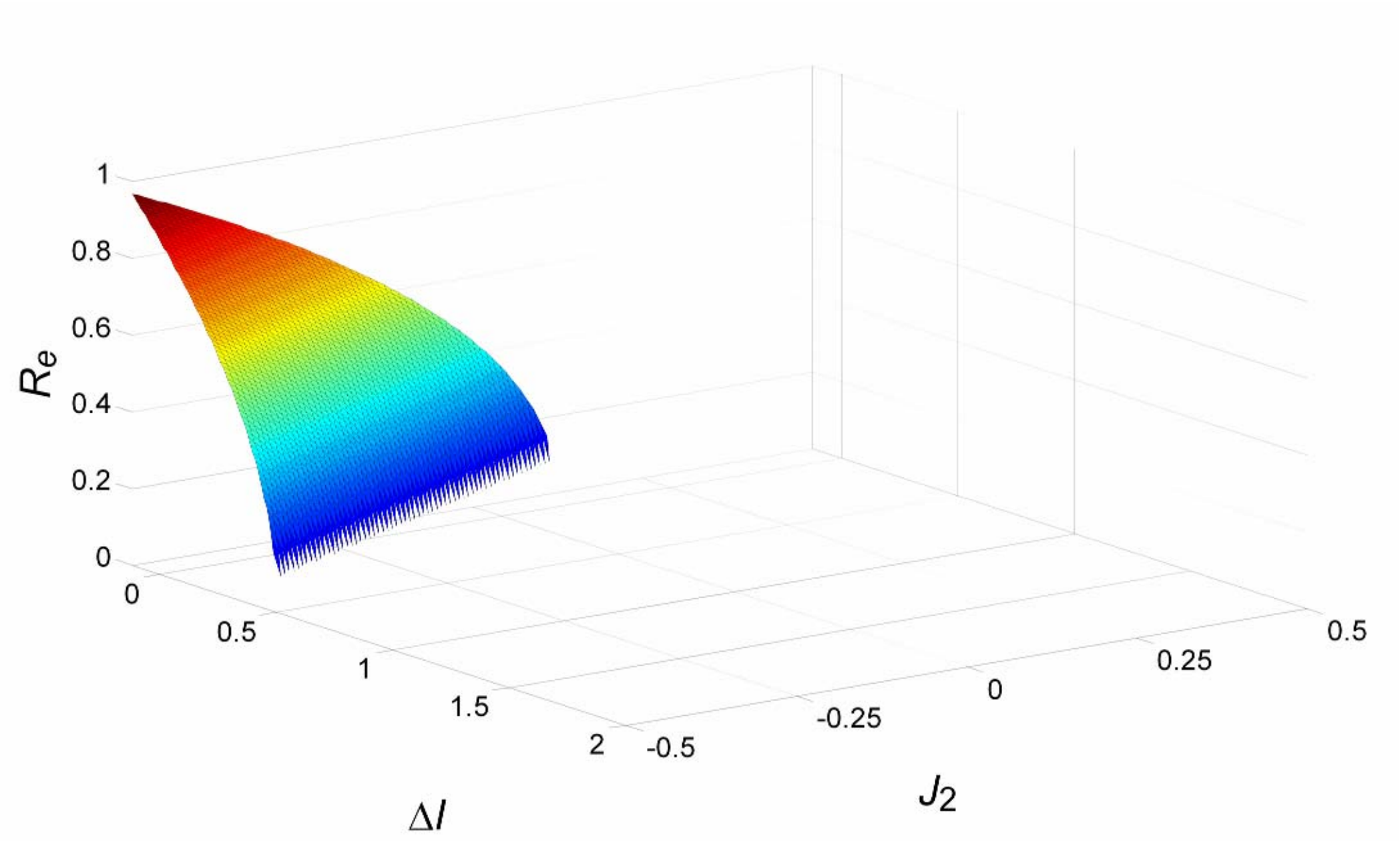

**Fig. 4.** The boundary surface of the existence condition of the classical relative equilibria

According to Eq. (19), the equivalent zonal harmonic $J_{2-Eq}$ can be defined as:

$$J_{2-Eq} = \Delta I + J_2 . \quad (29)$$

The range of $J_{2-Eq}$ considered here can be given by

$$-0.625 < J_{2-Eq} < 2.5 . \quad (30)$$

According to the results by Howard (1990), through some calculation it is found that Eq. (19) has only one positive root for the orbital radius $R_e$ when

$$J_{2-Eq} > 0\,. \tag{31}$$

Equation (19) has two positive roots for $R_e$ in the following case:

$$-\frac{2}{5}\left(\frac{2}{5}\right)^{2/3}\left(\frac{1}{\Omega_e^2}\right)^{2/3} < J_{2-Eq} < 0\,, \tag{32}$$

and have no positive root for $R_e$ when

$$J_{2-Eq} < -\frac{2}{5}\left(\frac{2}{5}\right)^{2/3}\left(\frac{1}{\Omega_e^2}\right)^{2/3}. \tag{33}$$

According to Eqs. (32) and (33), in the case of a negative $J_{2-Eq}$, the bifurcation value of $\Omega_e$, where the number of roots of Eq. (19) increase from zero to one, then to two, i.e., the peak of the $\Omega_e$-$R_e$ curve, is given by:

$$\Omega_e = \left(\frac{2}{5}\right)^{5/4}\left(-\frac{1}{J_{2-Eq}}\right)^{3/4}. \tag{34}$$

The curves of the angular velocity $\Omega_e$ with respect to the orbital radius $R_e$ in the cases of different values of $J_{2-Eq}$ are given in Fig. 5. The eight curves, from bottom to up, are corresponding to eight different values of $J_{2-Eq}$: -0.625, -0.4, -0.2, 0, 0.5, 1, 1.5, 2.5 respectively.

As shown by Fig. 5, the $\Omega_e$-$R_e$ curve is monotone decreasing in the case of a positive $J_{2-Eq}$, which means that Eq. (19) has only one positive root for $R_e$. This is consistent with the conclusion obtained above. It is easy to find that when the value of the equivalent zonal harmonic $J_{2-Eq}$ is -0.625, the $\Omega_e$-$R_e$ curve is not monotone decreasing. This means for some values of $\Omega_e$ Eq. (19) has two positive roots for $R_e$, and for other values of $\Omega_e$ Eq. (19) has no positive root. The bifurcation value of $\Omega_e$ at the peak of the curve is consistent with the result given by Eq. (34). The $\Omega_e$ - $R_e$ curves in the cases of $J_{2-Eq} = -0.4$ and $J_{2-Eq} = -0.2$ are not monotone decreasing either. However the peak of these two curves are not shown in Fig. 5, since they are located in the interval $R_e < 1$.

From Fig. 5, we can find that as the orbital radius $R_e$ increasing, the differences between the eight curves are getting smaller very rapidly. This is due to the fact that as the orbital radius $R_e$ increasing, the effect of the equivalent zonal harmonic, which is the sum of the zonal harmonic $J_2$ and the orbit-rotation coupling $\Delta I$, is being reduced at the rate of $R_e^{-2}$, as shown by Eq. (19).

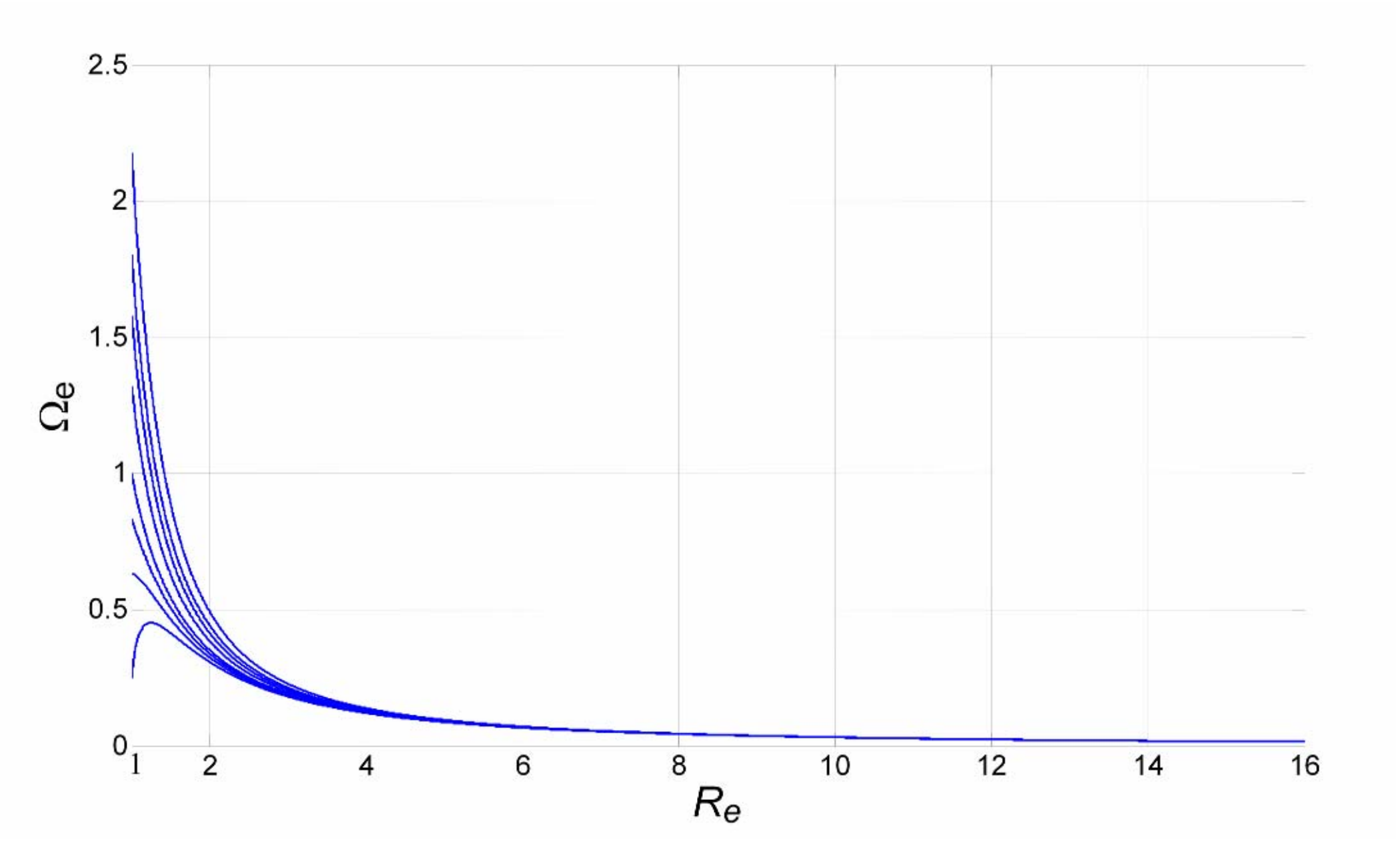

**Fig. 5.** The curves of the angular velocity $\Omega_e$ with respect to the orbital radius $R_e$

# 4 Non-classical Relative Equilibria

## 4.1 Existence condition of non-classical relative equilibria

We have also found a non-classical type of relative equilibria based on the equations of motion under the second-order gravitational potential in Wang and Xu (2013). At this non-classical type of relative equilibria, the orbit of the mass center of the rigid body is a circle with its center located on $\boldsymbol{e}_3$ but not coinciding with the origin *O*, and the orbital plane is parallel to, but not in, the equatorial plane of the body *P*. The rigid body rotates uniformly around $\boldsymbol{e}_3$ in the inertial frame *S* in an angular velocity that is equal to the orbital angular velocity $\boldsymbol{\Omega}_e$. The linear momentum $\boldsymbol{P}_e$ is parallel to a principal axis of the rigid body, whereas neither $\boldsymbol{\gamma}_e$ nor the position vector $\boldsymbol{R}_e$ is parallel to a principal axis of the rigid body. The plane spanned by $\boldsymbol{\gamma}_e$ and $\boldsymbol{R}_e$ is parallel to a principal plane of the rigid body, which is perpendicular to $\boldsymbol{P}_e$.

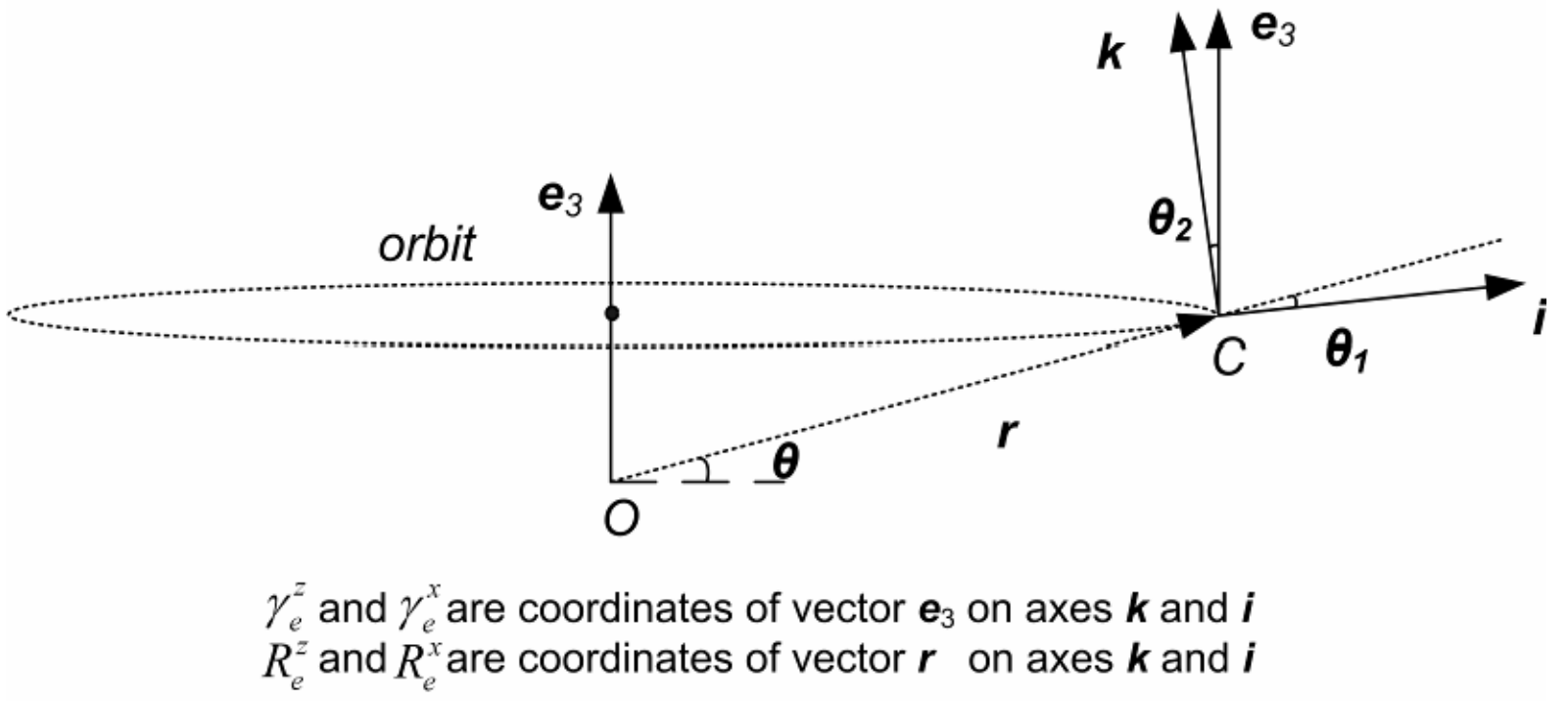

**Fig. 6.** The geometry of the non-classical type of relative equilibria

Without of loss of generality, we assume that $\boldsymbol{P}_e$ is parallel to the principal axis $\boldsymbol{j}$

$$\boldsymbol{R}_e = \begin{bmatrix} R_e^x & 0 & R_e^z \end{bmatrix}^T, \ \boldsymbol{\gamma}_e = \begin{bmatrix} \gamma_e^x & 0 & \gamma_e^z \end{bmatrix}^T,$$

$$\boldsymbol{\Omega}_e = \Omega_e \begin{bmatrix} \gamma_e^x & 0 & \gamma_e^z \end{bmatrix}^T, \ \boldsymbol{P}_e = m\Omega_e \begin{bmatrix} 0 & R_e^x\gamma_e^z - R_e^z\gamma_e^x & 0 \end{bmatrix}^T.$$

We assume that $\gamma_e^x > 0$, $\gamma_e^z > 0$, $R_e^x > 0$ and $R_e^z > 0$ further. Then, the geometry of this non-classical type of relative equilibria can be described by Fig. 6, where $\theta$ is the angle between $\boldsymbol{r}_e(t)$ and the equatorial plane of body *P*, $\theta_1$ is the angle between $\boldsymbol{r}_e(t)$ and $\boldsymbol{i}$, and $\theta_2$ is the angle between $\boldsymbol{e}_3$ and $\boldsymbol{k}$ (Wang and Xu

2013). The angles $\theta$, $\theta_1$ and $\theta_2$ are given by:

$$\tan\theta_1 = \frac{R_e^z}{R_e^x}, \quad \tan\theta_2 = \frac{\gamma_e^x}{\gamma_e^z}, \quad \theta = \theta_1 + \theta_2. \tag{35}$$

According to Wang and Xu (2013), with the orbital angular velocity $\Omega_e$ given, the non-classical relative equilibria, i.e., $R_e^x$, $R_e^z$, $\gamma_e^x$ and $\gamma_e^z$, can be given by solving the following algebraic equations:

$$\gamma_e^x = \left(\frac{1}{2} \mp \sqrt{\frac{1}{4} - \left(\frac{3}{\Omega_e^2 R_e^5} R_e^x R_e^z\right)^2}\right)^{\frac{1}{2}}, \gamma_e^z = \left(\frac{1}{2} \pm \sqrt{\frac{1}{4} - \left(\frac{3}{\Omega_e^2 R_e^5} R_e^x R_e^z\right)^2}\right)^{\frac{1}{2}}, \tag{36}$$

$$\Omega_e^2\left(\left(\gamma_e^z\right)^2 - \frac{3}{R_e^5\Omega_e^2}\left(R_e^z\right)^2\right)R_e^x - \frac{1}{R_e^3}R_e^x - \frac{3}{2R_e^7}\left[R_e^2 tr(\frac{\boldsymbol{I}}{m}) - 5(R_e^x)^2\frac{I_{xx}}{m} - 5(R_e^z)^2\frac{I_{zz}}{m}\right.$$
$$\left.+J_2\left(R_e^2 - 5\left(\gamma_e^x R_e^x + \gamma_e^z R_e^z\right)^2\right)\right]R_e^x - \frac{3}{R_e^5}\frac{I_{xx}}{m}R_e^x - \frac{3J_2\left(\gamma_e^x R_e^x + \gamma_e^z R_e^z\right)}{R_e^5}\gamma_e^x = 0, \tag{37}$$

$$\Omega_e^2\left(\left(\gamma_e^x\right)^2 - \frac{3}{R_e^5\Omega_e^2}\left(R_e^x\right)^2\right)R_e^z - \frac{1}{R_e^3}R_e^z - \frac{3}{2R_e^7}\left[R_e^2 tr(\frac{\boldsymbol{I}}{m}) - 5(R_e^x)^2\frac{I_{xx}}{m} - 5(R_e^z)^2\frac{I_{zz}}{m}\right.$$
$$\left.+J_2\left(R_e^2 - 5\left(\gamma_e^x R_e^x + \gamma_e^z R_e^z\right)^2\right)\right]R_e^z - \frac{3}{R_e^5}\frac{I_{zz}}{m}R_e^z - \frac{3J_2\left(\gamma_e^x R_e^x + \gamma_e^z R_e^z\right)}{R_e^5}\gamma_e^z = 0. \tag{38}$$

The equation (36) contains two cases

$$\gamma_e^z > \gamma_e^x \text{ and } \gamma_e^x > \gamma_e^z, \tag{39}$$

only one of which can by solved with Eqs. (37) and (38). The existence condition of this non-classical type of relative equilibria is equivalence to the solvable condition of the algebraic equations Eqs. (36)-(38).

### 4.2 Existence regions of non-classical relative equilibria

However, it is difficult to analyze the existence of the non-classical relative equilibria through theoretical studies of this system described by nonlinear algebraic equations Eqs. (36)-(38). Therefore, we will try to solve Eqs. (36)-(38) using numerical method with different values of the parameters of the system. Notice that all the five parameters of the system, i.e., $J_2$, $\Omega_e$, $I_{xx}/m$, $\sigma_x$ and $\sigma_y$, need to be discussed. It is impossible to carry out the numerical studies for every combination of the values of the five system parameters, since it will lead to large consumption of computation time and the results will be difficult to display in the form of charts.

Actually, to investigate the existence of the non-classical relative equilibria and the effects of the system parameters on the existence, we only need to carry out the numerical studies for some chosen values of the system parameters. We will choose some different values for the zonal harmonic $J_2$, the characteristic dimension of the rigid body $d_C$ and the orbital angular velocity $\Omega_e$. Then for each combination of the values of $J_2$, $d_C$ (or $I_{xx}/m$) and $\Omega_e$, we try to solve the algebraic equations Eqs. (36)-(38) for each point $(\sigma_y, \sigma_x)$ on the $\sigma_y$-$\sigma_x$ plane. If the point $(\sigma_y, \sigma_x)$ can guarantee the solvable condition of the algebraic equations Eqs. (36)-(38), that is to say, guarantee the existence of the non-classical relative equilibria, we plot the point $(\sigma_y, \sigma_x)$ on the $\sigma_y$-$\sigma_x$ plane.

With this method, we can obtain the existence regions of the non-classical relative equilibria on the $\sigma_y$-$\sigma_x$ plane with different values of the zonal harmonic $J_2$, the characteristic dimension of the rigid body $d_C$ and the orbital angular velocity $\Omega_e$. Through comparisons between existence regions with different values of $J_2$, $d_C$ (or $I_{xx}/m$) and $\Omega_e$, we can find out the individual effect of the system parameters on the existence of the non-classical relative equilibria.

The unit of the angular velocity is $\sqrt{GM_1/a_E^3}$ that is the angular velocity of a point mass orbiting the body *P* on it's the surface along the equator with the effect of $J_2$ neglected. If the $\Omega_e$ is larger than 1, it means that the point mass will move within the surface of the central body *P*, which is excluded from the real physical system. Considering that the angular velocity will be modified by the effect of $J_2$ and the orbit-rotation coupling of the rigid body, we relax the upper limit of $\Omega_e$ from 1 to 1.5. The upper limit $\Omega_e = 1.5$ could cover all the real physical situations. The upper limit of the characteristic dimension of the rigid body $I_{xx}/m$ is chosen to be 0.125, equal to that in the studies of the classical relative equilibria.

The different values of $J_2$, $I_{xx}/m$ and $\Omega_e$ in the numerical studies are chosen as follows:

$$J_2 = -0.5, -0.2, -0.1, -0.05, 0, 0.05, 0.1, 0.2, 0.5\,; \tag{40}$$

$$\frac{I_{xx}}{m} = 0.125, 0.5\mathrm{e}-2, 0.5\mathrm{e}-6\,; \tag{41}$$

$$\Omega_e = 1.5, 1, 0.5, 0.1\,. \tag{42}$$

We find that in the cases of $J_2 = 0, 0.05, 0.1, 0.2, 0.5$, there is no existence region on the $\sigma_y$-$\sigma_x$ plane for all the values of $I_{xx}/m$ and $\Omega_e$ given by Eqs. (41) and (42). Then, the numerical studies are carried out with $J_2$=0.01, 0.3 and 0.4 further, and there is no existence region on the $\sigma_y$-$\sigma_x$ plane either. Then the numerical results suggest that the non-classical relative equilibria cannot exist in the case of a positive $J_2$ in the real physical situation.

Whereas the numerical results show that there exist existence regions of the non-classical relative equilibria on the $\sigma_y$-$\sigma_x$ plane in the cases of $J_2 = -0.5$, $-0.2$, $-0.1$ and $-0.05$, which are given in Tables 1, 2, 3 and 4 respectively. In these figures, the interval $-0.01 < \sigma_y < 0.01$ is not considered, since $\sigma_y = 0$ means $I_{zz} = I_{xx}$ that is the singular point of the existence condition of the non-classical relative equilibria, as shown by Wang and Xu (2013).

In Wang and Xu (2013), we have known that the existence of our physically realistic non-classical relative equilibria is due to the combined effects of the second zonal harmonic of the central body *P* and the orbit-rotation coupling of the rigid body *B*. We will investigate the effects of all the five parameters of the system, i.e., $J_2$, $\Omega_e$, $I_{xx}/m$, $\sigma_x$ and $\sigma_y$, on the existence of the non-classical relative equilibria in the following analysis. Especially, we will focus on the individual effect of the second zonal harmonic and the orbit-rotation coupling in details.

**Table 1**

The existence regions of the non-classical relative equilibria with $J_2 = -0.5$

| | $\frac{I_{xx}}{m} = 0.125$ | $\frac{I_{xx}}{m} = 0.5\text{e}-2$ | $\frac{I_{xx}}{m} = 0.5\text{e}-6$ |
|---|---|---|---|
| $\Omega_e = 1.5$ | 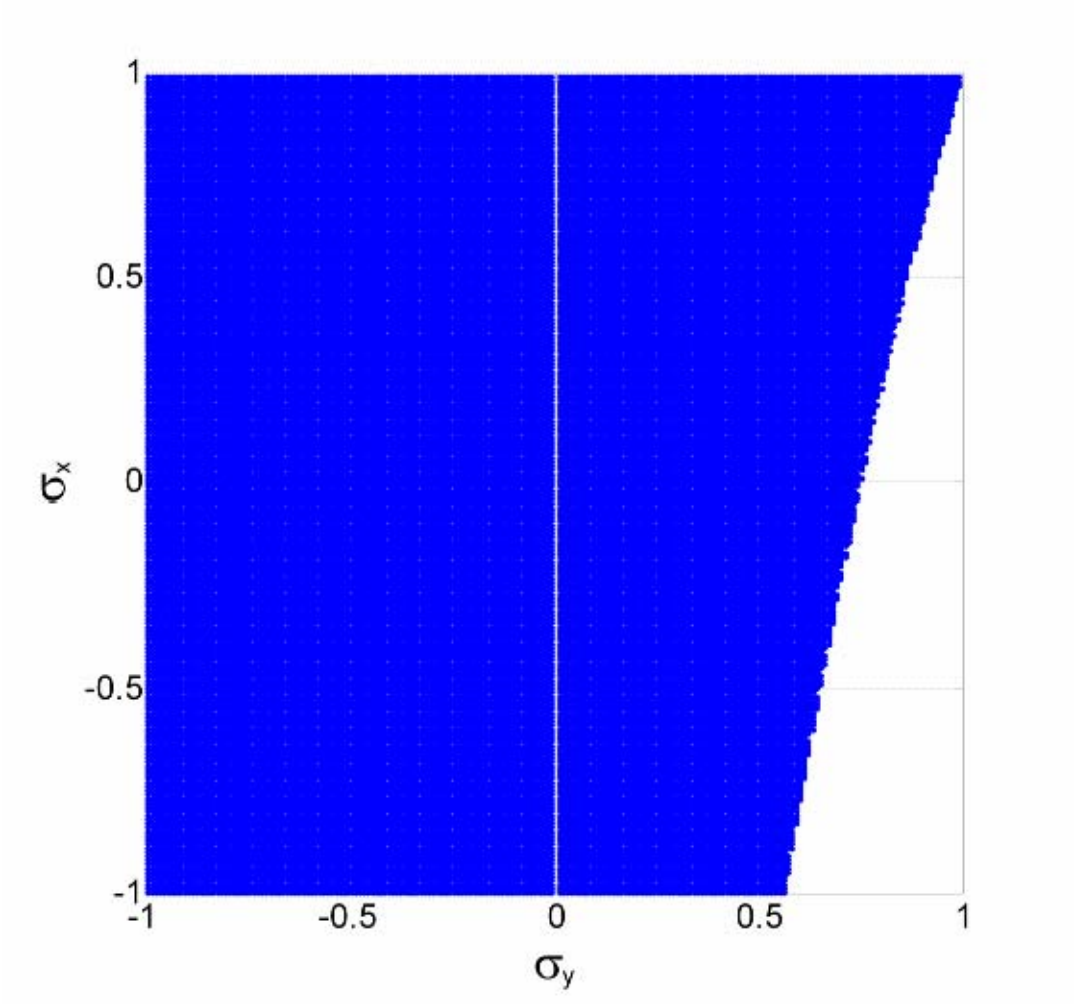 | 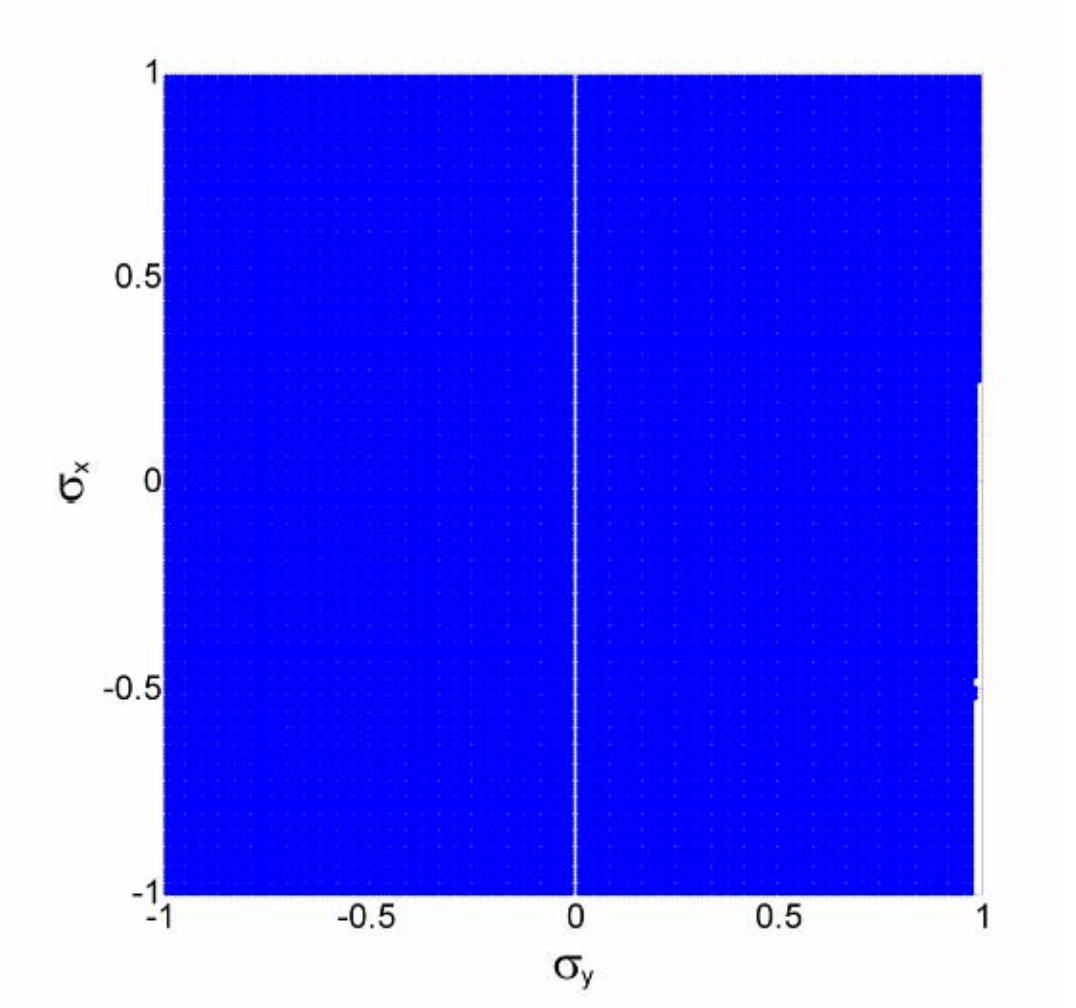 | 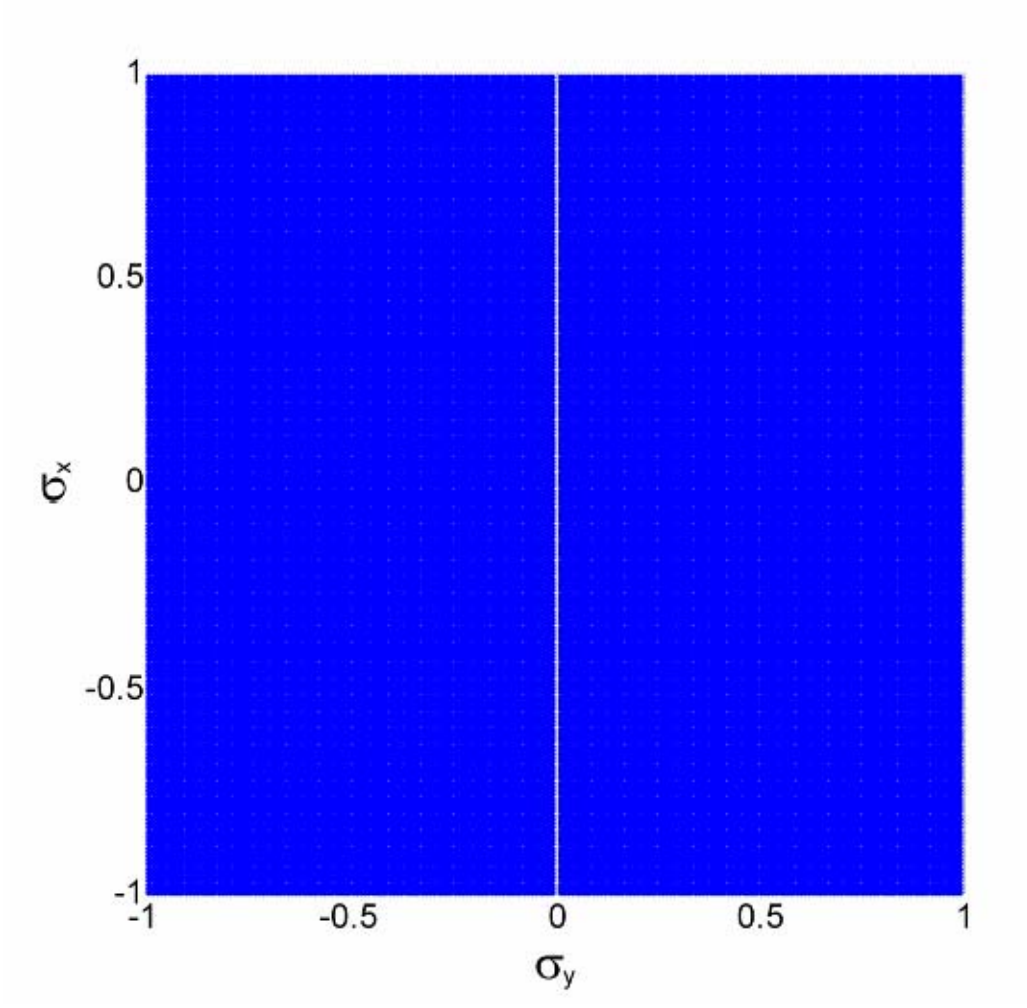 |

<table>
<tr>
<td>$\Omega_e = 1$</td>
<td>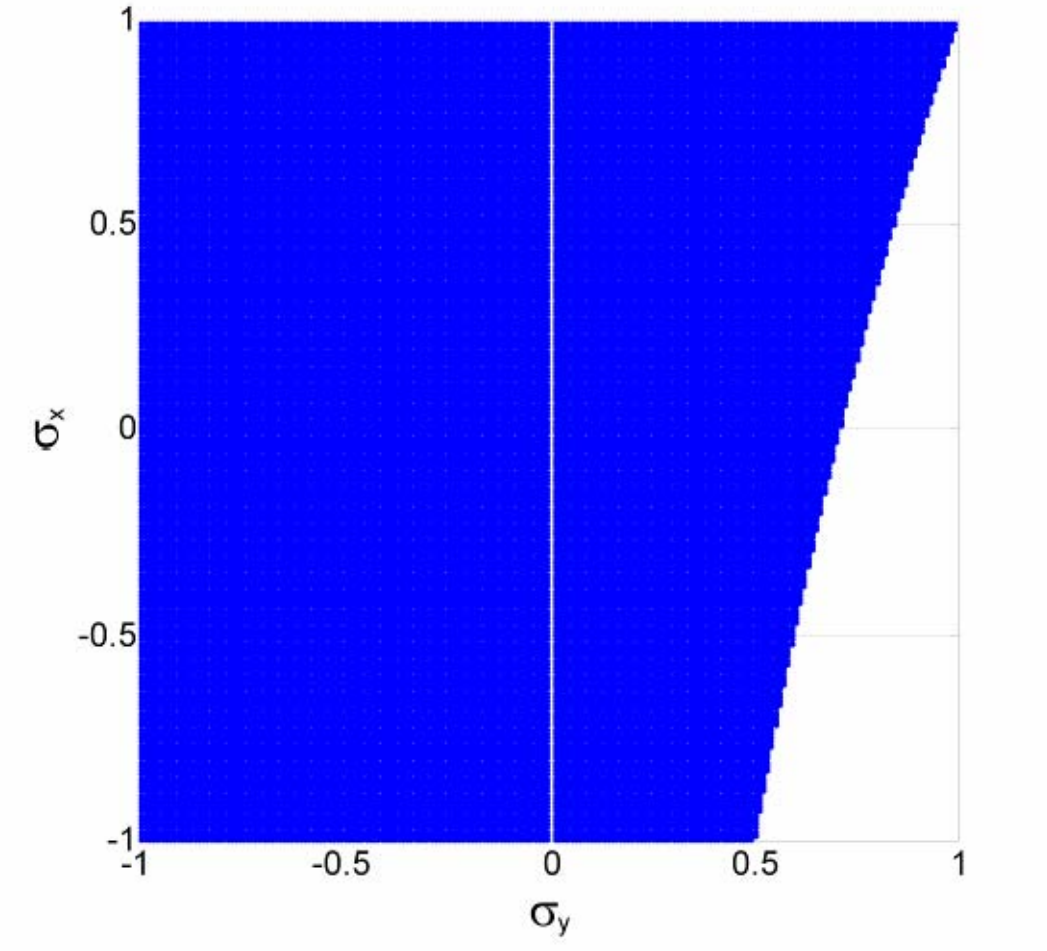
</td>
<td>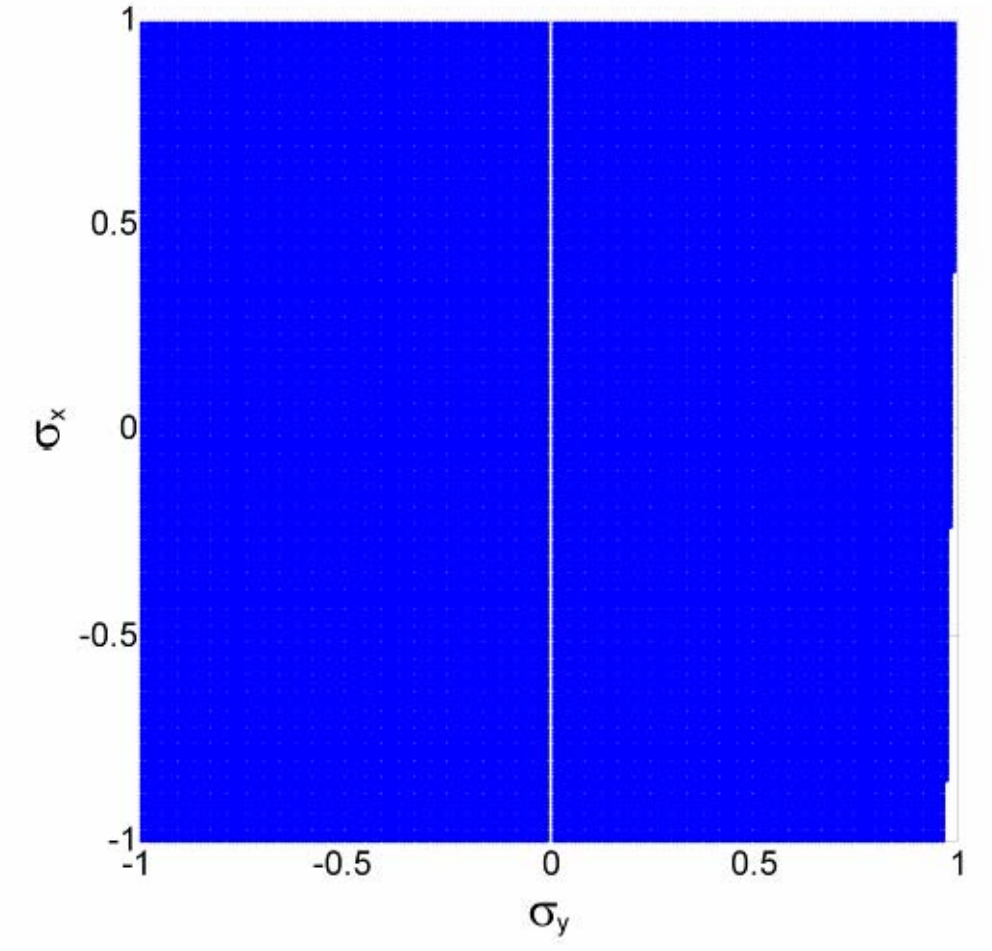
</td>
<td>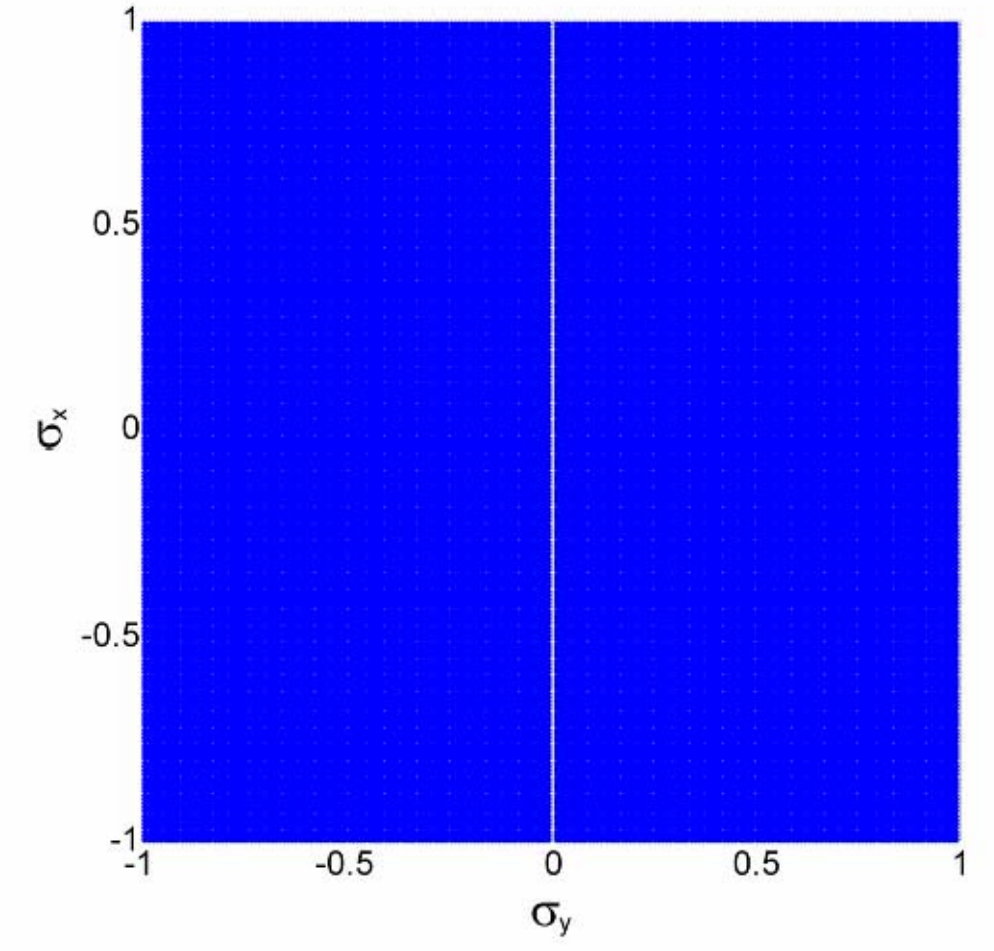
</td>
</tr>
</table>

| | | | |
|---|---|---|---|
| $\Omega_e = 0.5$ | 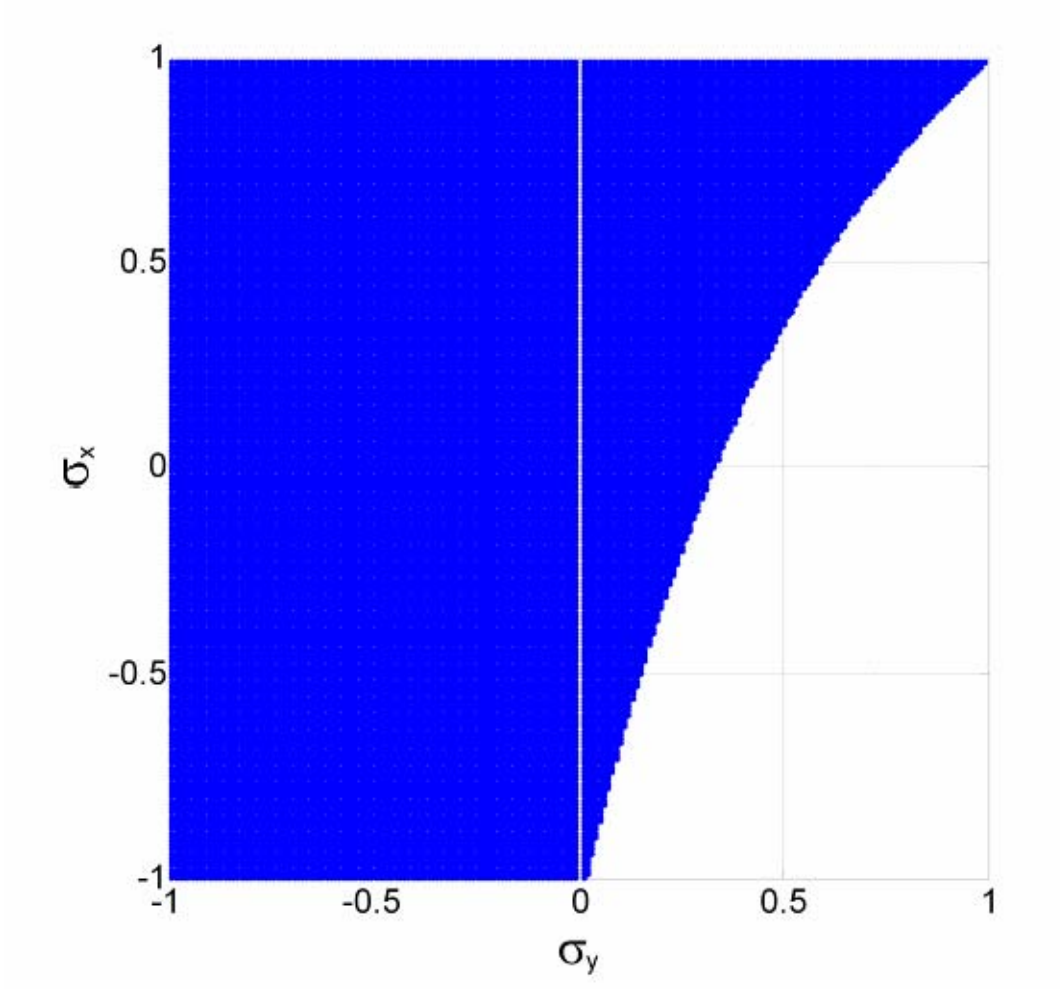σx<br>σy<br>1<br>0.5<br>0<br>-0.5<br>-1 | 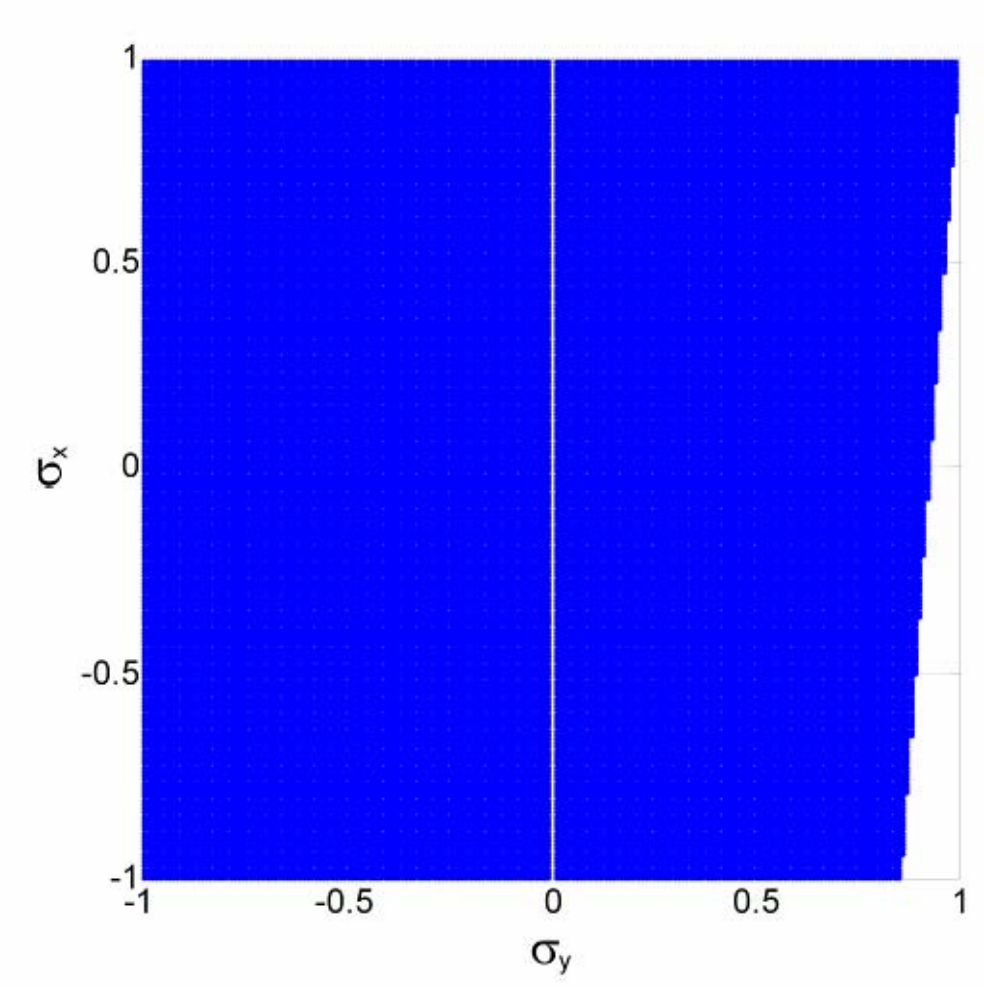σx<br>σy<br>1<br>0.5<br>0<br>-0.5<br>-1 | 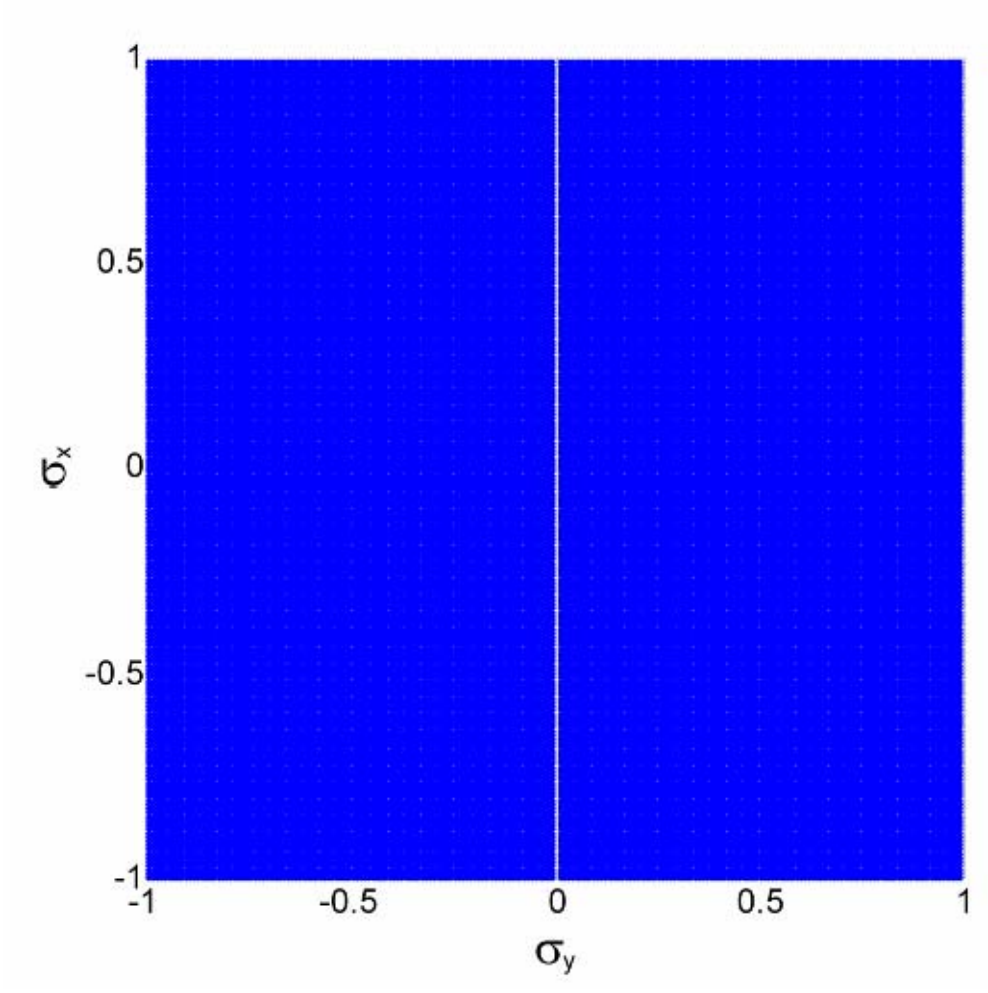σx<br>σy<br>1<br>0.5<br>0<br>-0.5<br>-1 |
| $\Omega_e = 0.1$ | 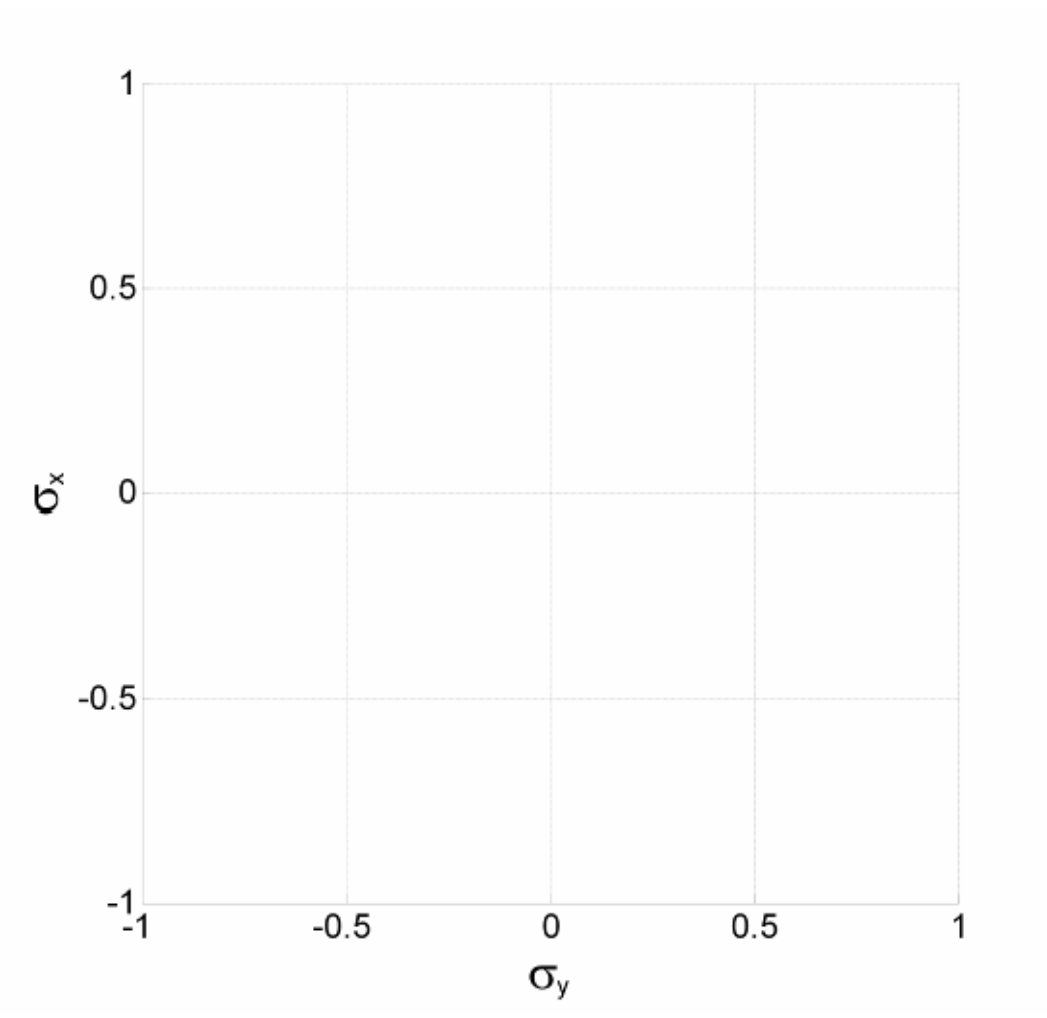1<br>0.5<br>0<br>-0.5<br>-1<br>σx<br>-1<br>-0.5<br>0<br>0.5<br>1<br>σy | 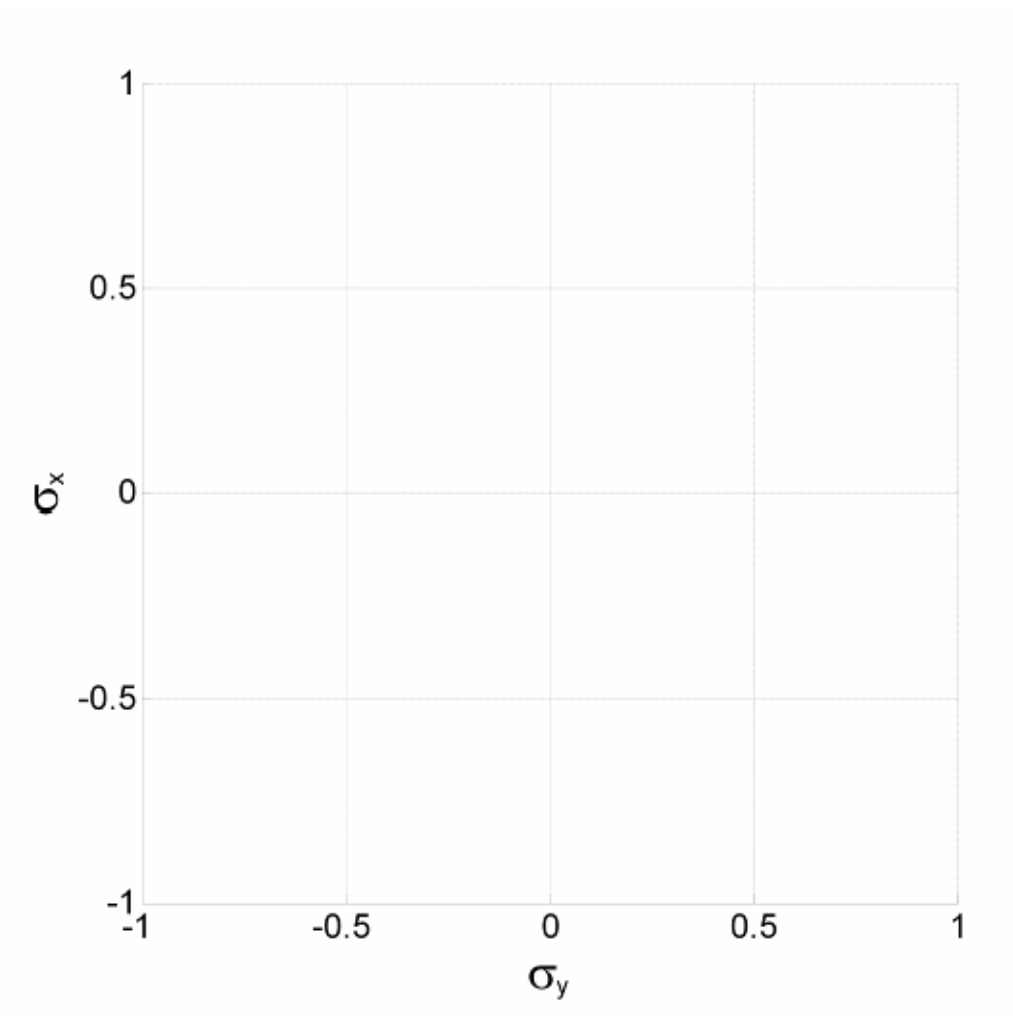1<br>0.5<br>0<br>-0.5<br>-1<br>σx<br>-1<br>-0.5<br>0<br>0.5<br>1<br>σy | 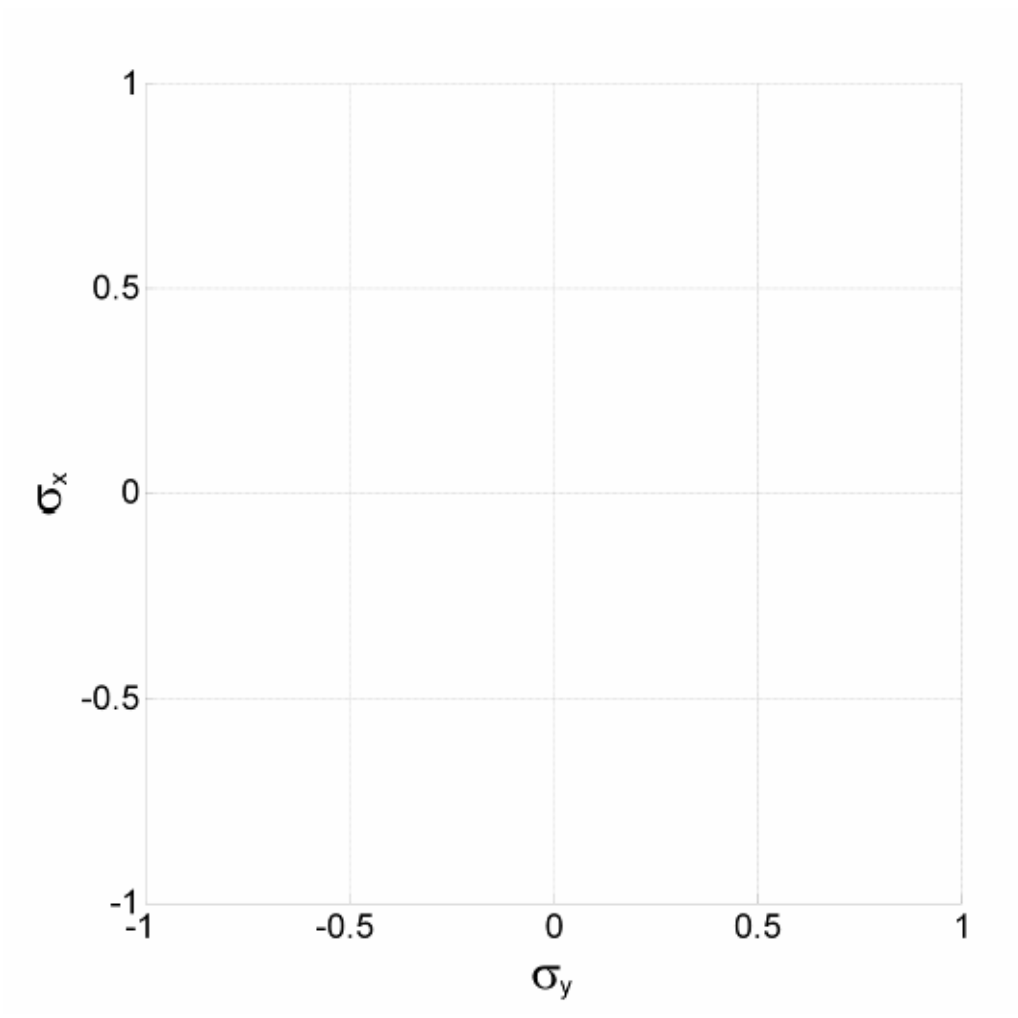1<br>0.5<br>0<br>-0.5<br>-1<br>σx<br>-1<br>-0.5<br>0<br>0.5<br>1<br>σy |

**Table 2**

The existence regions of the non-classical relative equilibria with $J_2 = -0.2$

| | $\frac{I_{xx}}{m} = 0.125$ | $\frac{I_{xx}}{m} = 0.5\mathrm{e}-2$ | $\frac{I_{xx}}{m} = 0.5\mathrm{e}-6$ |
|---|---|---|---|
| $\Omega_e = 1.5$ | 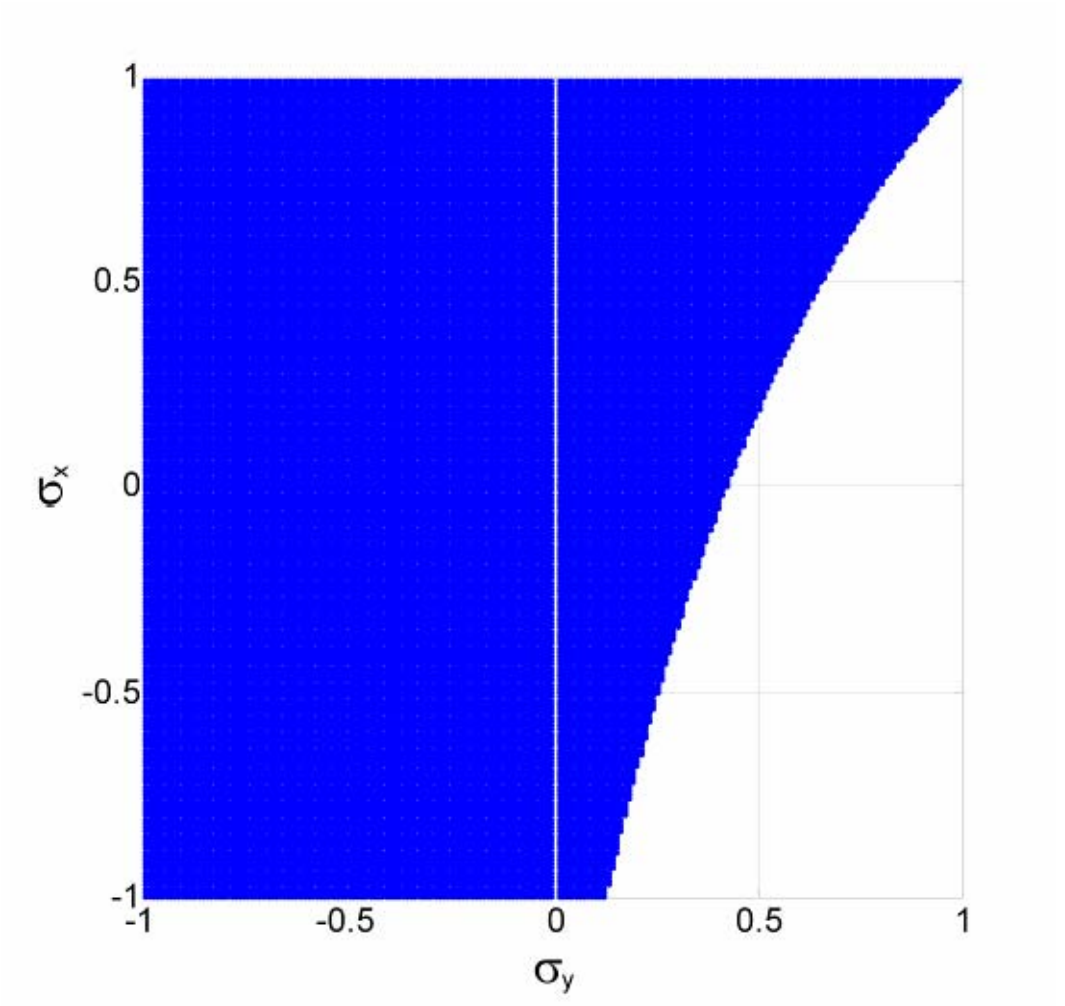 | 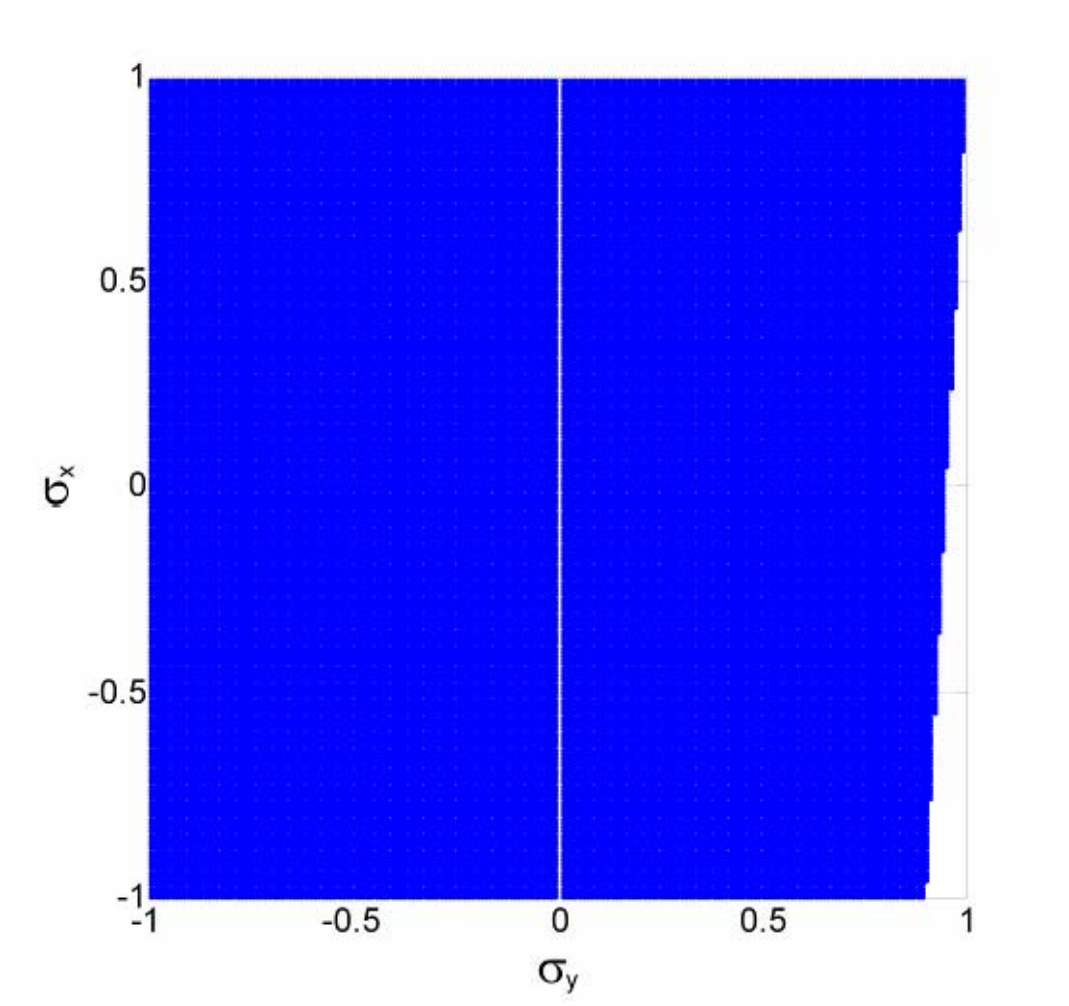 | 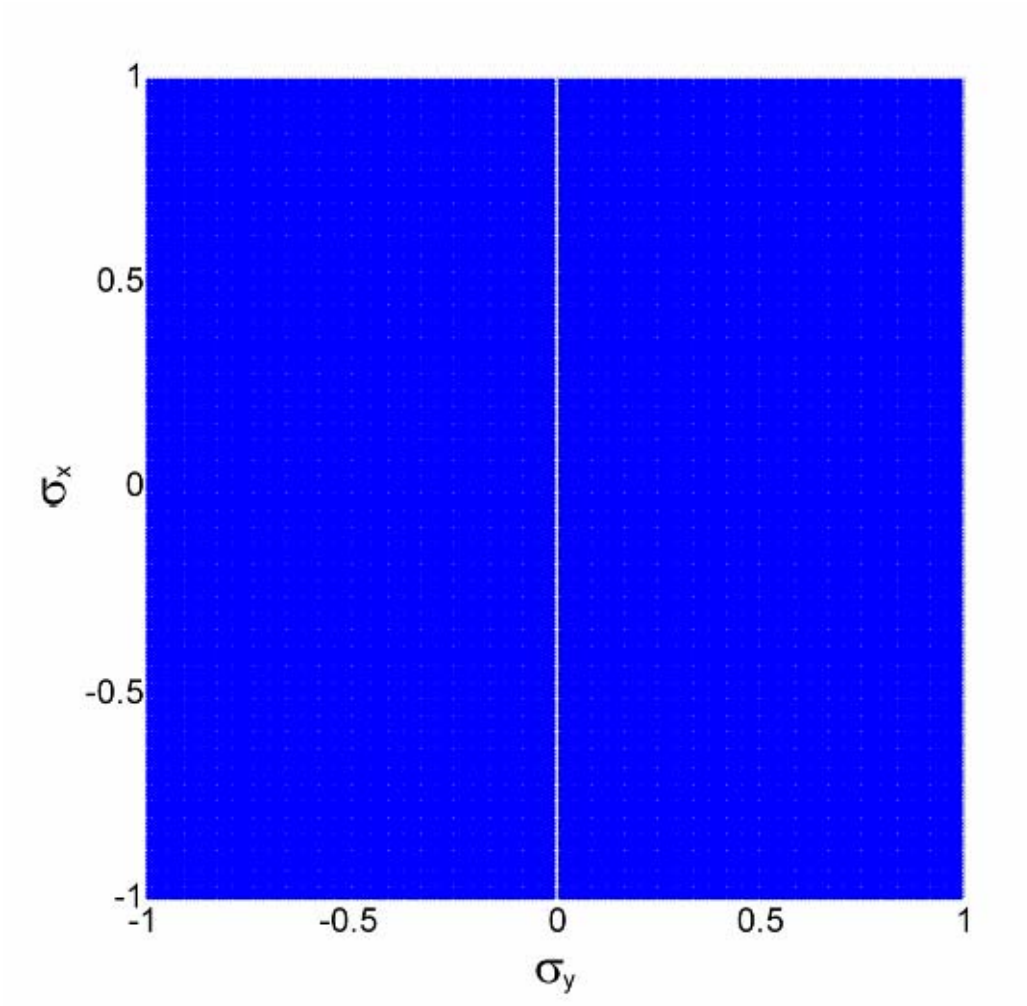 |
| $\Omega_e = 1$ | 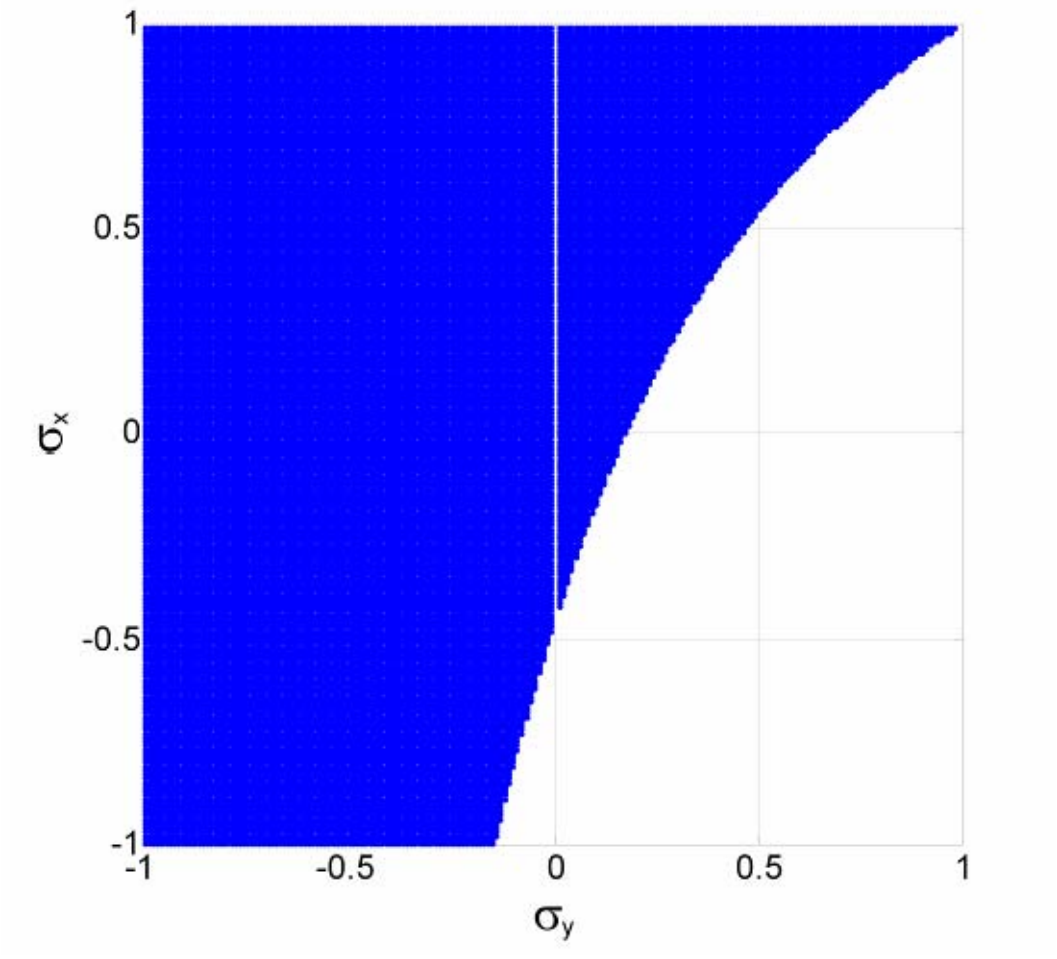σx<br>σy<br>1 0.5 0 -0.5 -1<br>-1 -0.5 0 0.5 1 | 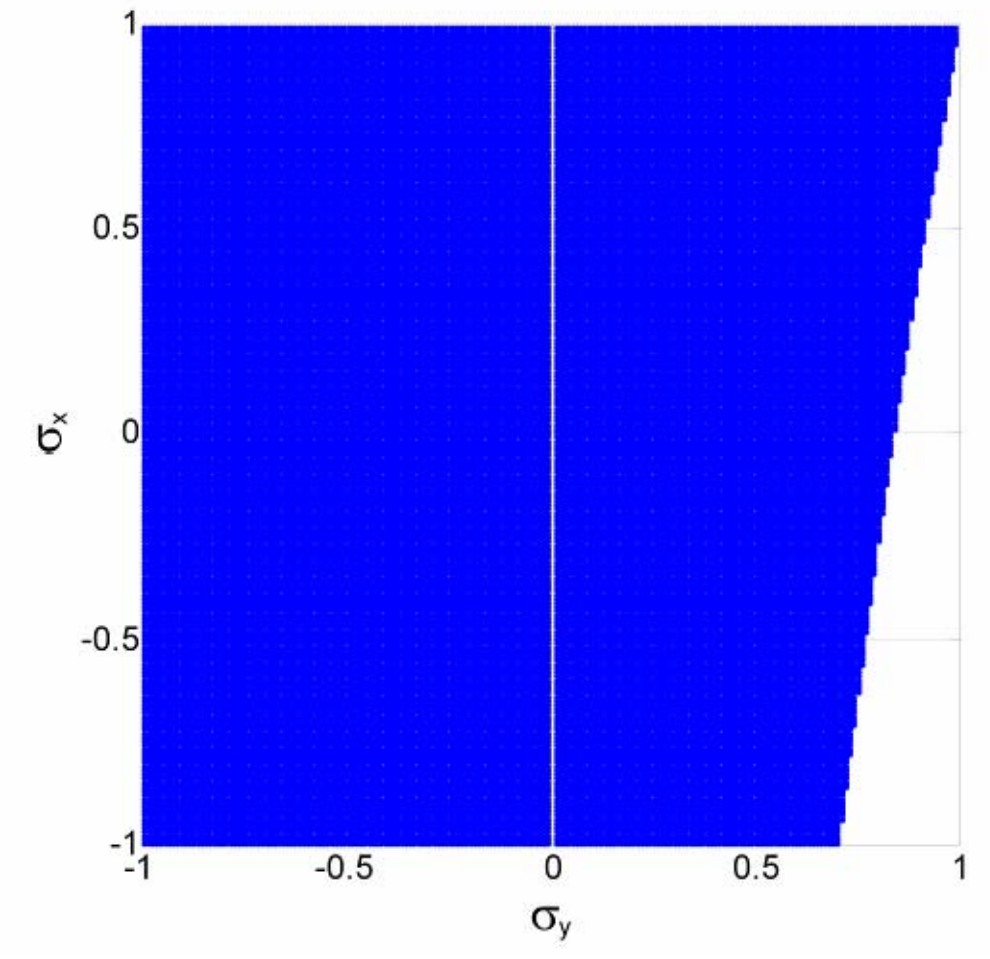σx<br>σy<br>1 0.5 0 -0.5 -1<br>-1 -0.5 0 0.5 1 | 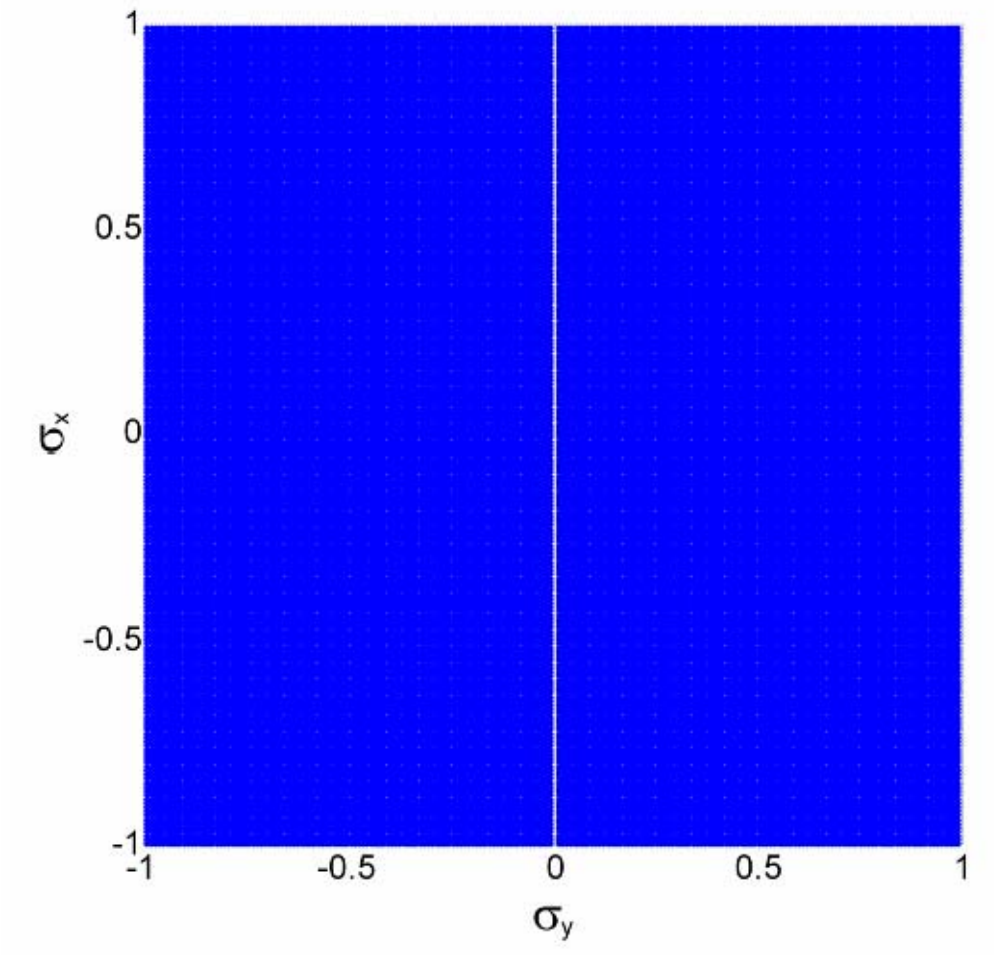σx<br>σy<br>1 0.5 0 -0.5 -1<br>-1 -0.5 0 0.5 1 |

<table>
<tr>
<td>$\Omega_e = 0.5$</td>
<td>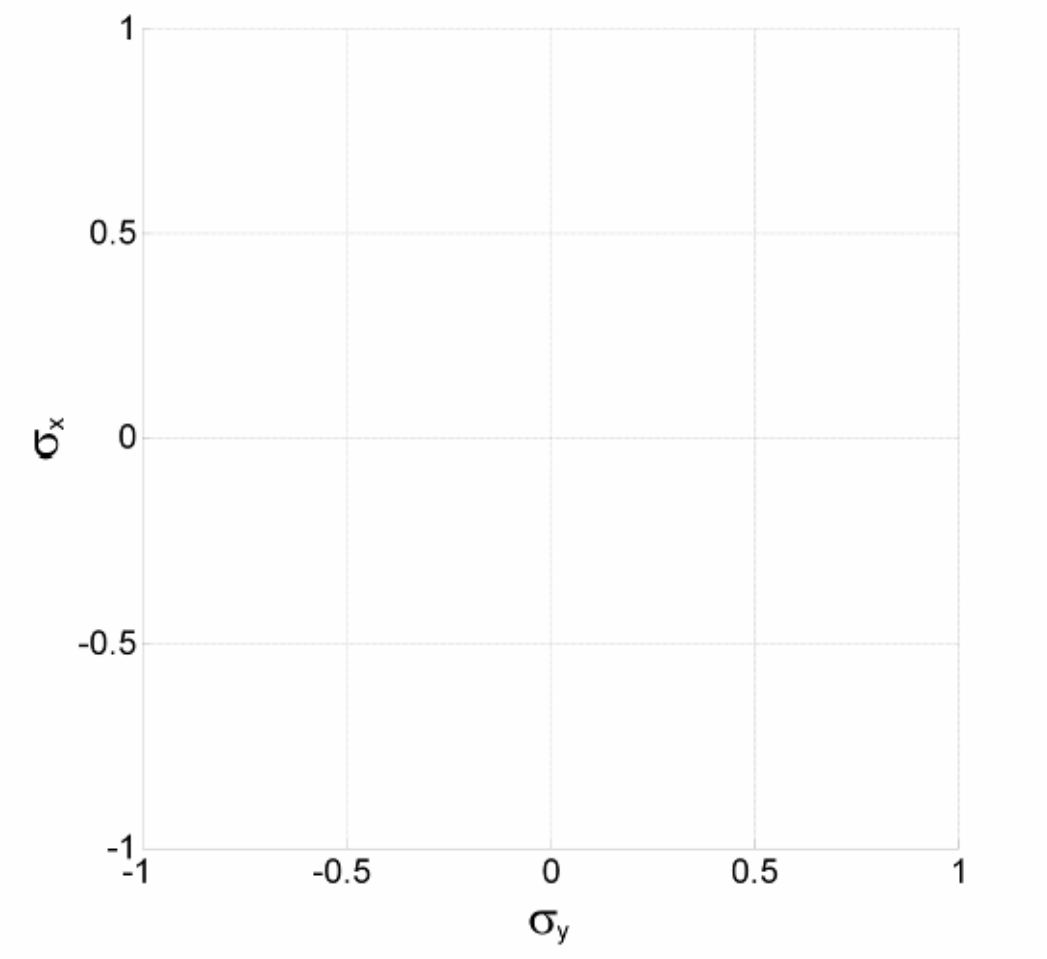
</td>
<td>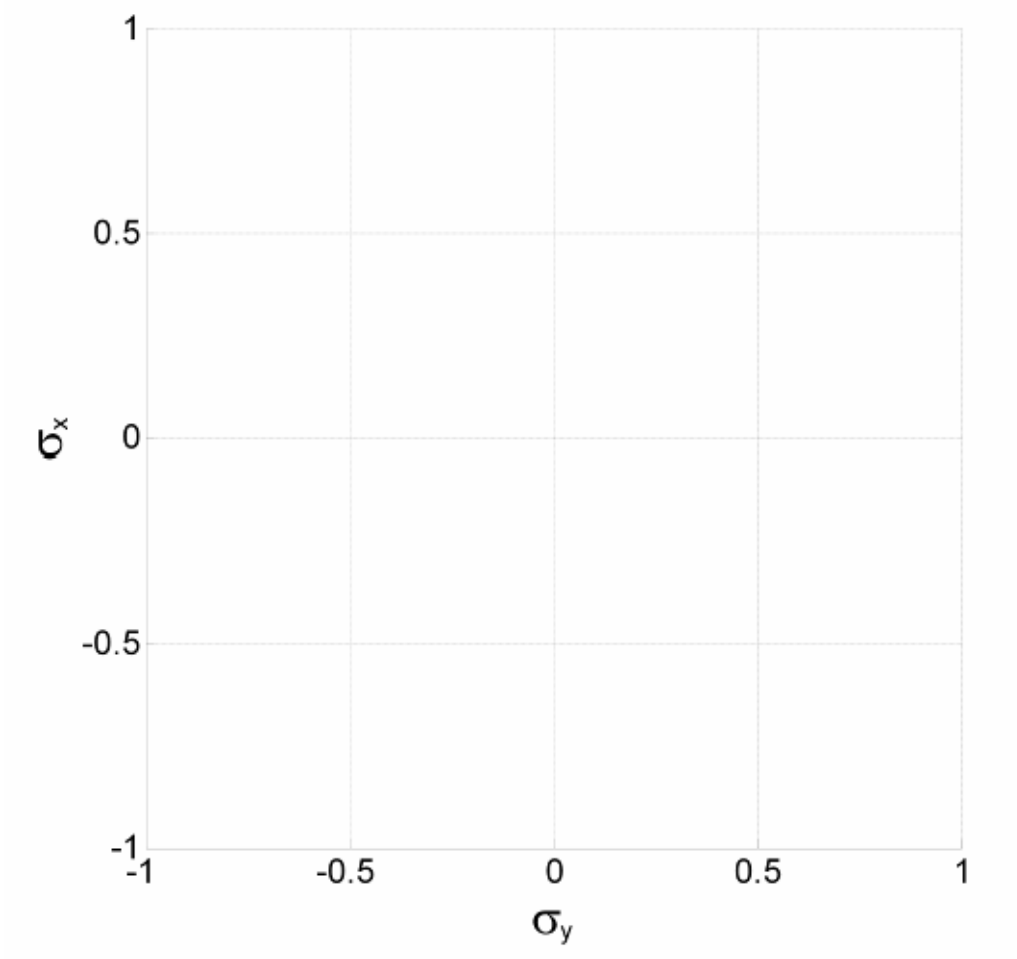
</td>
<td>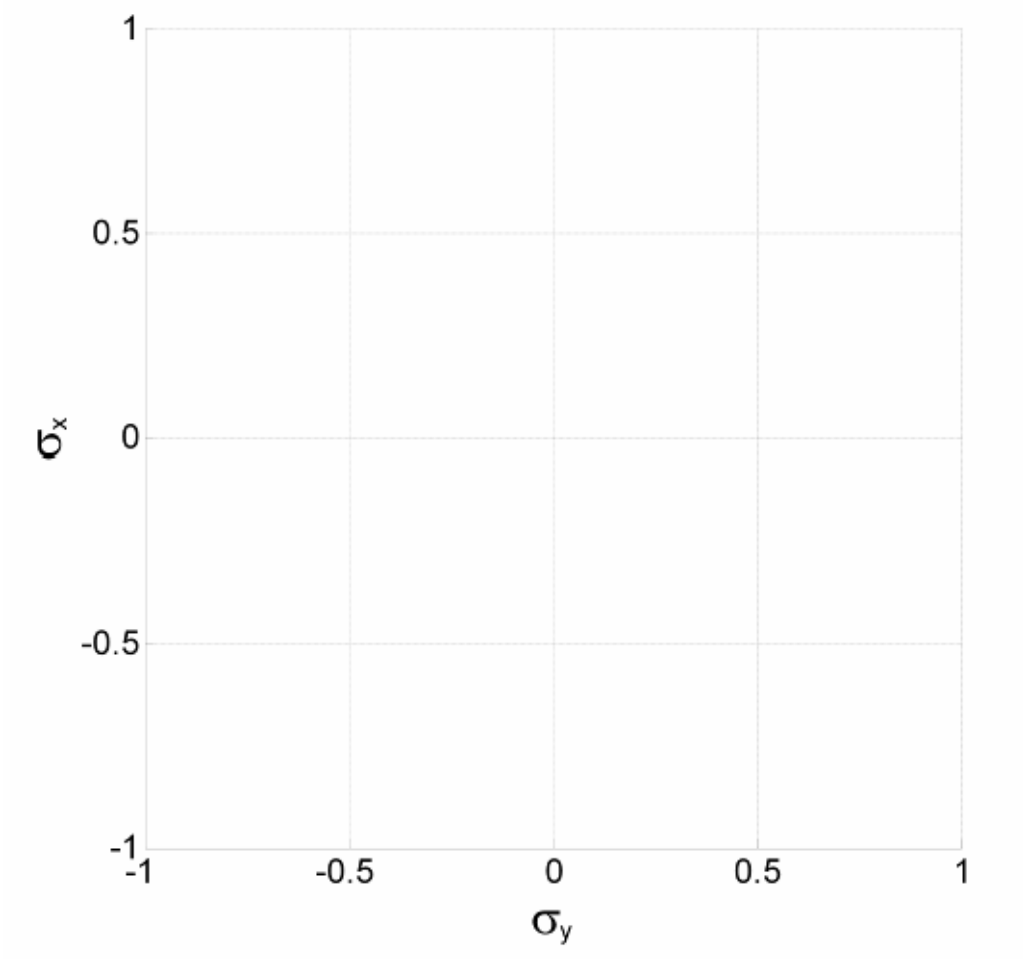
</td>
</tr>
</table>

| | | | |
|---|---|---|---|
| $\Omega_e = 0.1$ | 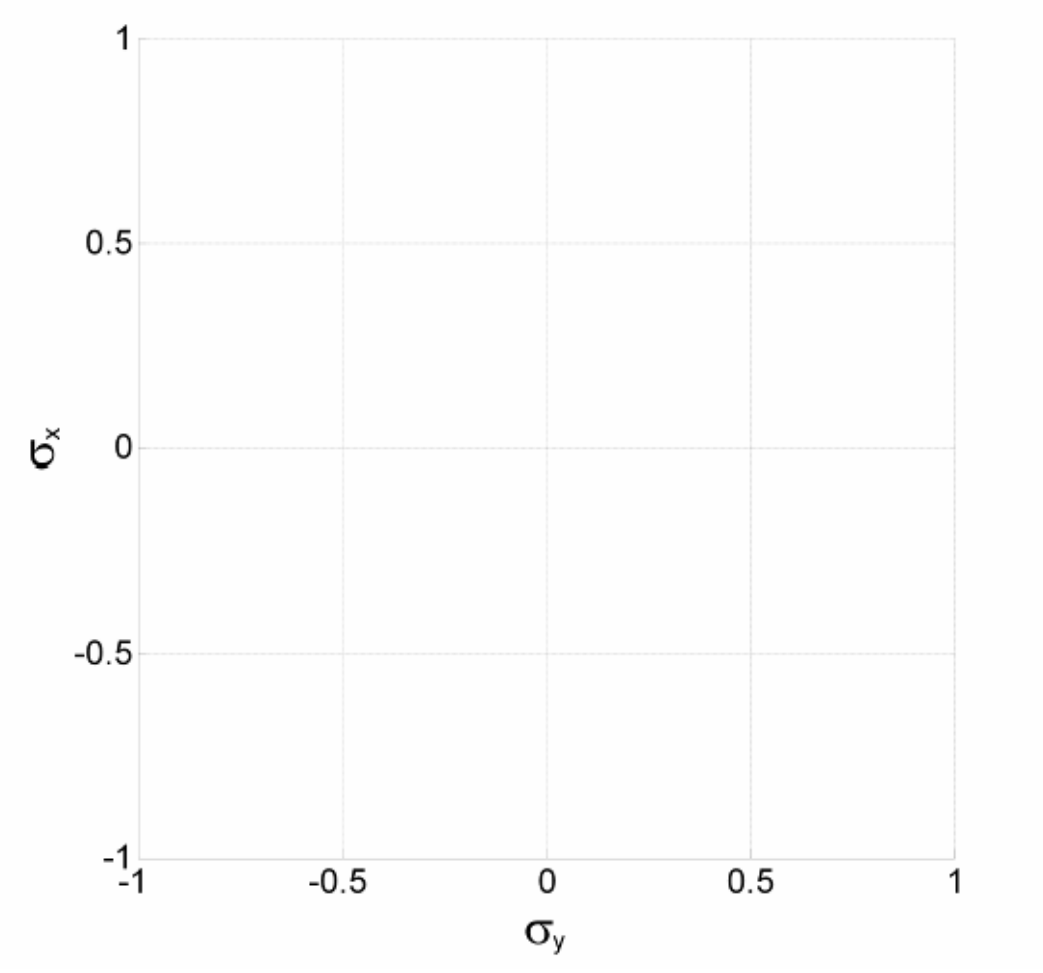1 0.5 0 -0.5 -1<br>σx<br>-1 -0.5 0 0.5 1<br>σy | 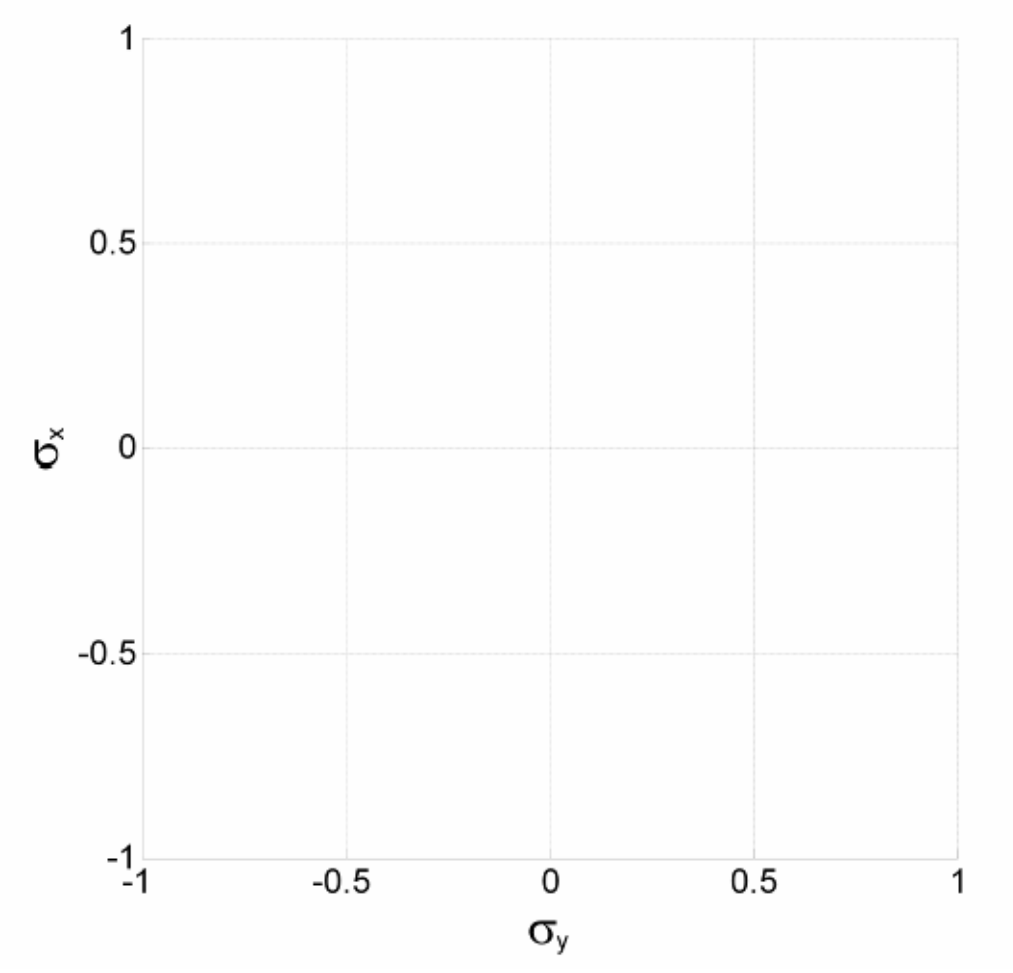1 0.5 0 -0.5 -1<br>σx<br>-1 -0.5 0 0.5 1<br>σy | 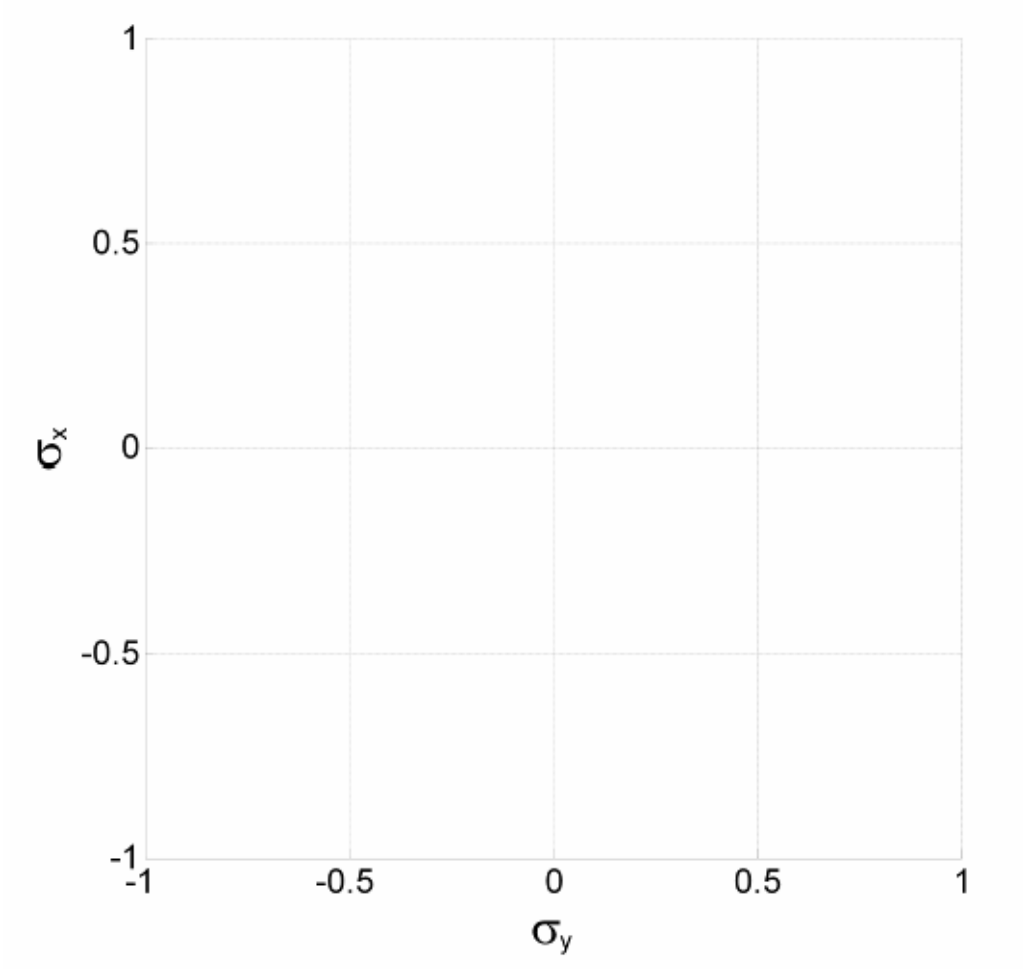1 0.5 0 -0.5 -1<br>σx<br>-1 -0.5 0 0.5 1<br>σy |

**Table 3**

The existence regions of the non-classical relative equilibria with $J_2 = -0.1$

<table>
<tr><td></td><td>$\frac{I_{xx}}{m} = 0.125$</td><td>$\frac{I_{xx}}{m} = 0.5\mathrm{e}-2$</td><td>$\frac{I_{xx}}{m} = 0.5\mathrm{e}-6$</td></tr>
<tr><td>$\Omega_e = 1.5$</td><td>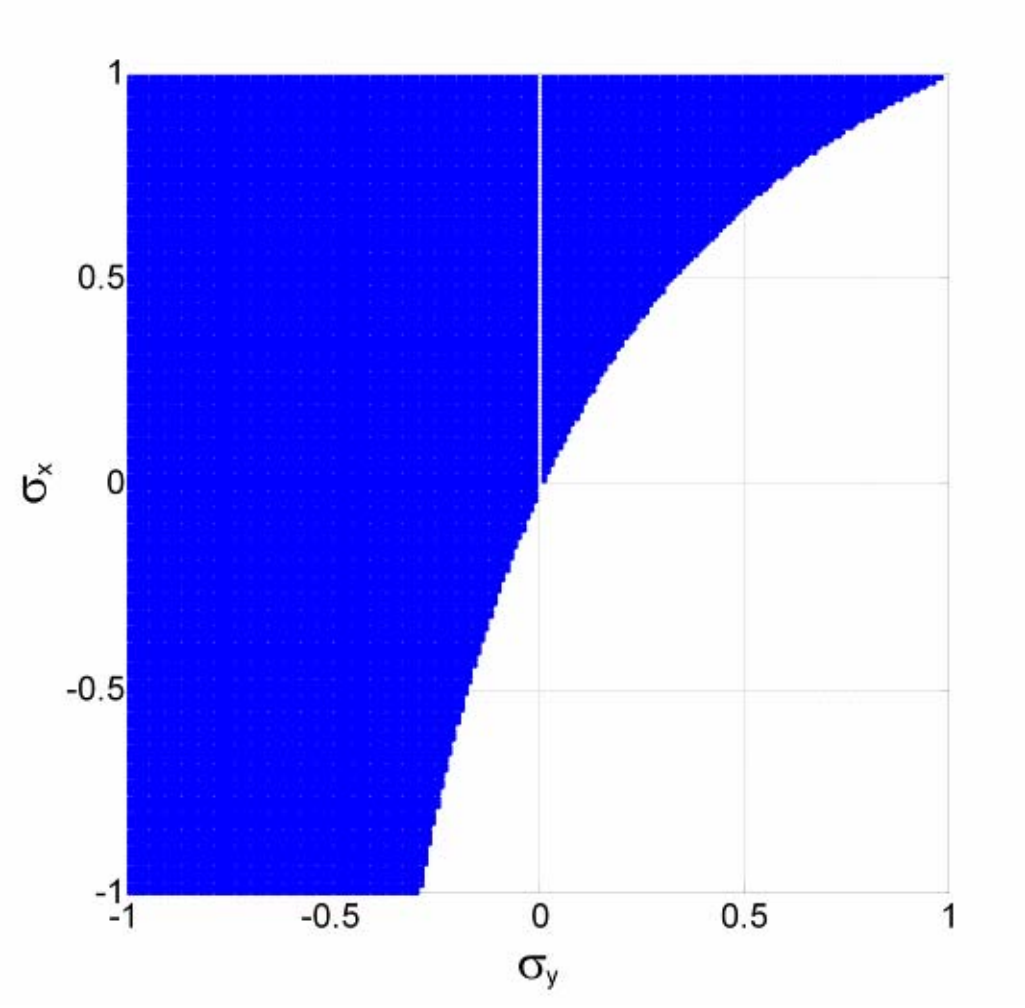
</td><td>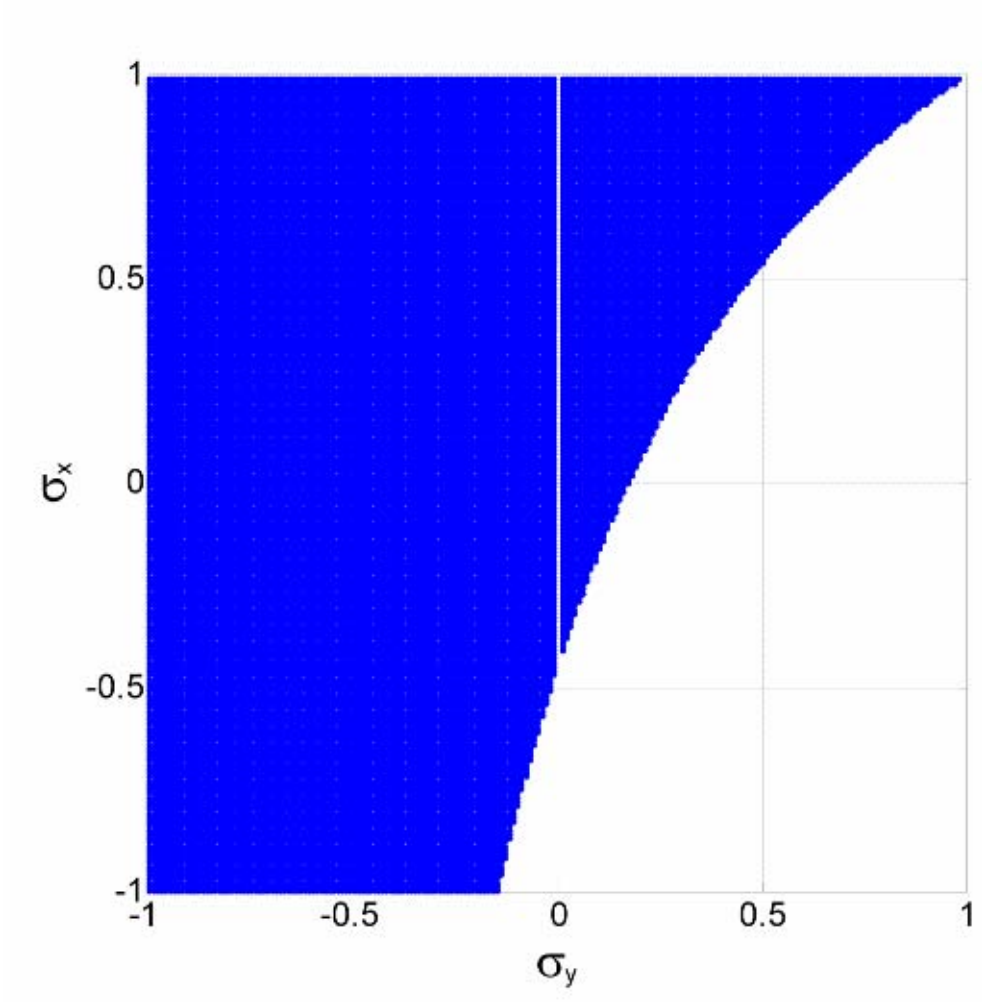
</td><td>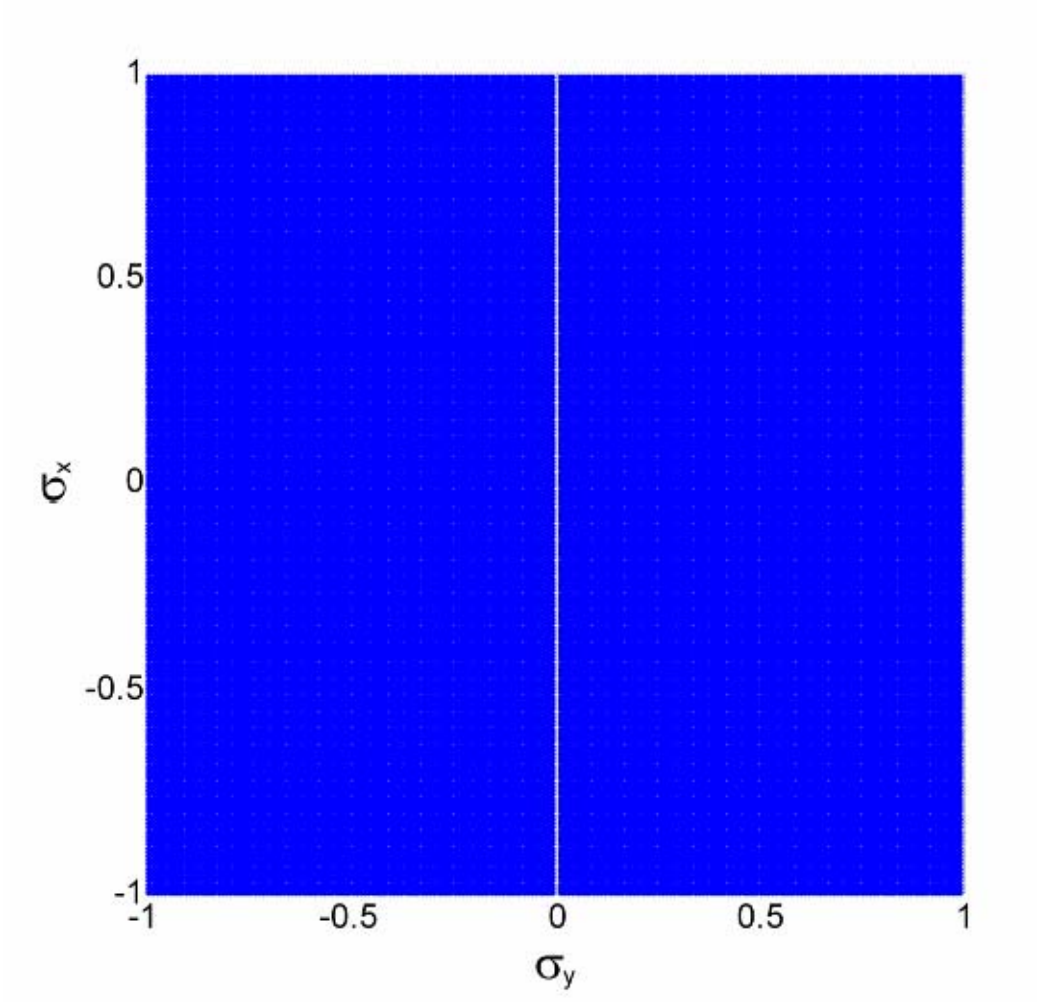
</td></tr>
</table>

$\Omega_e = 1$

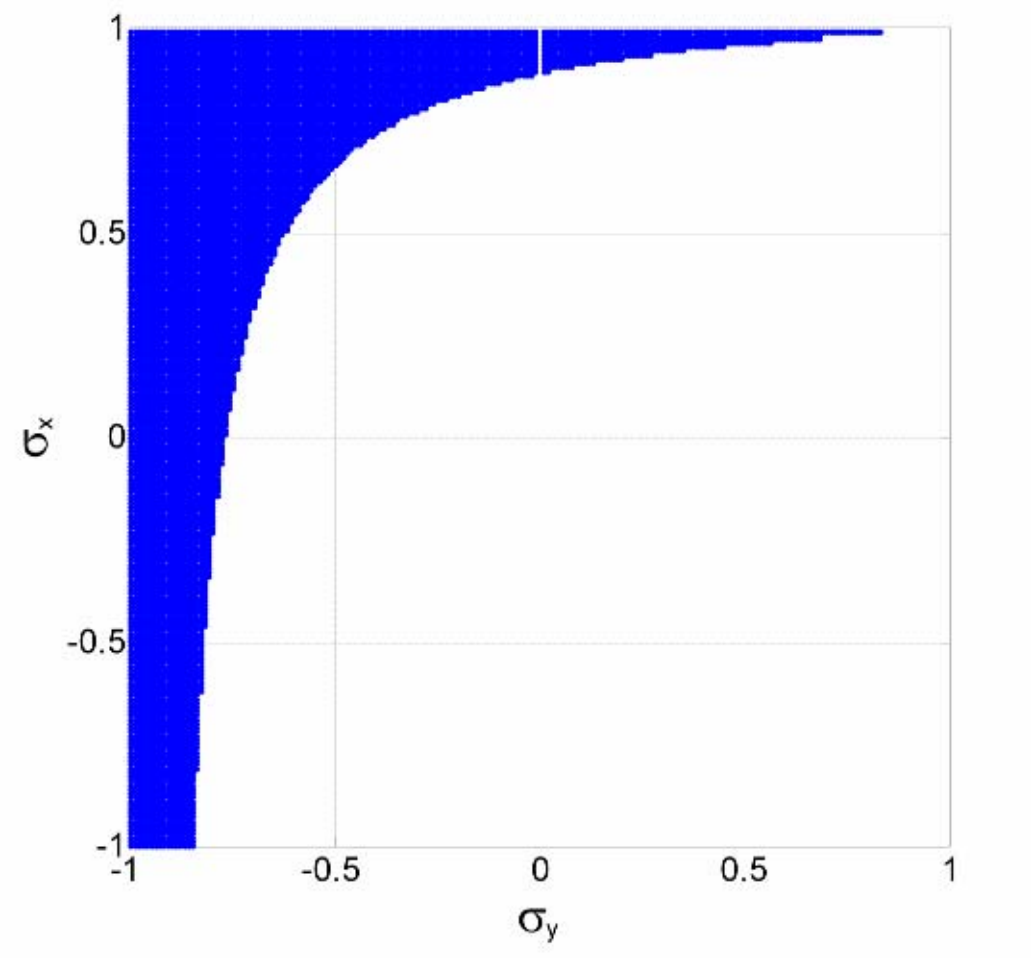

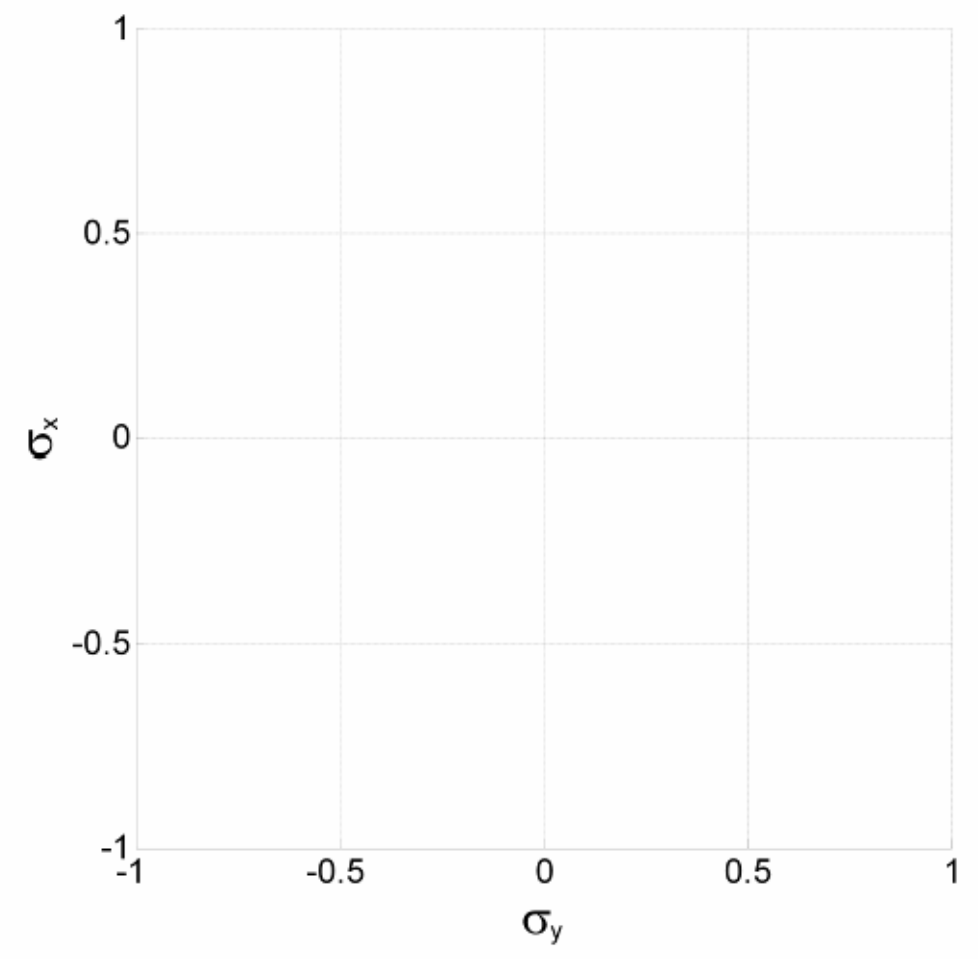

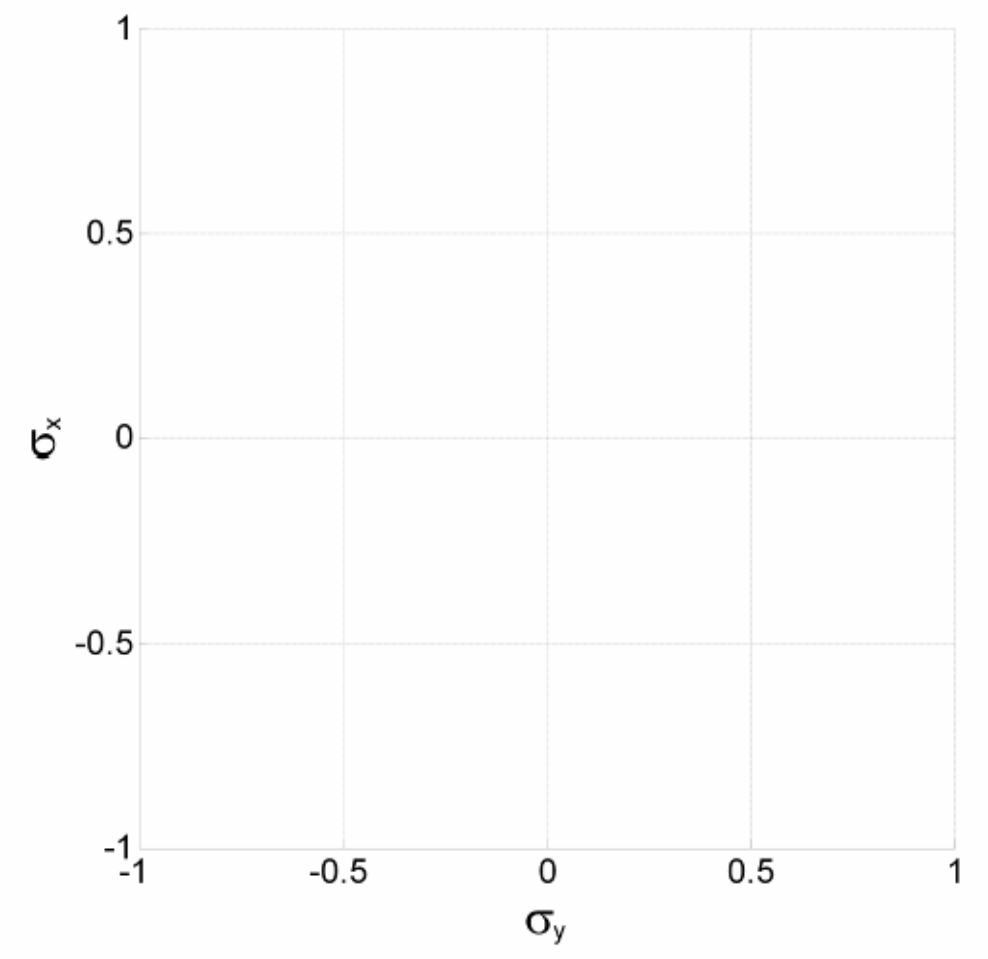

$\Omega_e = 0.5$

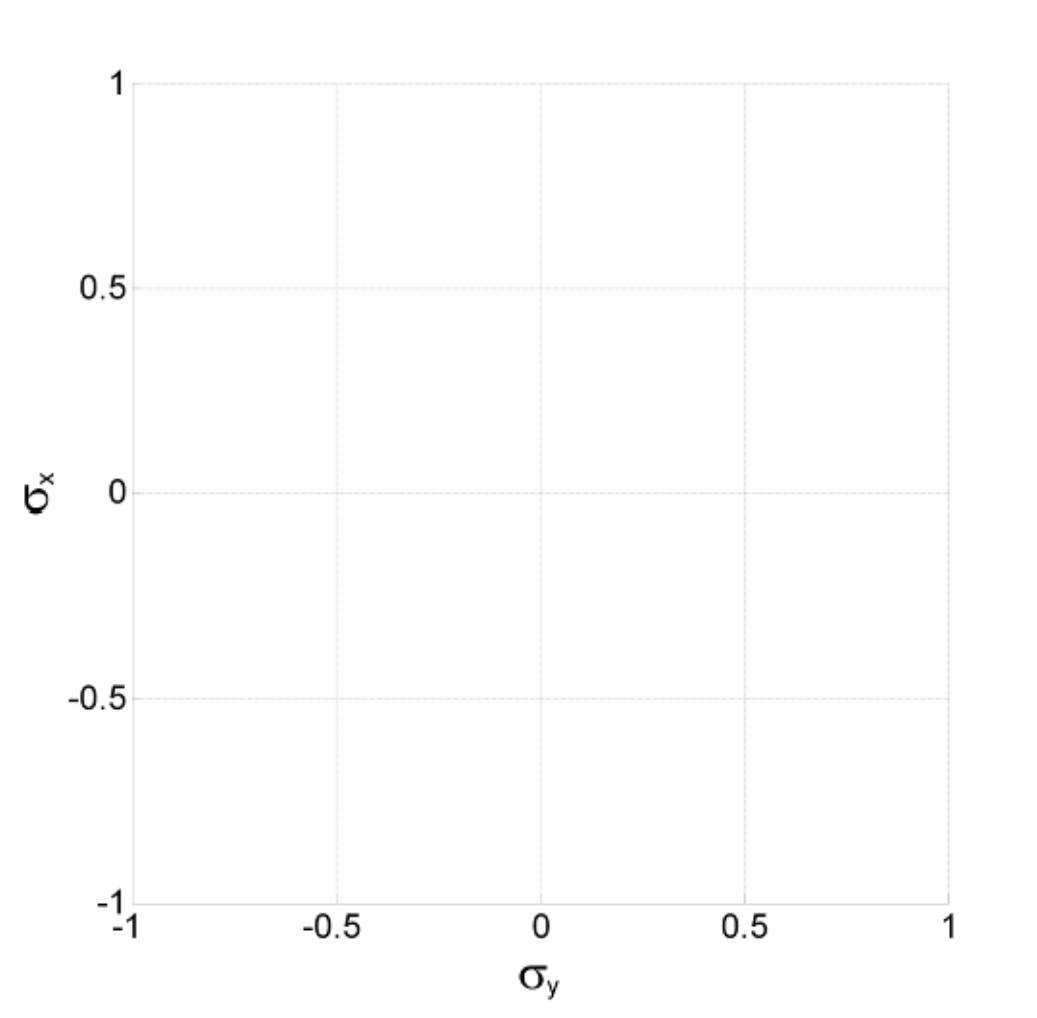

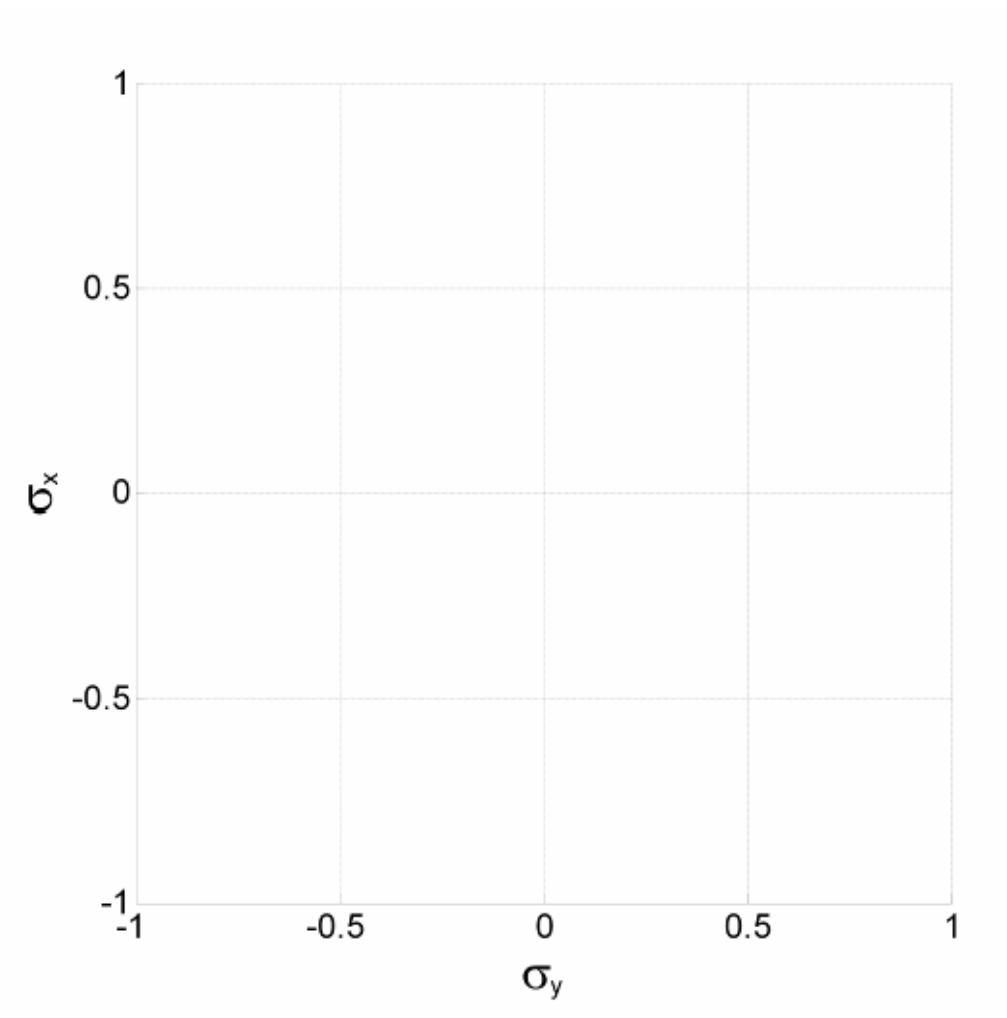

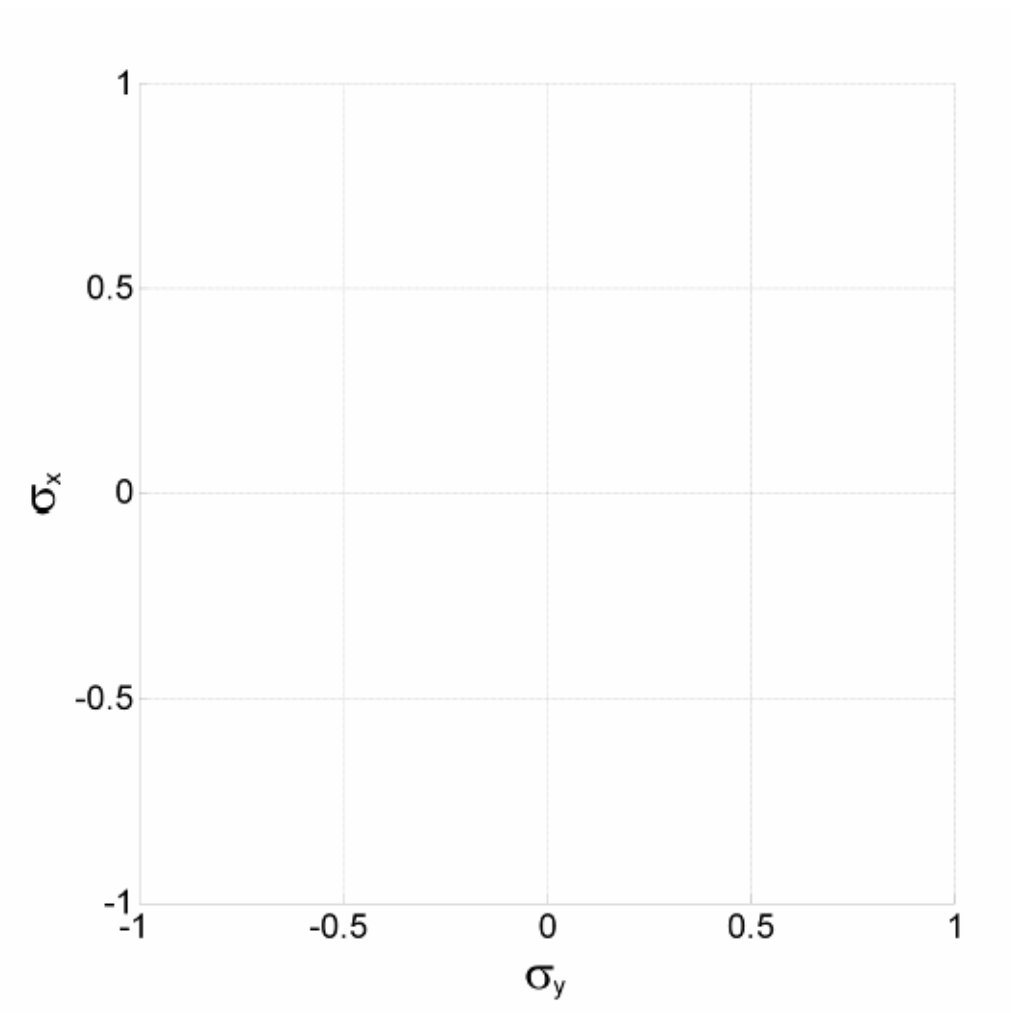

| | | | |
|---|---|---|---|
| $\Omega_e = 0.1$ | 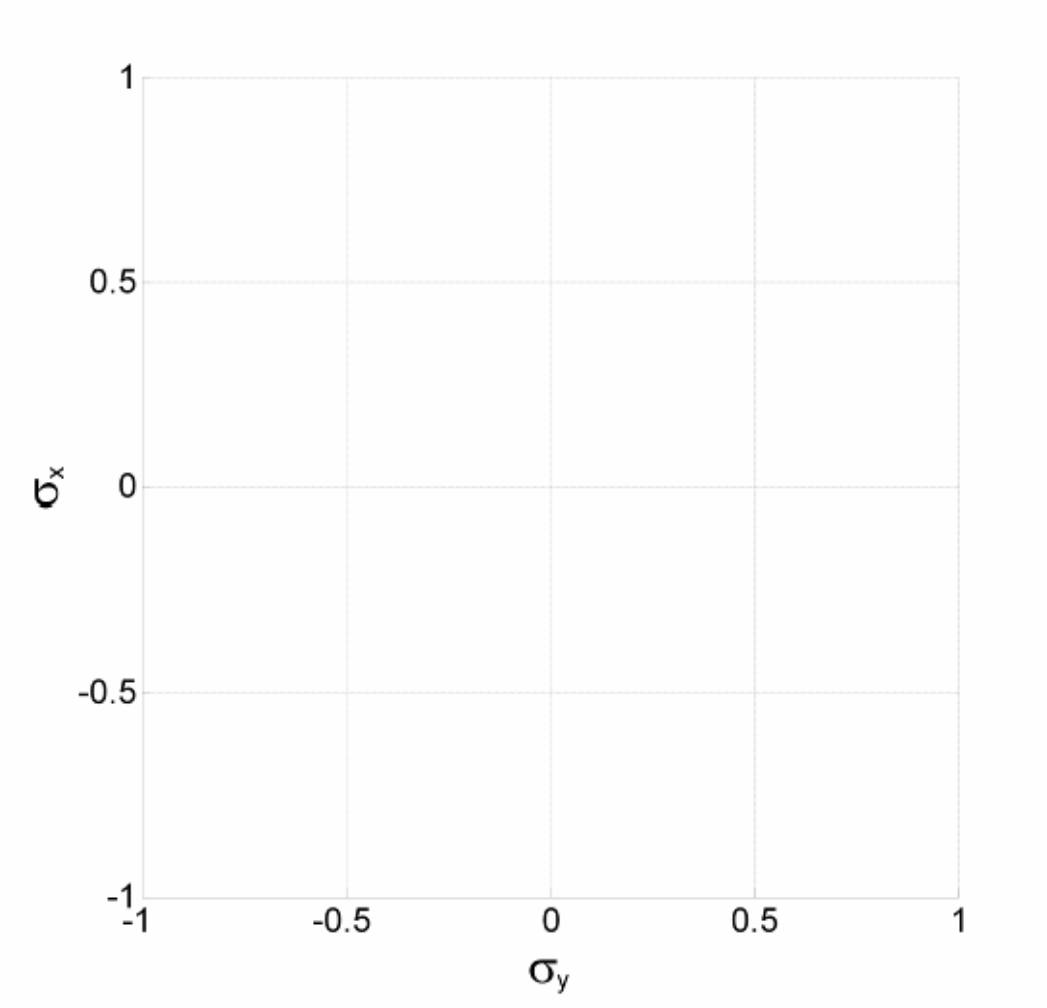1 0.5 0 -0.5 -1 σx -1 -0.5 0 0.5 1 σy | 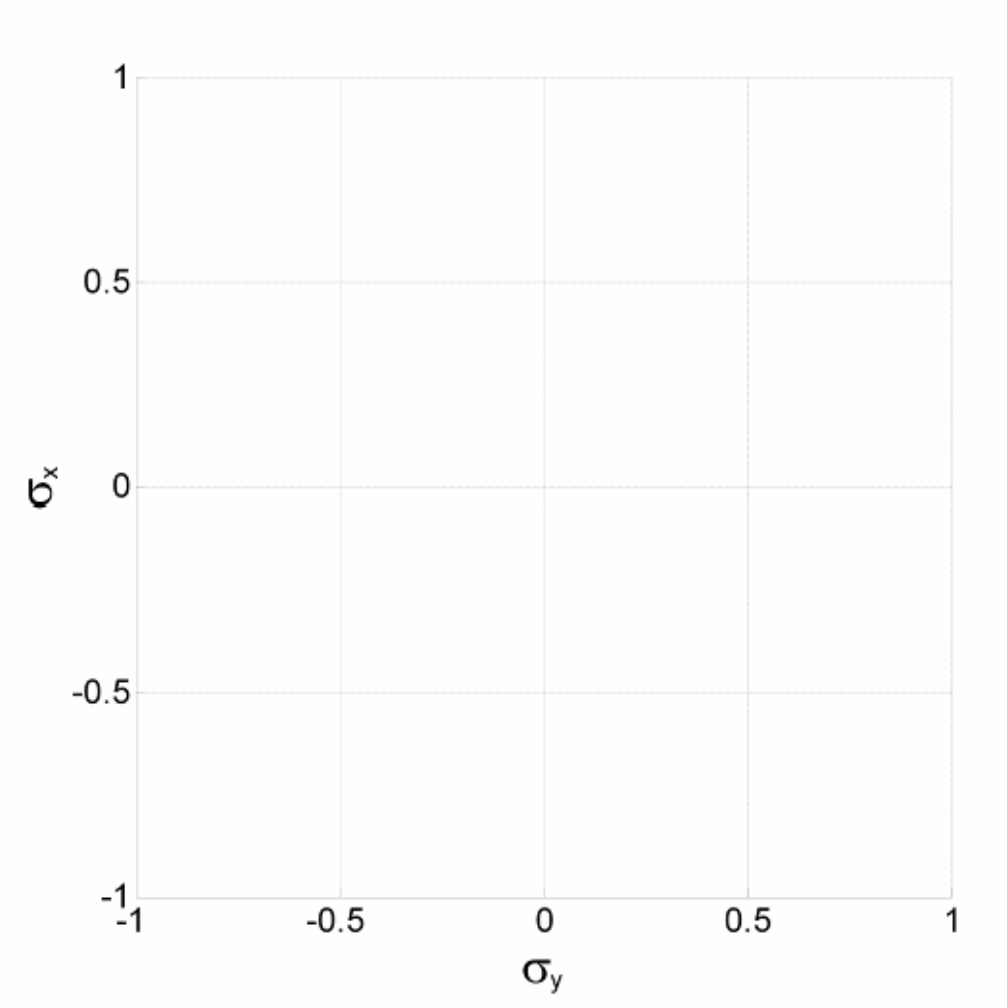1 0.5 0 -0.5 -1 σx -1 -0.5 0 0.5 1 σy | 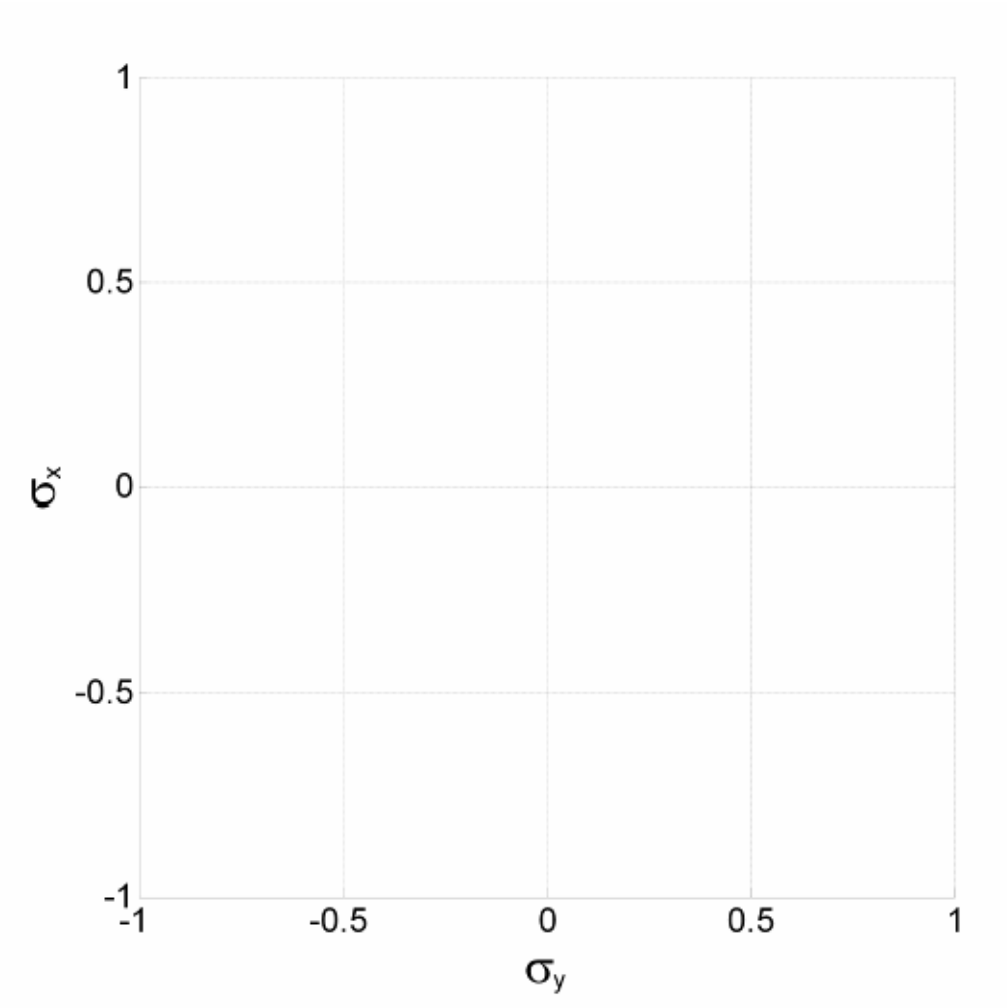1 0.5 0 -0.5 -1 σx -1 -0.5 0 0.5 1 σy |

**Table 4**

The existence regions of the non-classical relative equilibria with $J_2 = -0.05$

| | $\frac{I_{xx}}{m} = 0.125$ | $\frac{I_{xx}}{m} = 0.5\text{e}-2$ | $\frac{I_{xx}}{m} = 0.5\text{e}-6$ |
|---|---|---|---|
| $\Omega_e = 1.5$ | 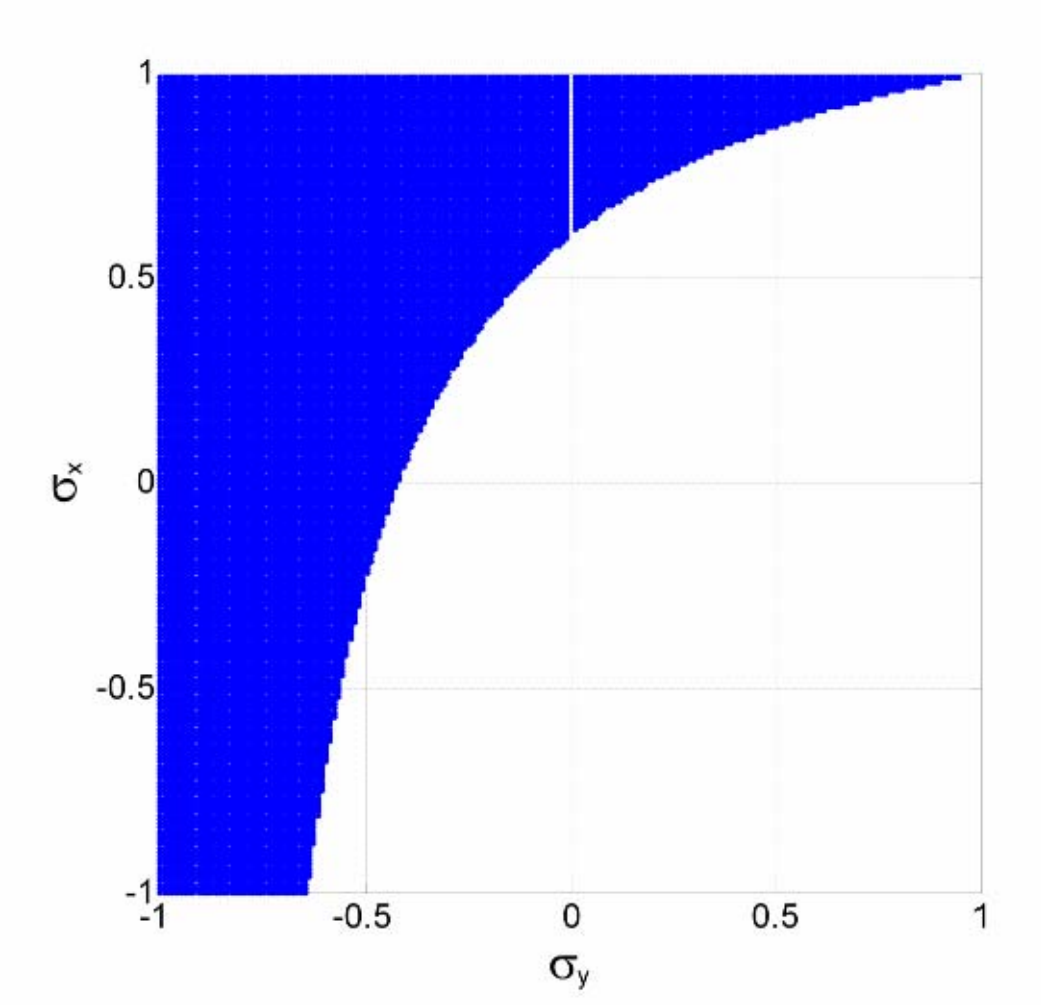 | 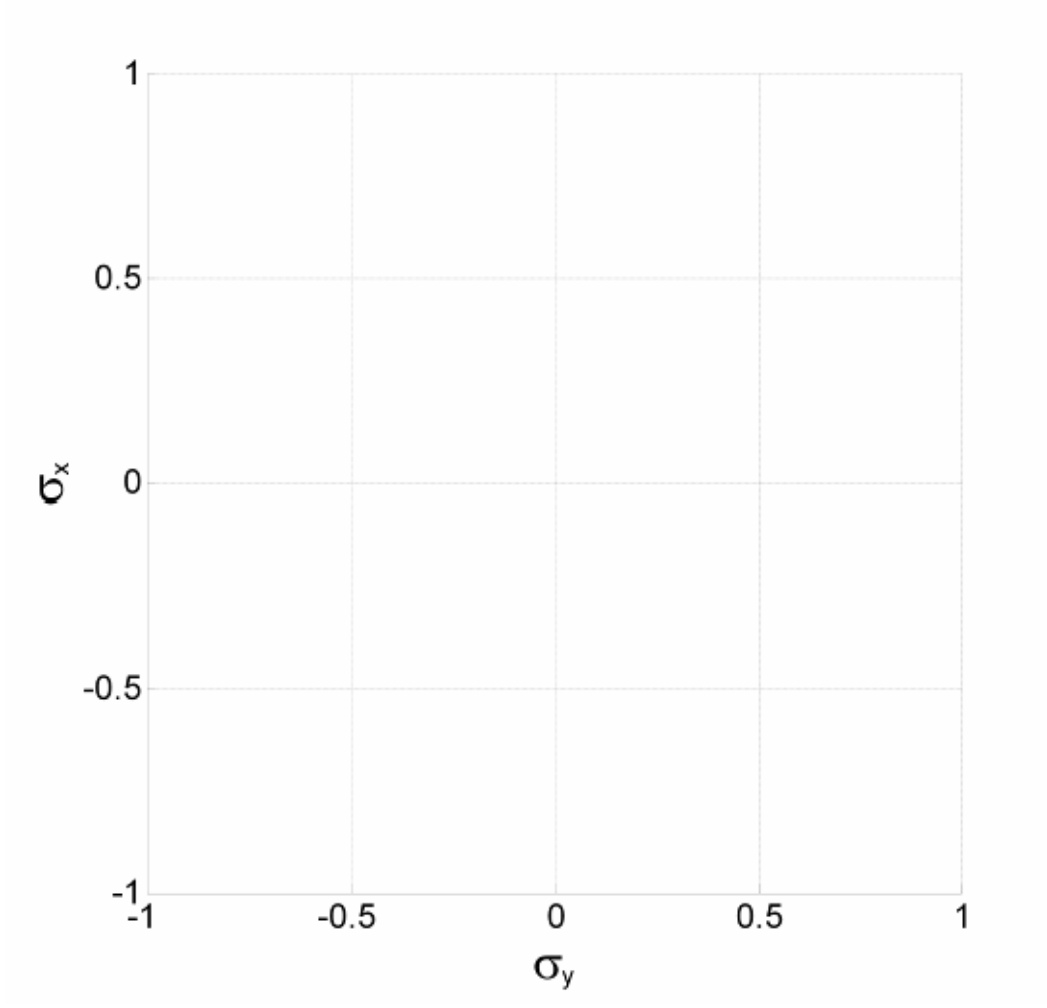 | 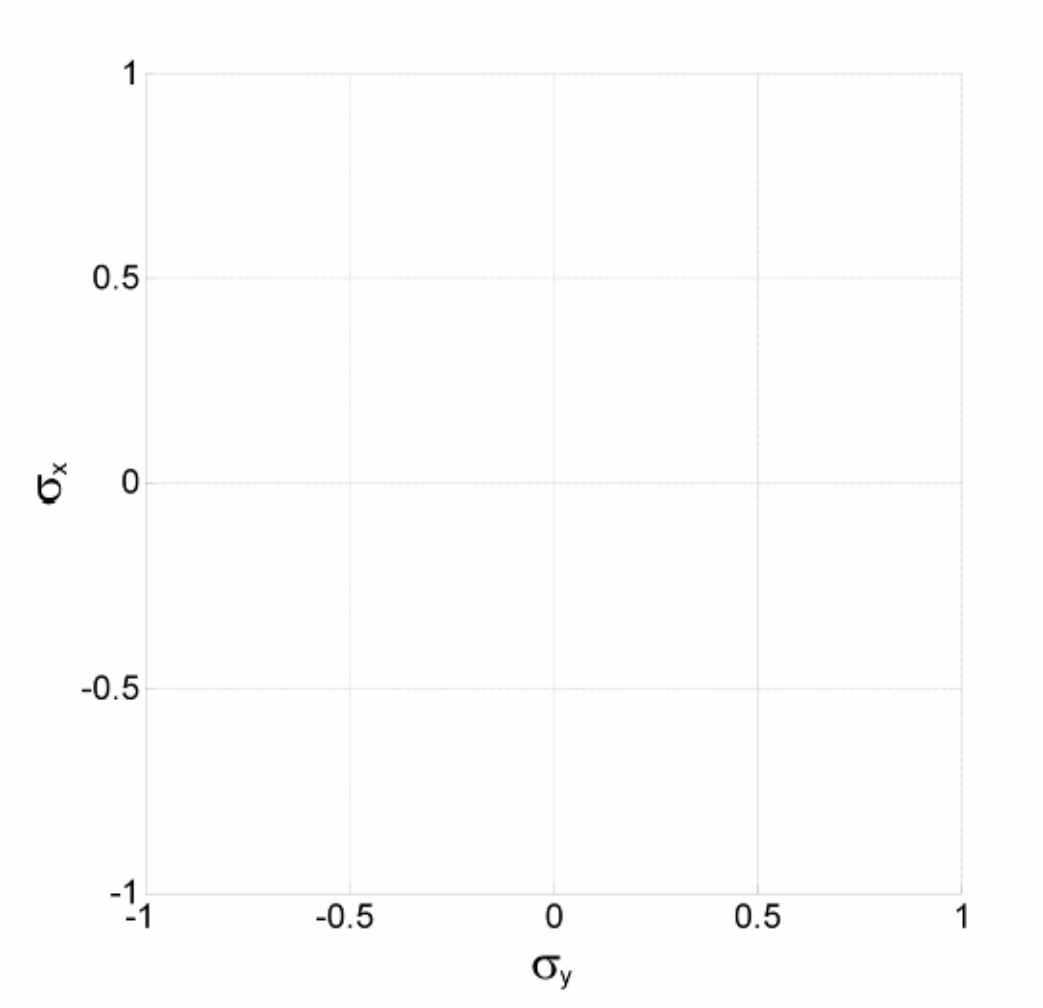 |

<table>
<tr>
<td>$\Omega_e = 1$</td>
<td>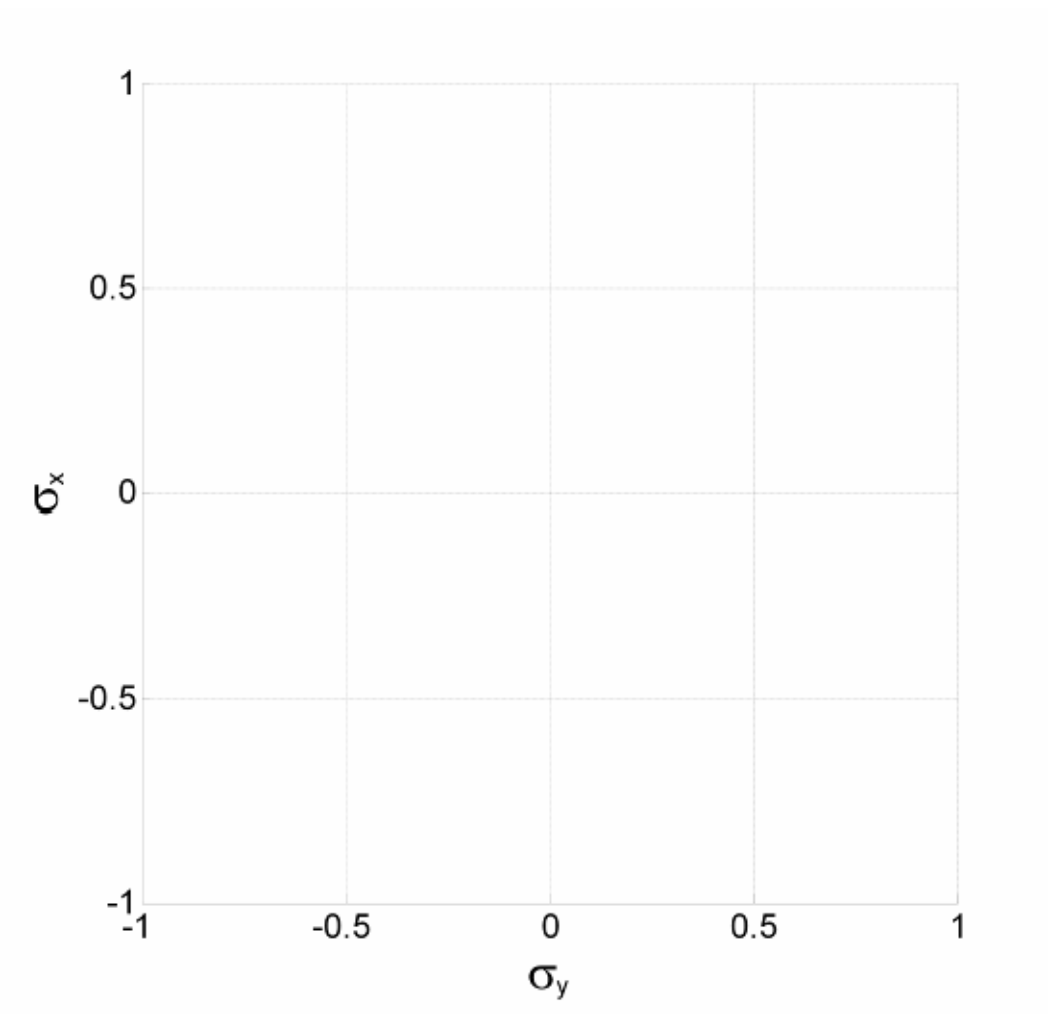
</td>
<td>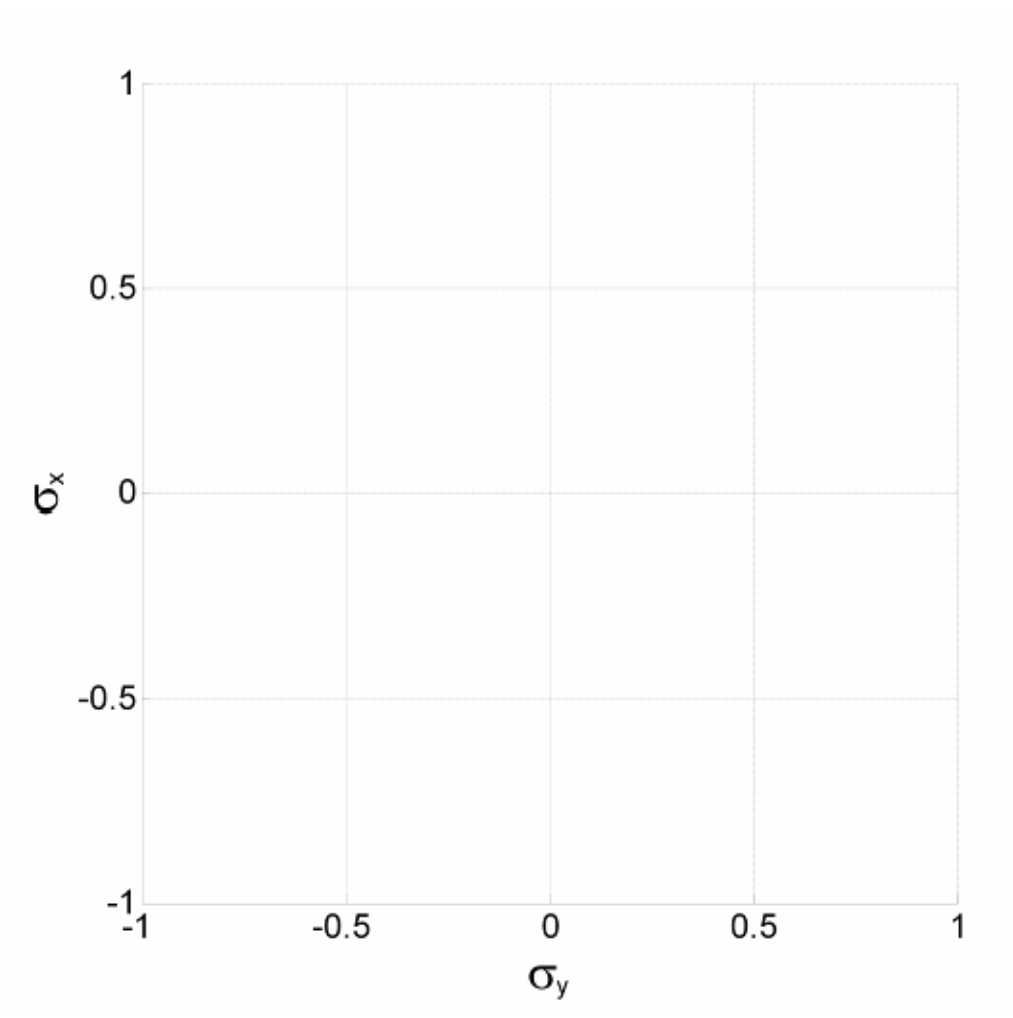
</td>
<td>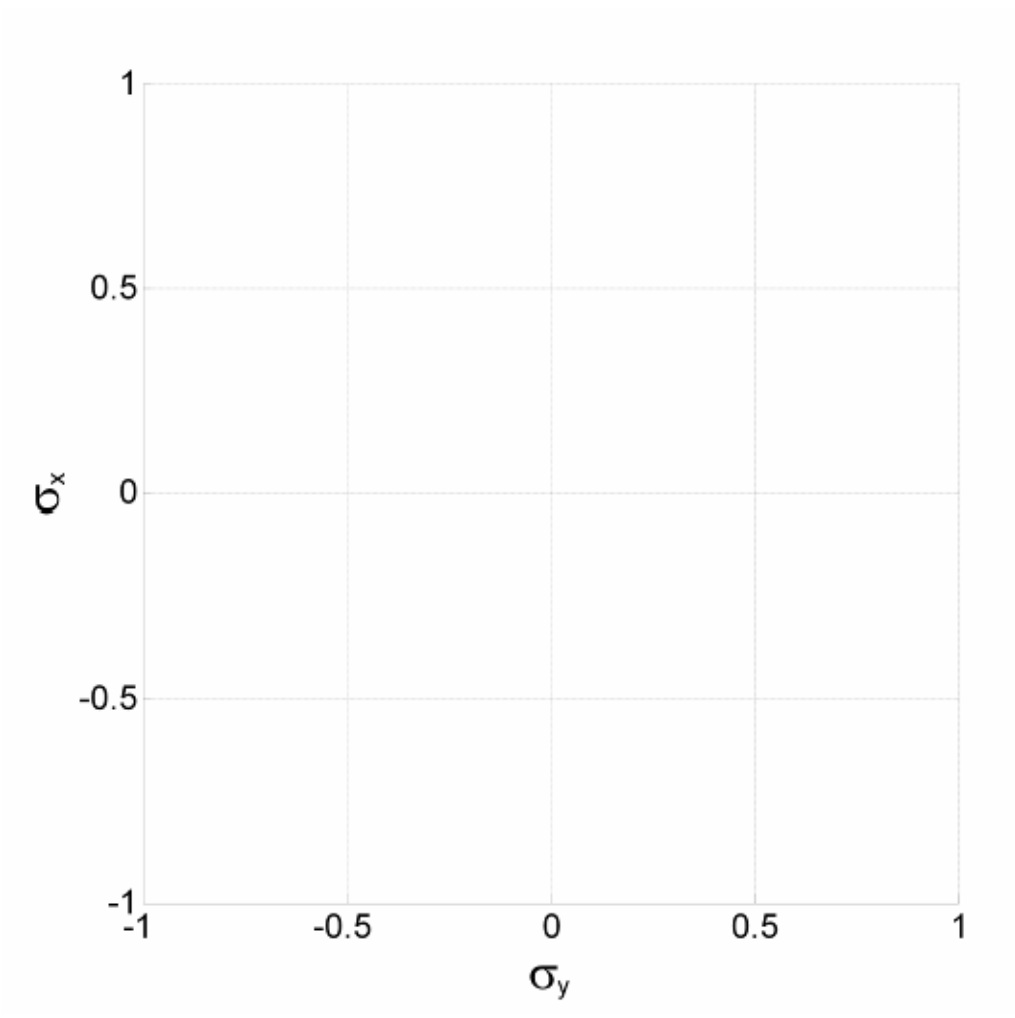
</td>
</tr>
</table>

| | | | |
|---|---|---|---|
| $\Omega_e = 0.5$ | 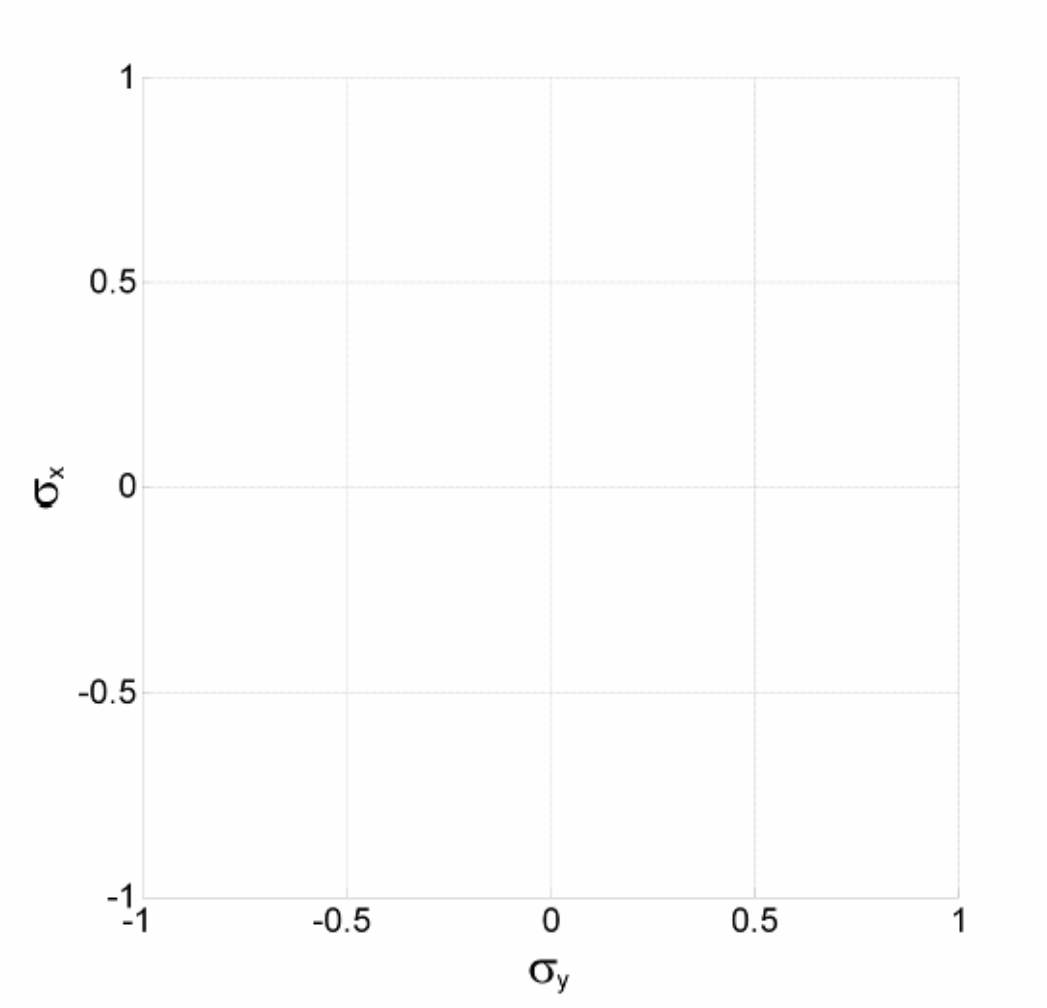1 0.5 0 -0.5 -1 σx -1 -0.5 0 0.5 1 σy | 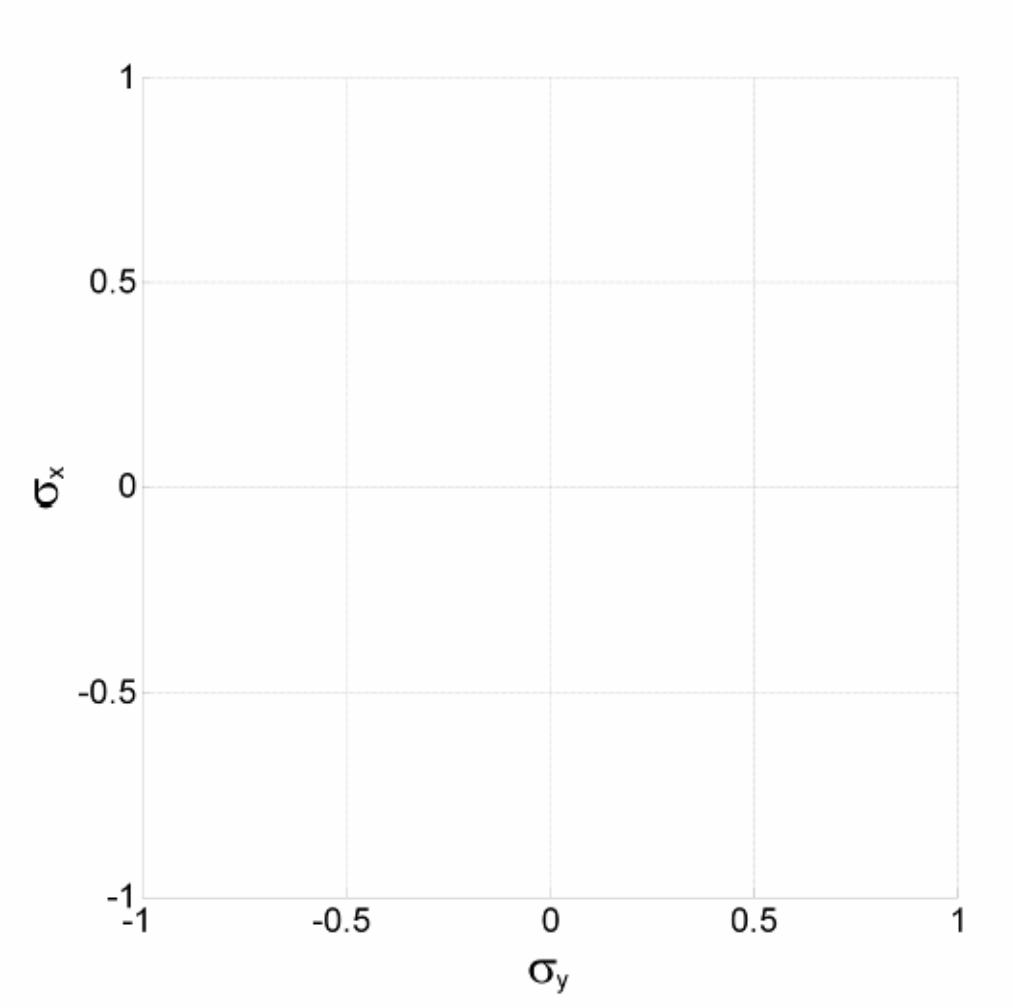1 0.5 0 -0.5 -1 σx -1 -0.5 0 0.5 1 σy | 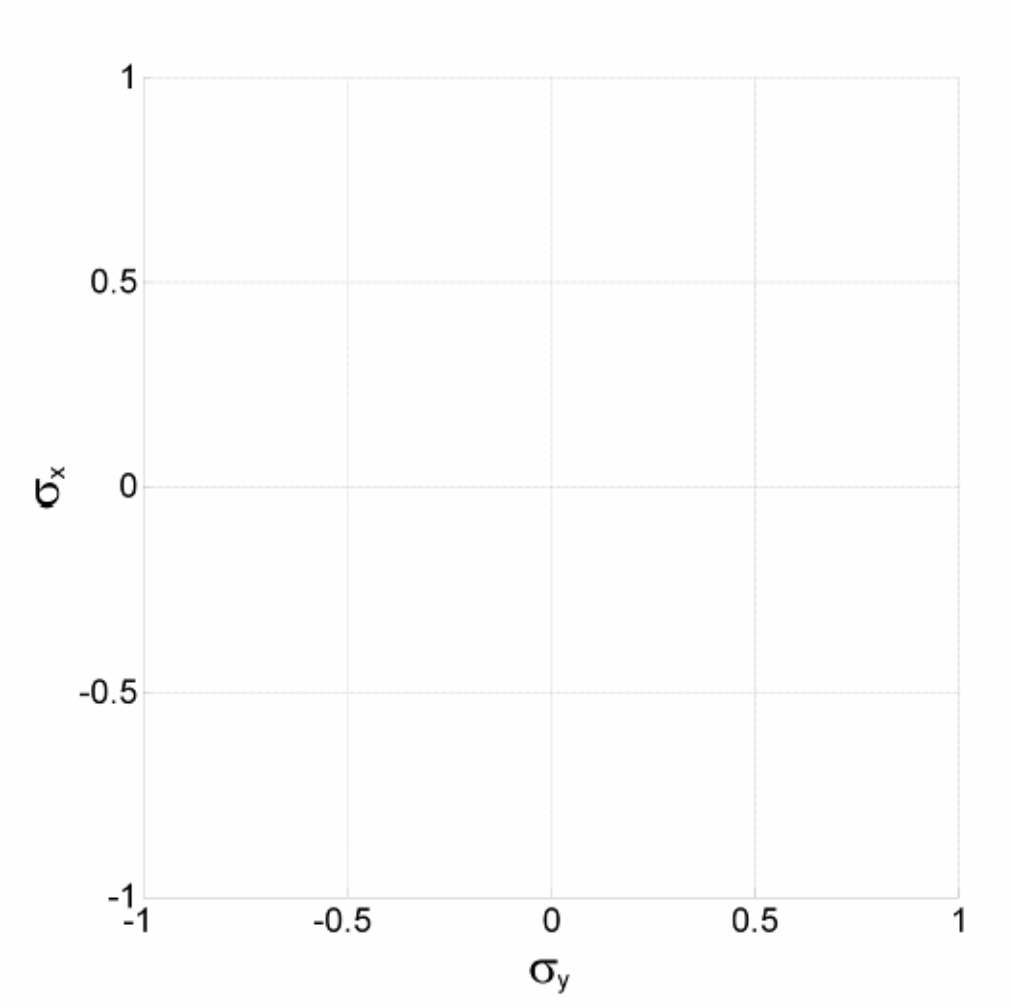1 0.5 0 -0.5 -1 σx -1 -0.5 0 0.5 1 σy |

<table>
<tr>
<td>$\Omega_e = 0.1$</td>
<td>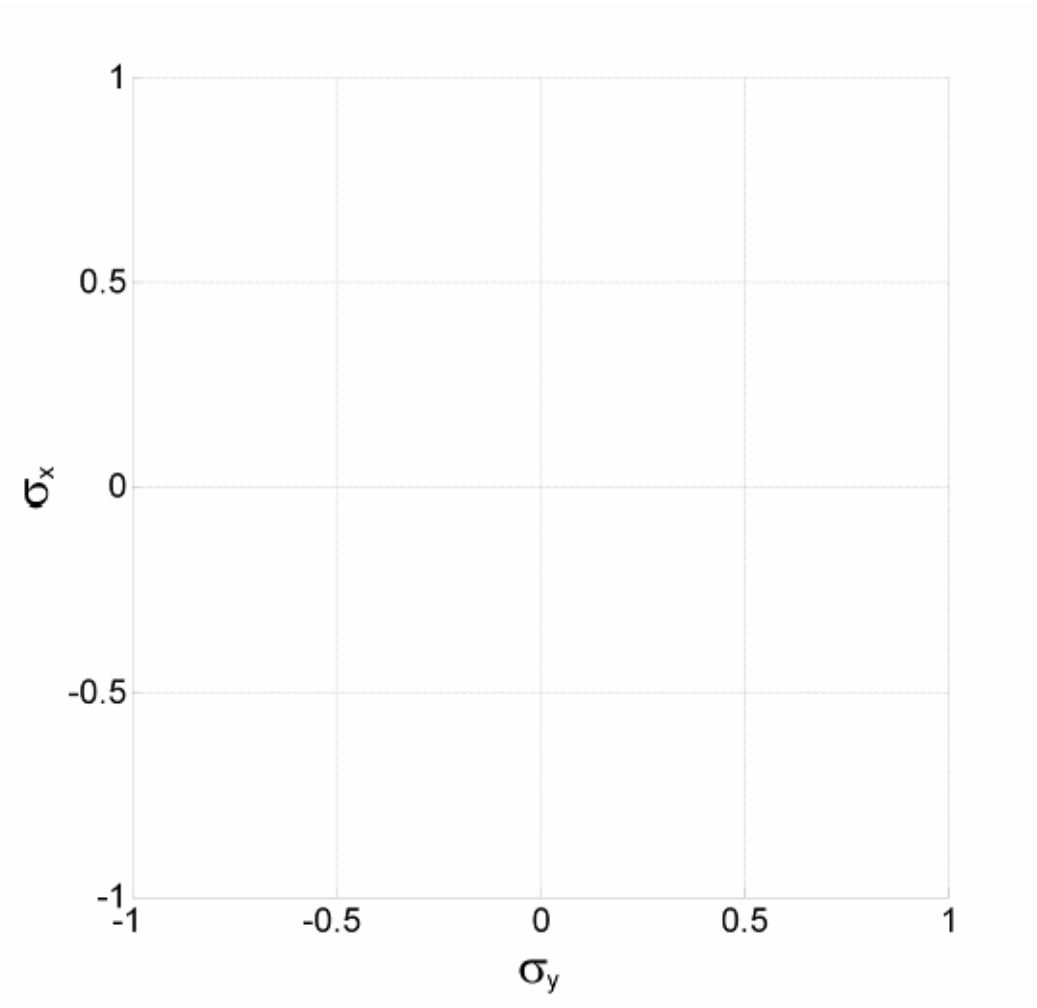
</td>
<td>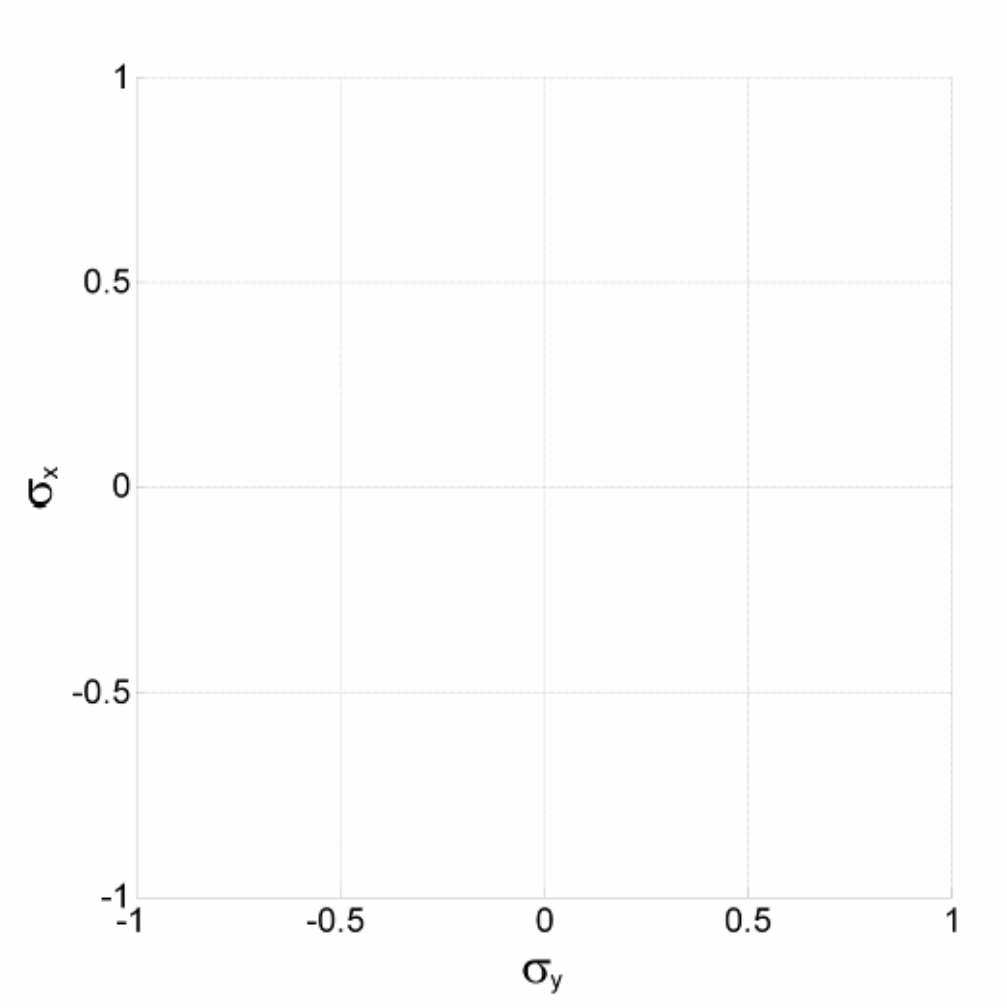
</td>
<td>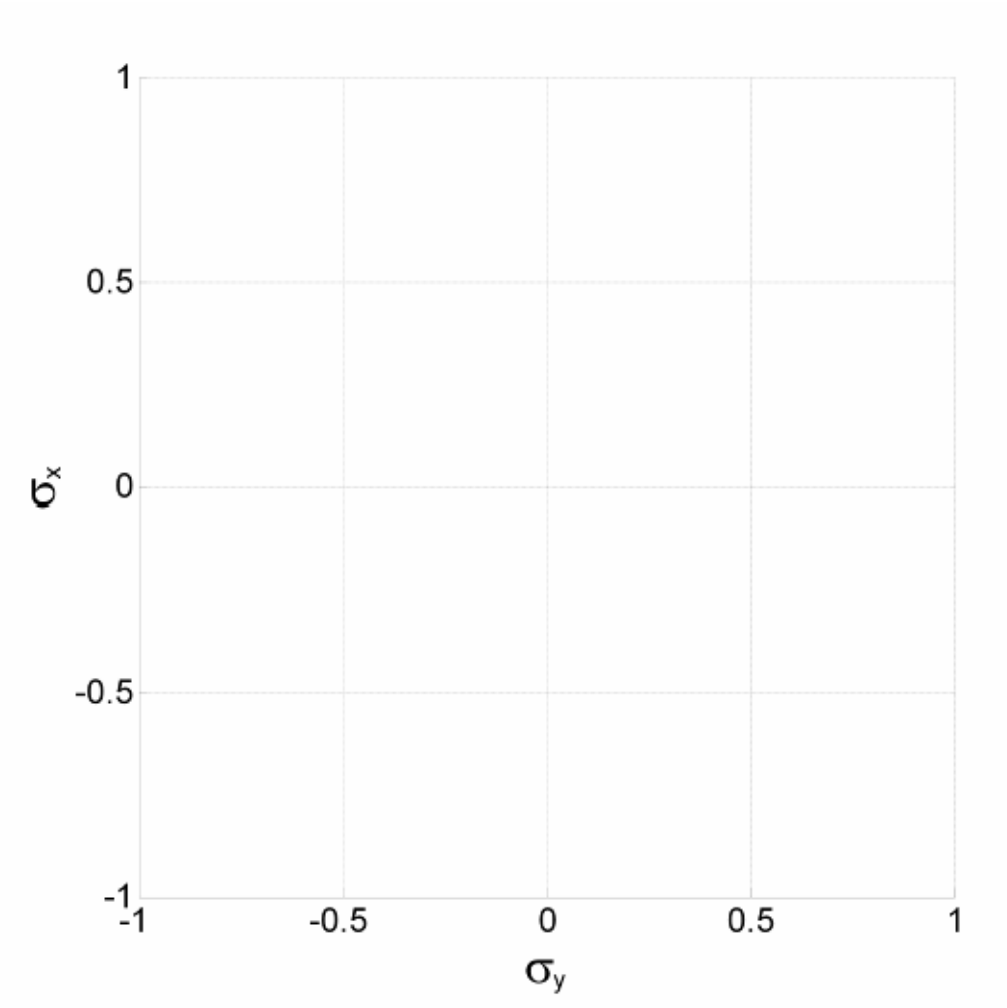
</td>
</tr>
</table>

Through comparisons between existence regions on the $\sigma_y$-$\sigma_x$ plane in Tables 1, 2, 3 and 4 with different values of $J_2$, $I_{xx}/m$ and $\Omega_e$, we can find out the individual effect of the system parameters on the existence of the non-classical relative equilibria. Several important conclusions can be achieved as follows:

**(a). The effect of the second zonal harmonic $J_2$**

The numerical results suggest that the non-classical relative equilibria cannot exist in the case of a positive $J_2$, i.e., the central body *P* is oblate, in the real physical situation. The non-classical relative equilibria can only exist in the case of a negative $J_2$ in the real physical situation when the central body *P* is elongated. If we fix the values of $I_{xx}/m$ and $\Omega_e$, through comparisons between existence regions on the $\sigma_y$-$\sigma_x$ plane with different values of $J_2$, we can easily find that the elongatedness of the central body *P* has a positive effect on the existence of the non-classical relative equilibria. That is to say, with a larger absolute value of the negative $J_2$, the existence region of the non-classical relative equilibria, if exists, is larger.

**(b). The effect of the angular velocity $\Omega_e$**

For each value of the negative $J_2$, with the value of the characteristic dimension of the rigid body $I_{xx}/m$ given, as the orbital angular velocity $\Omega_e$ decreasing, the existence region of the non-classical relative equilibria is getting smaller and smaller, and disappears eventually. This phenomenon is due to the fact that as the orbital angular velocity $\Omega_e$ decreasing, the orbital radius is getting larger, then both effects of the second zonal harmonic of the central body *P* and the orbit-rotation coupling of the rigid body *B* are getting weaker. The individual effect of the second zonal harmonic and the orbit-rotation coupling cannot be distinguished from this point.

**(c). The effect of the orbit-rotation coupling of the rigid body**

Notice that the orbit-rotation coupling is more significant when the ratio of the characteristic dimension of the rigid body to the orbit radius is larger. The effect of the orbit-rotation coupling of the rigid body can be discussed through comparisons between existence regions with different values of $I_{xx}/m$.

As shown by Tables 1, 2, 3 and 4, the effect of the orbit-rotation coupling of the rigid body on the existence of the non-classical relative equilibria is complex, since it can be positive or negative, which depends on the values of $J_2$ and $\Omega_e$.

When the values of $J_2$ and $\Omega_e$ favor the existence of the non-classical relative equilibria, that is to say, a negative $J_2$ with a large absolute value and a large $\Omega_e$, including $J_2 = -0.5$, $\Omega_e$=1.5, 1 or 0.5; $J_2 = -0.2$, $\Omega_e$=1.5, or 1; $J_2 = -0.1$, $\Omega_e$=1.5, the orbit-rotation coupling of the rigid body has a negative effect on the existence of the non-classical relative equilibria. If we fix the value of the orbital angular velocity $\Omega_e$ at 1.5, 1 or 0.5 when $J_2 = -0.5$ (or at 1.5 or 1 when $J_2 = -0.2$; at 1.5 when $J_2 = -0.1$), as the characteristic dimension of the rigid body $I_{xx}/m$ decreasing, the existence region of the non-classical relative equilibria is getting larger and larger, and is equal to the whole $\sigma_y$-$\sigma_x$ plane eventually. That is to say, the orbit-rotation coupling of the rigid body *B* has a negative effect on the existence of the non-classical relative equilibria in these cases.

However, when the values of $J_2$ and $\Omega_e$ do not favor the existence of the non-classical relative equilibria, that is to say, a negative $J_2$ with a small absolute value and a small $\Omega_e$, including $J_2=-0.1$, $\Omega_e=1$; $J_2=-0.05$, $\Omega_e=1.5$, the orbit-rotation coupling of the rigid body has a positive effect on the existence of the non-classical relative equilibria. If we fix the value of the orbital angular velocity $\Omega_e$ at 1 when $J_2=-0.1$ (or at 1.5 when $J_2=-0.05$), as the characteristic dimension of the rigid body $I_{xx}/m$ decreasing, the existence region of the non-classical relative equilibria disappears. That is to say, the orbit-rotation coupling of the rigid body $B$ has a positive effect on the existence of the non-classical relative equilibria in these cases.

When the characteristic dimension of the rigid body $I_{xx}/m$ is very small, such as $I_{xx}/m=0.5\text{e}-6$, the effect of the orbit-rotation coupling of the rigid body is very weak and the mass distribution of the rigid body $\sigma_x$ and $\sigma_y$ has no influence on the existence of the non-classical relative equilibria. The existence of the non-classical relative equilibria will be determined by the values of $J_2$ and $\Omega_e$. In this case, the non-classical relative equilibria can exist on the whole $\sigma_y$-$\sigma_x$ plane when the values of $J_2$ and $\Omega_e$ favor the existence of the non-classical relative equilibria, such as $J_2=-0.5$, $\Omega_e$=1.5; the non-classical relative equilibria cannot exist on the whole $\sigma_y$-$\sigma_x$ plane when the values of $J_2$ and $\Omega_e$ do not favor the existence of the non-classical relative equilibria, such as $J_2=-0.1$, $\Omega_e$=1.

When the characteristic dimension of the rigid body is large, such as $I_{xx}/m=0.125$, the effect of the orbit-rotation coupling is significant. The orbit-rotation coupling can destroy the existence of the non-classical relative equilibria in some regions on the $\sigma_y$-$\sigma_x$ plane when the values of $J_2$ and $\Omega_e$ favor the existence of the non-classical relative equilibria, such as $J_2 = -0.5$, $\Omega_e$=1.5; the orbit-rotation coupling can lead to the existence of the non-classical relative equilibria on the $\sigma_y$-$\sigma_x$ plane when the values of $J_2$ and $\Omega_e$ do not favor the existence of the non-classical relative equilibria, such as $J_2 = -0.1$, $\Omega_e$=1.

When the characteristic dimension of the rigid body is very small, the orbit-rotation coupling is insignificant and the rigid body can be considered as point mass. According to our conclusions stated above, the displaced orbit can exist for a point mass in a $J_2$ gravity field with a negative $J_2$ when the orbital angular velocity $\Omega_e$ is large. This is consistent with the conclusion in Howard (1990) that the nonequatorial stationary orbits can exist with a negative $J_2$ when the orbital angular velocity is large enough, and the lower limit of the orbital angular velocity for the existence of the nonequatorial stationary orbits is smaller with a larger absolute value of the negative $J_2$.

**(d). The effect of the mass distribution parameters $\sigma_x$ and $\sigma_y$**

From the existence regions on the $\sigma_y$-$\sigma_x$ plane in Tables 1, 2, 3 and 4, we can easily find that the mass distribution in the left upper side of the $\sigma_y$-$\sigma_x$ plane, which means the mass distribution of the rigid body is near to a rod along the ***j***-axis or ***k***-axis, favor the existence of the non-classical relative equilibria. Whereas the mass distribution in the right lower side of the $\sigma_y$-$\sigma_x$ plane, which means the mass distribution of the rigid body is near to a rod along the ***i***-axis, do not favor the existence of the non-classical relative equilibria.

Here we have given an throughout analysis on the existence of the non-classical relative equilibria with respect to all the five parameters of the system, i.e., $J_2$, $\Omega_e$, $I_{xx}/m$, $\sigma_x$ and $\sigma_y$. The individual effect of the parameters of the system has been discussed. Our analysis here is more systematical than that in Wang and Xu (2013).

### 4.3 Details of non-classical relative equilibria

We will investigate the details of the non-classical relative equilibria in the following four different cases for a rigid body with $\sigma_x = 0.5$ and $\sigma_y = -0.5$:

$J_2 = -0.5$:

(1). $I_{xx}/m = 0.125$; (2). $I_{xx}/m = 0.5\mathrm{e}-6$;

$J_2 = -0.2$:

(3). $I_{xx}/m = 0.125$; (4). $I_{xx}/m = 0.5\mathrm{e}-6$.

The curves of the parameters of the non-classical relative equilibria including $R_e^x$, $R_e^z$, $\theta$, $\theta_1$, $\theta_2$, $r_e^x$ and $r_e^z$ with respect to the orbital angular velocity $\Omega_e$ are given in Figs. 7-12, where $r_e^x$ and $r_e^z$ are the components of the position vector of the mass center of the rigid body in and perpendicular to the equatorial plane of the body *P* respectively. $r_e^x$ and $r_e^z$ are given by:

$$r_e^x = \cos\theta\sqrt{\left(R_e^x\right)^2 + \left(R_e^z\right)^2}\ ,\ \ r_e^z = \sin\theta\sqrt{\left(R_e^x\right)^2 + \left(R_e^z\right)^2}\ . \tag{43}$$

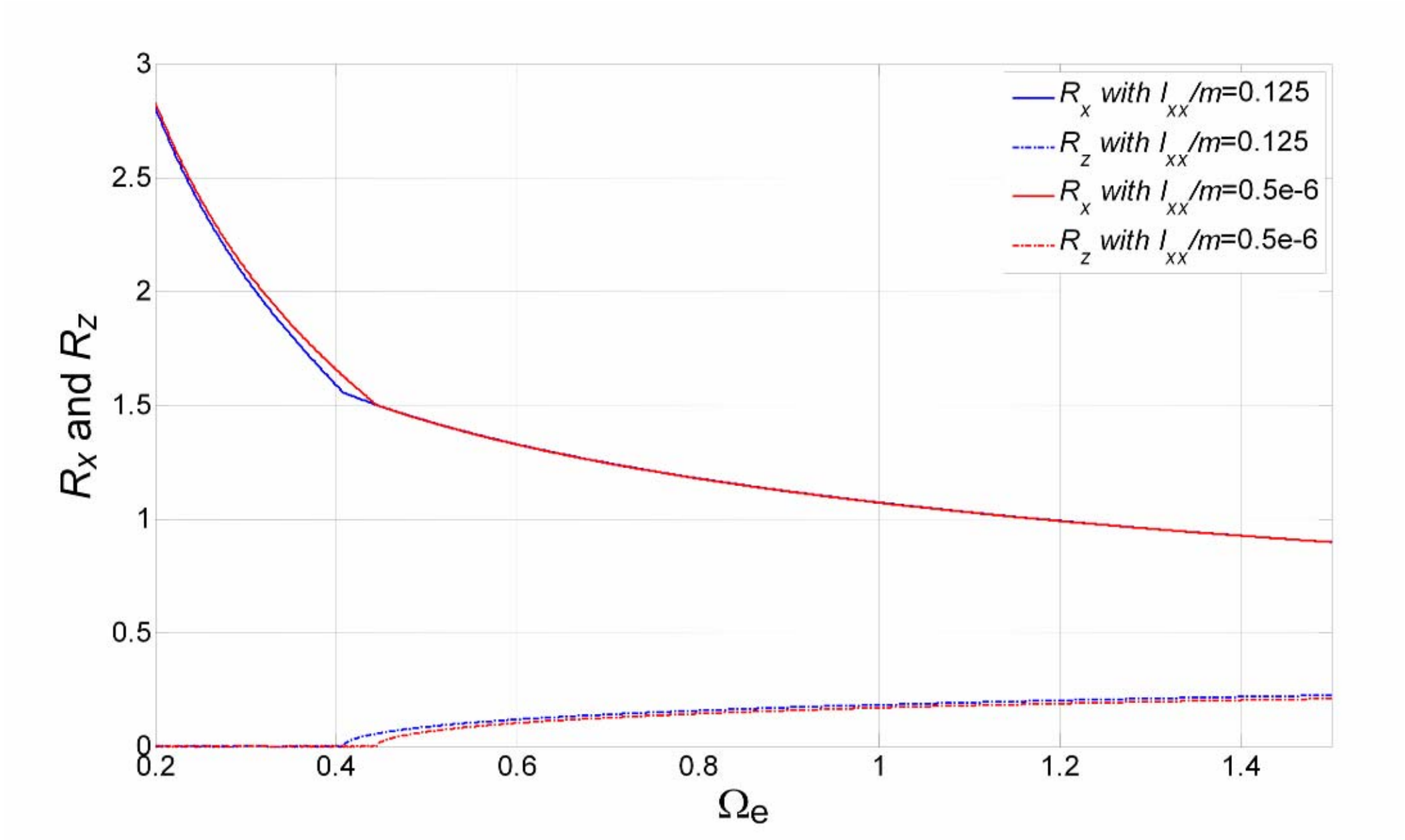

**Fig. 7.** The curves of $R_e^x$ and $R_e^z$ with respect to the angular velocity $\Omega_e$ with $J_2 = -0.5$

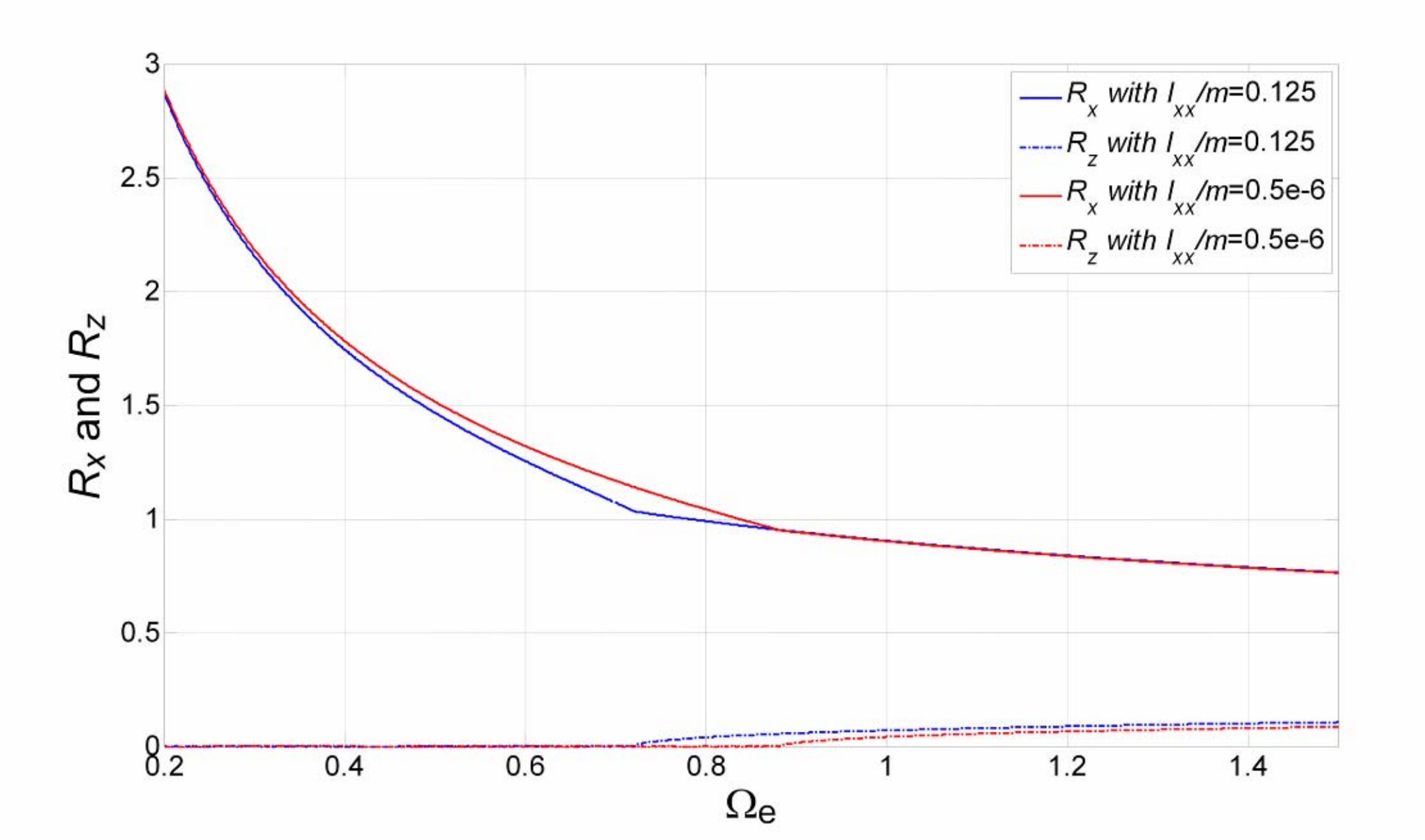

**Fig. 8.** The curves of $R_e^x$ and $R_e^z$ with respect to the angular velocity $\Omega_e$ with $J_2 = -0.2$

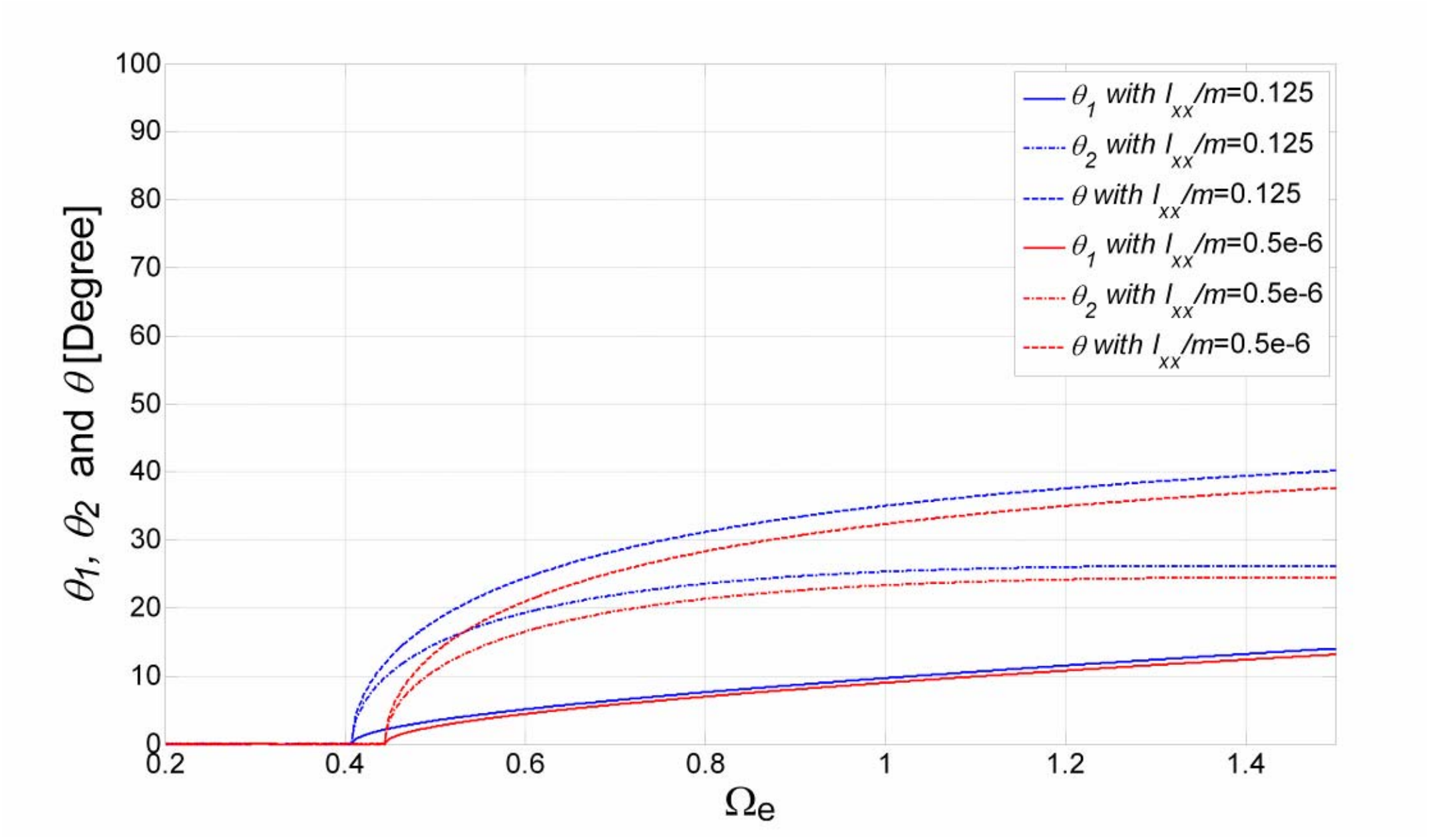

**Fig. 9.** The curves of $\theta$, $\theta_1$ and $\theta_2$ with respect to the angular velocity $\Omega_e$ with $J_2 = -0.5$

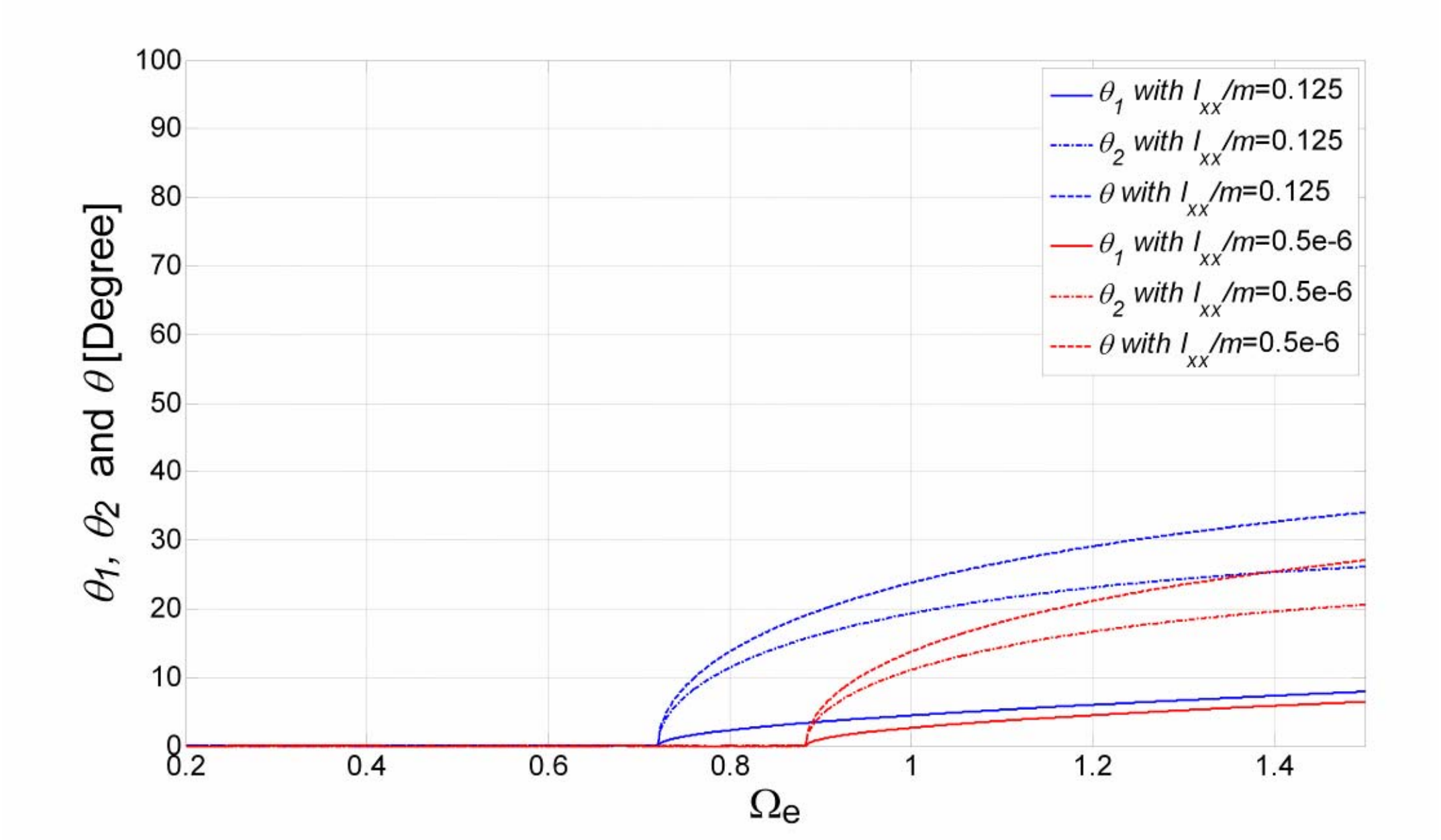

**Fig. 10.** The curves of $\theta$, $\theta_1$ and $\theta_2$ with respect to the angular velocity $\Omega_e$ with $J_2 = -0.2$

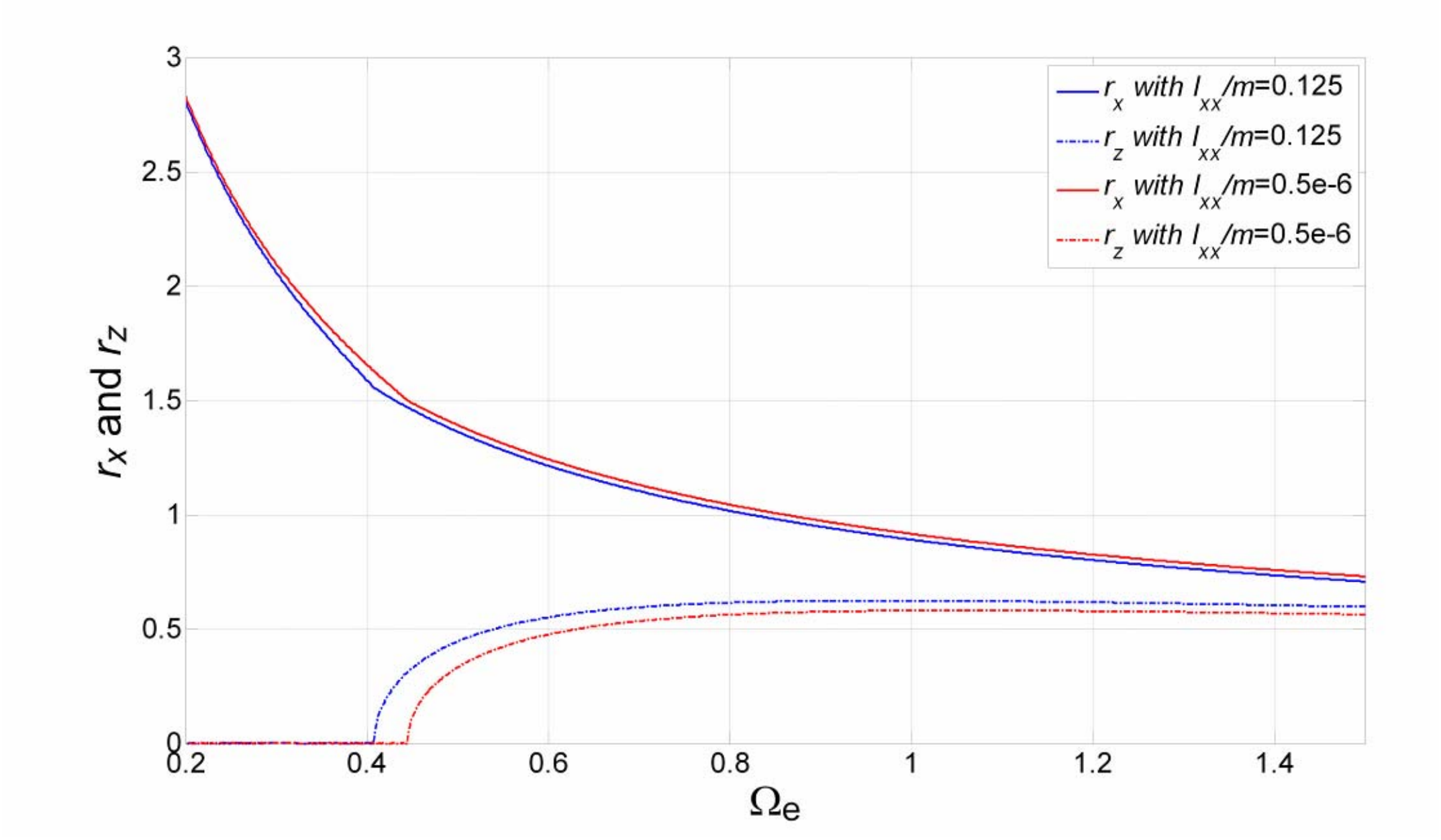

**Fig. 11.** The curves of $r_e^x$ and $r_e^z$ with respect to the angular velocity $\Omega_e$ with $J_2 = -0.5$

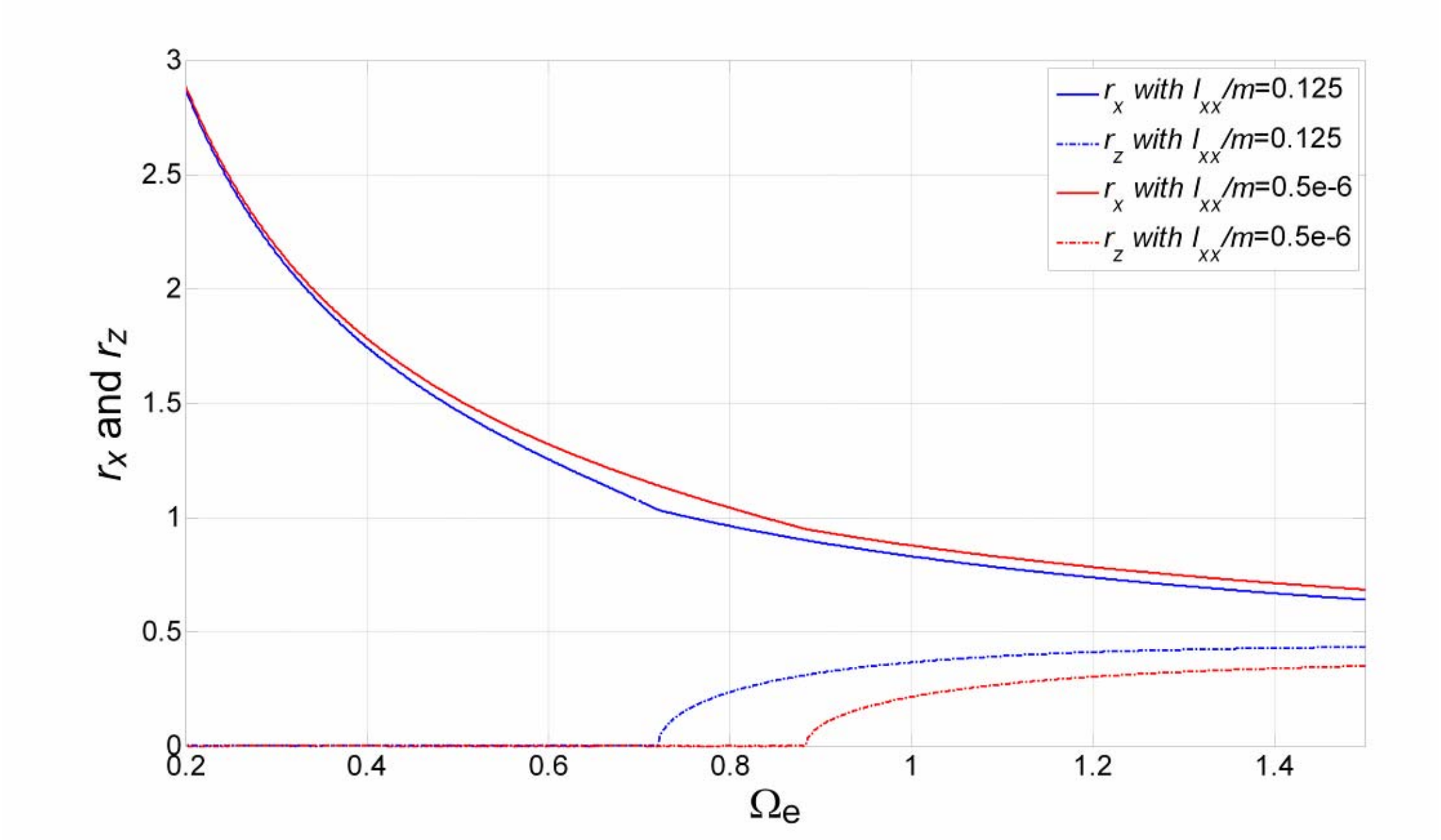

**Fig. 12.** The curves of $r_e^x$ and $r_e^z$ with respect to the angular velocity $\Omega_e$ with $J_2 = -0.2$

From the results in Figs. 7-12, we can easily find that as the orbital angular velocity $\Omega_e$ increasing, there is a bifurcation from the classical relative equilibria, at which the non-classical relative equilibria appear. The bifurcation takes place at a smaller value of the orbital angular velocity $\Omega_e$ in the case of a larger absolute value of the negative $J_2$. This is consistence with our conclusions in the Section 4.2 that the elongatedness of the central body $P$ has a positive effect on the existence of the non-classical relative equilibria.

It is also easily found that at the same value of $J_2$, the characteristic dimension of the rigid body $I_{xx}/m$ has a significant effect on the bifurcation and properties of the non-classical relative equilibria. The effect of the characteristic dimension $I_{xx}/m$ is more significant in the case of $J_2 = -0.2$ than in the case of $J_2 = -0.5$ that is due to the fact that the effect of the second zonal harmonic $J_2$ is more dominative in the case of $J_2 = -0.5$.

In our studies, the range of the zonal harmonic is chosen as $-0.5 < J_2 < 0.5$, same as in Broucke (1994). The body $P$ with a negative second zonal harmonic $J_2$ has been considered. However, the bodies with a negative $J_2$ have not been discovered in our Solar System yet, since a negative $J_2$ means that the body is rotating uniformly around its minimum-moment principal axis. This rotation state is unstable and will evolve to rotating around the maximum-moment principal axis eventually in the presence of energy dissipation. Although the existence of the body with a negative $J_2$ is doubtful at present, our results are of interest and value for the theoretical studies on the related problems in the celestial mechanics and astrophysics. Perhaps, the body with a negative $J_2$ could appear in an artificial system or in other places of the universe in the future astrophysics.

The linear and nonlinear stability of the classical type of relative equilibria of the rigid body have been investigated in the framework of geometric mechanics in Wang and Xu (in press b). The linear and nonlinear stability of the non-classical type of relative equilibria will be studied in the future. The dynamical behaviors near the non-classical type of relative equilibria, such as the displaced orbit with a little inclination and its precession, are also of great interest and worthy of detailed studies in the future.

## 5 Conclusions

The existence and properties of both the classical and non-classical relative equilibria of a rigid body in a $J_2$ gravity field have been investigated in details in the present paper.

The existence condition of the classical relative equilibria and the range of the parameters of the system have been discussed comprehensively. We have found that the classical relative equilibria can always exist in the real physical situation. The curves of the angular velocity with respect to the orbital radius with different values of the system parameters have also been given.

The existence and properties of the non-classical relative equilibria have also been studied in details. The numerical results suggested that the non-classical relative equilibria can only exist in the case of a negative $J_2$, i.e., the central body is elongated; they cannot exist in the case of a positive $J_2$ when the central body is oblate. In the case of a negative $J_2$, the effect of the orbit-rotation coupling of the rigid body on the existence of the non-classical relative equilibria can be positive or negative, which depends on the values of $J_2$ and $\Omega_e$. When the values of $J_2$ and $\Omega_e$ favor the existence of the non-classical relative equilibria, i.e., a negative $J_2$ with a large absolute value and a large $\Omega_e$, the orbit-rotation coupling has a negative effect on the existence of the non-classical relative equilibria. Whereas when the values of $J_2$ and $\Omega_e$ do not favor the existence of the non-classical relative equilibria, i.e., a negative $J_2$ with a small absolute value and a small $\Omega_e$, the orbit-rotation coupling has a positive effect on the existence of the non-classical relative equilibria.

We have also found that the mass distribution of the rigid body near to a rod along the ***j***-axis or ***k***-axis favor the existence of the non-classical relative equilibria, whereas the mass distribution of the rigid body near to a rod along the ***i***-axis, do not favor the existence of the non-classical relative equilibria.

The details of the non-classical relative equilibria have also been given, including the curves of $R_e^x$, $R_e^z$, $\theta$, $\theta_1$, $\theta_2$, $r_e^x$ and $r_e^z$ with respect to the orbital angular velocity $\Omega_e$. The bifurcation from the classical relative equilibria, at which the non-classical relative equilibria appear, has been shown clearly in these curves. It has been found that the bifurcation takes place at a smaller value of the orbital angular velocity $\Omega_e$ in the case of a larger absolute value of the negative $J_2$, and the characteristic dimension of the rigid body $I_{xx}/m$ has a significant effect on the bifurcation and properties of the non-classical relative equilibria.

Our results in the present paper are consistent with the previous results on the equilibrium point of a point mass in a non-central gravity field. We have also extended the previous results on the relative equilibria of a rigid body in a central gravity field by taking into account the oblateness of the central body.

The stability of the non-classical relative equilibria, as well as the dynamical behaviors near the non-classical relative equilibria, such as the displaced orbit with a little inclination and its precession, are of great interest and worthy of detailed studies in the future.

## Acknowledgements

This work is supported by the Innovation Foundation of BUAA for PhD Graduates, the Graduate Innovation Practice Foundation of BUAA under Grant YCSJ-01-201306.